\documentclass[twocolumn,aps,superscriptaddress,prb]{revtex4-1}
\setcitestyle{numbers,square}
\usepackage{amsmath,amssymb}
\usepackage{tabularx}
\usepackage{graphicx}% Include figure files \usepackage{dcolumn}% Align table columns on decimal point
\usepackage{bmpsize}
\usepackage{bm}% bold math
\usepackage{color}
\usepackage{amssymb}
\usepackage[version=3]{mhchem}
\usepackage{newtxmath}
\usepackage{braket}
\usepackage[normalem]{ulem} %for strikethrough text
\usepackage{float}
\usepackage{here}
\DeclareMathAlphabet{\mathpzc}{OT1}{pzc}{m}{it}
\newcommand{\nn}{\nonumber}
\def\diff{\mathrm d}
\def\sgn{\mathrm {sgn}}

\def\i{\mathrm{i}}
\def\e{\mathrm{e}}

\newcommand{\red}[1]{\textcolor{red}{ #1}}

\begin{document}
%\preprint{APS/123-QED}
%%%%%%%%%%%%%%%%%%%%%%%%%%%%%%%%%%%%%%%%%%%%%%%%%%%%%%%%%%%%%%%%%%%%%%%%%%%%%%%%%%%%%%%%%%%%%%%%%%%%%%%%%%%%%%%%%%%%%%%
\title{Moir\'e Band Engineering in Twisted Trilayer  WSe$_{2}$}
%\title{\blue{Structural Relaxation and Moir\'e Band Engineering in Twisted Trilayer  WSe$_{2}$}}
%\title{
%{The electronic properties and lattice relaxation for general twisted trilayer WSe$_{2}$}
%}

\author{Naoto Nakatsuji}
\affiliation{Department of Physics, Osaka University, Toyonaka, Osaka 560-0043, Japan}
\author{Takuto Kawakami}
\affiliation{Department of Physics, Osaka University, Toyonaka, Osaka 560-0043, Japan}
\author{Hayato Tateishi}
\affiliation{Department of Applied Chemistry, Kyushu University, Motooka, Fukuoka 819-0395, Japan}
\author{Koichiro Kato}
\affiliation{Department of Applied Chemistry, Kyushu University, Motooka, Fukuoka 819-0395, Japan}
\author{Mikito Koshino}
\affiliation{Department of Physics, Osaka University, Toyonaka, Osaka 560-0043, Japan}
\date{\today}
%%%%%%%%%%%%%%%%%%%%%%%%%%%%%%%%%%%%%%%%%%%%%%%%%%%%%%%%%%%%%%%%%%%%%%%%%%%%%%%%%%%%%%%%%%%%%%%%%%%%%%%%%%%%%%%%%%%%%%%

%%%%%%%%%%%%%%%%%%%%%%%%%%%%%%%%%%%%%%%%%%%%%%%%%%%%%%%%%%%%%%%%%%%%%%%%%%%%%%%%%%%%%%%%%%%%%%%%%%%%%%%%%%%%%%%%%%%%%%%
\begin{abstract}

We present a systematic theoretical study on the structural and electronic properties of twisted trilayer transition metal dichalcogenide (TMD) WSe$_2$, where two independent moiré patterns form between adjacent layers. Using a continuum approach, we investigate the optimized lattice structure and the resulting energy band structure, revealing fundamentally different electronic behaviors between helical and alternating twist configurations.
In helical trilayers, lattice relaxation induces $\alpha\beta$ and $\beta\alpha$ domains, where the two moiré patterns shift to minimize overlap, while in alternating trilayers, $\alpha\alpha'$ domains emerge with aligned moiré patterns. 
%The electronic structure is primarily governed by intralayer potentials rather than interlayer hybridization, resulting in characteristic layer-polarized states near the valence band edge.
A key feature of trilayer TMDs is the summation of moiré potentials from the top and bottom layers onto the middle layer, effectively doubling the potential depth. In helical trilayers, this mechanism generates a Kagome lattice potential in the $\alpha\beta$ domains, giving rise to flat bands characteristic of Kagome physics. In alternating trilayers, the enhanced potential confinement forms deep triangular quantum wells, distinct from those found in bilayer systems.
Beyond these bulk states arising from the commensurate domains, the global moiré-of-moiré electronic spectrum exhibits isolated boundary modes localized at domain intersections.
%The global moiré-of-moiré electronic spectrum consists of bulk states arising from commensurate domains and isolated boundary modes localized at domain intersections.
Furthermore, we demonstrate that a moderate perpendicular electric field can switch the layer polarization near the valence band edge, providing an additional degree of tunability. In particular, it enables tuning of the hybridization between orbitals on different layers, allowing for the engineering of diverse and controllable electronic band structures.
Our findings highlight the unique role of moiré potential summation in trilayer systems, offering a broader platform for designing moiré-based electronic and excitonic phenomena beyond those achievable in bilayer TMDs.

\end{abstract}
%%%%%%%%%%%%%%%%%%%%%%%%%%%%%%%%%%%%%%%%%%%%%%%%%%%%%%%%%%%%%%%%%%%%%%%%%%%%%%%%%%%%%%%%%%%%%%%%%%%%%%%%%%%%%%%%%%%%%%%

\maketitle
%%%%%%%%%%%%%%%%%%%%%%%%%%%%%%%%%%%%%%%%%%%%%%%%%%%%%%%%%%%%%%%%%%%%%%%%%%%%%%%%%%%%%%%%%%%%%%%%%%%%%%%%%%%%%%%%%%%%%%%
\section{introduction}
\label{sec_intro}

Moir\'e twisted materials have garnered significant attention as platforms for novel quantum phenomena. Twisted bilayer graphene (tBG) serves as a fundamental example, where the moir\'e-scale modulation of interlayer coupling hybridizes the Dirac cones of monolayer graphene, leading to the formation of flat bands \cite{bistritzer2011moirepnas,PhysRevB.84.035440,PhysRevB.86.155449}. These flat bands drive various correlated states 
\cite{cao2018_80,cao2018_43,doi:10.1126/science.aav1910,Kerelsky2019,xie2019spectroscopic,jiang2019charge,polshyn2019large,Choi2019,doi:10.1126/science.aaw3780,lu2019superconductors,PhysRevLett.124.076801,doi:10.1126/science.aay5533,chen2020tunable,saito2020independent,zondiner2020cascade,wong2020cascade,stepanov2020untying,arora2020superconductivity,PhysRevLett.127.197701}, including superconductivity, correlated insulator phases, and the anomalous Hall effect.

Beyond graphene-based systems, extensive studies have focused on semiconductor moir\'e systems, particularly transition metal dichalcogenides (TMDs) \cite{PhysRevLett.121.026402,PhysRevResearch.2.033087,PhysRevB.102.201104,PhysRevLett.127.096802,PhysRevB.104.075150,PhysRevB.104.115154,PhysRevX.12.021064,PhysRevResearch.5.L012005,PhysRevLett.132.146401,PhysRevB.110.165128,PhysRevB.100.060506,PhysRevB.106.235135,PhysRevResearch.5.L012034,PhysRevLett.130.126001,PhysRevLett.131.056001,PhysRevB.108.064506,PhysRevB.108.155111,PhysRevB.110.035143,kim2025theory,PhysRevB.104.195134,PhysRevB.106.235135,PhysRevResearch.5.L012034,PhysRevLett.130.126001,PhysRevLett.131.056001,PhysRevB.108.064506,PhysRevB.108.155111,PhysRevB.110.035143,wang2020correlated,D0NH00248H,xia2025superconductivity,guo2025superconductivity,kim2025theory,PhysRevLett.118.147401,yu2017moire,PhysRevB.97.035306,seyler2019signatures,seyler2019signatures,jin2019observation,tran2019evidence,PhysRevB.99.125424,huang2022excitons,doi:10.1021/acs.nanolett.3c04427,PhysRevLett.122.086402,10.1093/nsr/nwz117,li2021quantum,PhysRevResearch.2.033087,devakul2021magic,PhysRevLett.128.026402,cai2023signatures,zeng2023thermodynamic,park2023observation,PhysRevX.13.031037,PhysRevLett.132.096602,doi:10.1126/science.adi4728,PhysRevB.109.205121,zhang2024polarization,PhysRevB.110.035130,PhysRevResearch.3.L032070,PhysRevB.107.L201109,cai2023signatures,zeng2023thermodynamic,PhysRevB.108.085117,park2023observation,PhysRevResearch.5.L032022,PhysRevX.13.031037,PhysRevLett.131.136502,PhysRevLett.132.036501,PhysRevB.109.045147,doi:10.1073/pnas.2316749121,PhysRevB.109.085143,PhysRevB.109.L121107,chen2025robust,PhysRevB.111.125122,paul2025shining,kang2025time,kousa2025theory,shi2025effects,zaklama2024structure,reddy2024anti,tang2020simulation,regan2020mott,zhang2020flat,PhysRevLett.124.206101,xu2020correlated,doi:10.1073/pnas.2021826118,li2021imaging,wang2022interfacial,enaldiev2022scalable,doi:10.1126/science.adg4268,PhysRevB.110.115114,song2022deep,enaldiev2022self,PhysRevB.108.165152,soltero2024competition,cavicchi2024optical}. In contrast to tBG, moir\'e TMD bilayers exhibit a dominant moir\'e intralayer potential, which effectively transforms the system into a triangular array of weakly coupled quantum dots. This provides an ideal platform for investigating Hubbard model physics \cite{PhysRevLett.121.026402,PhysRevResearch.2.033087,PhysRevB.102.201104,PhysRevLett.127.096802,PhysRevB.104.075150,PhysRevB.104.115154,PhysRevX.12.021064,PhysRevResearch.5.L012005,PhysRevLett.132.146401,PhysRevB.110.165128,PhysRevB.100.060506,PhysRevB.106.235135,PhysRevResearch.5.L012034,PhysRevLett.130.126001,PhysRevLett.131.056001,PhysRevB.108.064506,PhysRevB.108.155111,PhysRevB.110.035143,kim2025theory}, superconductivity
 \cite{PhysRevB.100.060506,PhysRevB.104.195134,PhysRevB.106.235135,PhysRevResearch.5.L012034,PhysRevLett.130.126001,PhysRevLett.131.056001,PhysRevB.108.064506,PhysRevB.108.155111,PhysRevB.110.035143,wang2020correlated,D0NH00248H,xia2025superconductivity,guo2025superconductivity,kim2025theory}, 
and unique optical properties such as exciton confinement \cite{PhysRevLett.118.147401,yu2017moire,PhysRevB.97.035306,seyler2019signatures,seyler2019signatures,jin2019observation,tran2019evidence,PhysRevB.99.125424,huang2022excitons,doi:10.1021/acs.nanolett.3c04427}. Additionally, the interplay of moir\'e intralayer potential and interlayer hopping, as seen in MoTe$_2$, can generate an effective magnetic field from the non-trivial topological nature, leading to the anomalous Chern insulator \cite{PhysRevLett.122.086402,10.1093/nsr/nwz117,li2021quantum,PhysRevResearch.2.033087,devakul2021magic,PhysRevLett.128.026402,cai2023signatures,zeng2023thermodynamic,park2023observation,PhysRevX.13.031037,PhysRevLett.132.096602,doi:10.1126/science.adi4728,PhysRevB.109.205121,zhang2024polarization,PhysRevB.110.035130} and fractional Chern insulator phases \cite{PhysRevLett.132.096602,PhysRevB.110.035130,PhysRevResearch.3.L032070,PhysRevB.107.L201109,cai2023signatures,zeng2023thermodynamic,PhysRevB.108.085117,park2023observation,PhysRevResearch.5.L032022,PhysRevX.13.031037,PhysRevLett.131.136502,PhysRevLett.132.036501,PhysRevB.109.045147,doi:10.1073/pnas.2316749121,PhysRevB.109.085143,PhysRevB.109.L121107,chen2025robust,PhysRevB.111.125122,paul2025shining,kang2025time,kousa2025theory,shi2025effects,zaklama2024structure,reddy2024anti}.

    \begin{figure}
        \begin{center}
        \leavevmode\includegraphics[width=1. \hsize]{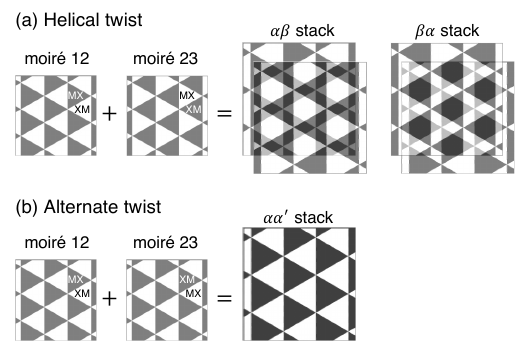}
    \caption{
        Schematic figure of the real-space distribution of the potential on the middle layer of twisted trilayer TMD 
        with (a) helical and (b) alternate twist.
        On each row, the left two panels show the contributions from the moir\'e pattern between layers 1 and 2 (moir\'e 12) and between layers 2 and 3 (moir\'e 23), where darker regions indicate higher potential energy.
The right panel displays the total potential in the commensurate domain, obtained by summing the moir\'e 12 and 23 potentials with the corresponding lateral shift.
        }
        \label{schematic_figure}
        \end{center}
    \end{figure}

Recent research has expanded to moir\'e trilayer materials, with twisted trilayer graphene (tTG) being extensively studied as a prototypical system \cite{PhysRevLett.123.026402,li2019electronic,doi:10.1021/acs.nanolett.9b04979,tritsaris2020electronic,PhysRevResearch.2.033357,PhysRevLett.127.026401,PhysRevB.104.035139,PhysRevB.104.L121116,PhysRevB.103.195411,PhysRevB.104.115167,PhysRevResearch.4.L012013,doi:10.1126/sciadv.adi6063,PhysRevX.12.021018,doi:10.1126/science.abn8585,doi:10.1126/science.abg0399,park2021tunable,Cao2021,Kim2022,PhysRevB.101.224107,PhysRevLett.125.116404,lin2020heteromoire,PhysRevLett.127.166802,gao2022symmetry,Ma2022,PhysRevB.105.195422,uri2023superconductivity,PhysRevB.107.125423,PhysRevB.108.L081124,PhysRevB.106.075423,meng2023commensurate}. In general trilayer systems, the moir\'e patterns formed between layers 1 and 2 (moir\'e 12) and layers 2 and 3 (moir\'e 23) interfere to create a higher-order moir\'e-of-moir\'e pattern with a typical length scale of tens to hundreds of nanometers \cite{PhysRevB.101.224107,lin2020heteromoire,PhysRevLett.127.166802,PhysRevLett.125.116404,doi:10.1126/science.abk1895,craig2024local}. In tTG, structural relaxation induces commensurate domains where the two moir\'e periodicities locally match, hosting nontrivial topological boundary modes \cite{doi:10.1126/science.abk1895,li2022symmetry,craig2024local,PhysRevB.101.224107,shin2021electron,PhysRevB.106.075423,meng2023commensurate,PhysRevX.13.041007,park2024tunable}.

Extending these concepts to trilayer TMDs, previous studies have investigated commensurate trilayers where the two moir\'e periods align \cite{PhysRevB.108.155106,PhysRevB.111.085412,PhysRevB.111.125410,fedorko2025engineeringmoirekagomesuperlattices}. Here, we examine twisted trilayers of WSe$_2$ with general twist angles, where two moir\'e patterns may have different periodicities. We obtain the optimized lattice structure under relaxation and calculate the electronic band structure near the valence band maximum. The system is characterized by twist angles $\theta_{12}$ and $\theta_{23}$, and we focus on nearly parallel trilayers with $|\theta_{12}|, |\theta_{23}| \sim 1^\circ$. The physical properties vary strikingly between the helical stack ($\theta_{12}\theta_{23} > 0$) and the alternating stack ($\theta_{12}\theta_{23} < 0$).

The lattice relaxation exhibits trends similar to tTG: in helical trilayers, $\alpha\beta$ and $\beta\alpha$ domains emerge where the moir\'e lattices shift to minimize overlap, whereas in alternating trilayers, $\alpha\alpha'$ domains form where the two moir\'e lattices align [See, Fig.~\ref{atomic_lattice_structure}]. However, the electronic structures differ significantly from tTG. Due to lattice relaxation, moir\'e interlayer hopping is suppressed, making the band structure primarily determined by the intralayer moir\'e potentials. The middle layer (layer 2) experiences the combined potential of layers 1 and 3, resulting in a potential twice as large as that of the outer layers. Consequently, the states near the valence band edge are strongly localized in the middle layer.

The spatial configuration of the middle-layer potential exhibits greater diversity than in bilayers. 
Figure \ref{schematic_figure} represents the schematic figure of the real-space distribution of the middle-layer potential for (a) helical and (b) alternate twist,
where dark region indicates a high potential region.
In the helical case, the summation of two triangular potentials from layers 1 and 3 with a spatial translation naturally generates a Kagome lattice potential in $\alpha\beta$, leading to flat bands characteristic of Kagome lattices. In alternating stack, the potentials from layers 1 and 3 coincide in $\alpha\alpha'$ domains, forming triangular quantum wells twice as deep as those in twisted bilayers.
In the full moir\'e-of-moir\'e band structure, the spectrum consists of cluster of bulk states from $\alpha\beta$, $\beta\alpha$, and $\alpha\alpha'$ domains, along with isolated boundary modes localized at domain corners. 

%\mage{
%Figure \ref{schematic_figure} represents the schematic figure of the real-space distribution of the intralayer potential of middle layer for $\alpha\beta$, $\beta\alpha$ and $\alpha\alpha'$ stacking, where dark region indicates a high potential region.
%Moir\'e 23 gives almost $180^\circ$ rotated potential for helical and alternate twist, and above local stacking has the different lateral shift, resulting the variaties of the intralayer potential of the middle layer as shown Fig.~\ref{schematic_figure}.
%}

%Furthermore, the layer polarization of valence band states can be switched by applying a moderate vertical electric field.

In addition to the intrinsic structural degrees of freedom, twisted trilayer TMDs offer rich opportunities for band structure engineering through external electric fields.
By applying a perpendicular electric field, the relative potential of each layer can be selectively tuned, allowing control over the layer polarization of electronic states and the interlayer hybridization between orbitals.
This enables the realization of diverse band structures, including graphene-like Dirac bands, flat bands, quadratic band touchings, and hybridized bands formed from %\mage{$s$-, $p-$ and $d$-like} 
various types of orbitals residing on different triangular lattices.
Such electrically tunable systems provide a versatile platform to explore correlated electronic phases and topological phenomena in moir\'e materials.

%The summation of two moir\'e potentials is a distinctive feature of trilayer systems, offering a broader platform beyond bilayer TMDs. This enables greater flexibility in designing spatial moir\'e potentials, including Kagome potentials, and enhances potential depth due to the sandwich-like structure, which may benefit moir\'e exciton coherence. These findings open new directions for exploring electronic and optical properties in twisted trilayer TMDs.

%\blue{
%The summation of two independent moir\'e potentials is a distinctive feature of twisted trilayer systems, offering a versatile platform far beyond bilayer TMDs.
%This trilayer architecture enables flexible design of spatial moir\'e potentials—such as triangular, honeycomb, or Kagome patterns—with enhanced potential depth.
%Moreover, the application of a perpendicular electric field introduces an additional tuning knob to control the layer polarization and hybridization of localized orbitals across different layers, giving rise to a rich variety of unconventional band structures, including Dirac cones, flat bands, and orbital-selective hybridization.
%}

 The paper is organized as follows.
In Sec.~\ref{sec_model}, we define the lattice geometry of twisted trilayer TMD and introduce the continuum method used to calculate the optimized lattice structure and electronic properties.
We calculate the optimized lattice structure and its electronic band structure for helical twist in Sec.~\ref{sec_helical}, and show the kagome flat band and higher topological flat bands appear on the local $\alpha\beta$ and $\beta\alpha$ stack.
In Sec.~\ref{sec_alternate} we study alternate twist case, and found the quantum dot states captured the trigonal shape potential.
The dependence of the perpendicular electric field is discussed in Sec.~\ref{sec_effect_electric_field}, and we found the various band structures formed from the $s$-, $p$- and $d$-like orbitals of triangular wells.
A brief conclusion is provided in Sec.~\ref{sec_con}.

\section{Model}
\label{sec_model}

\subsection{Geometry of twisted trilayer TMD}
\label{subsec_geometry}

\begin{figure*}
 \begin{center}
\leavevmode\includegraphics[width=0.8 \hsize]{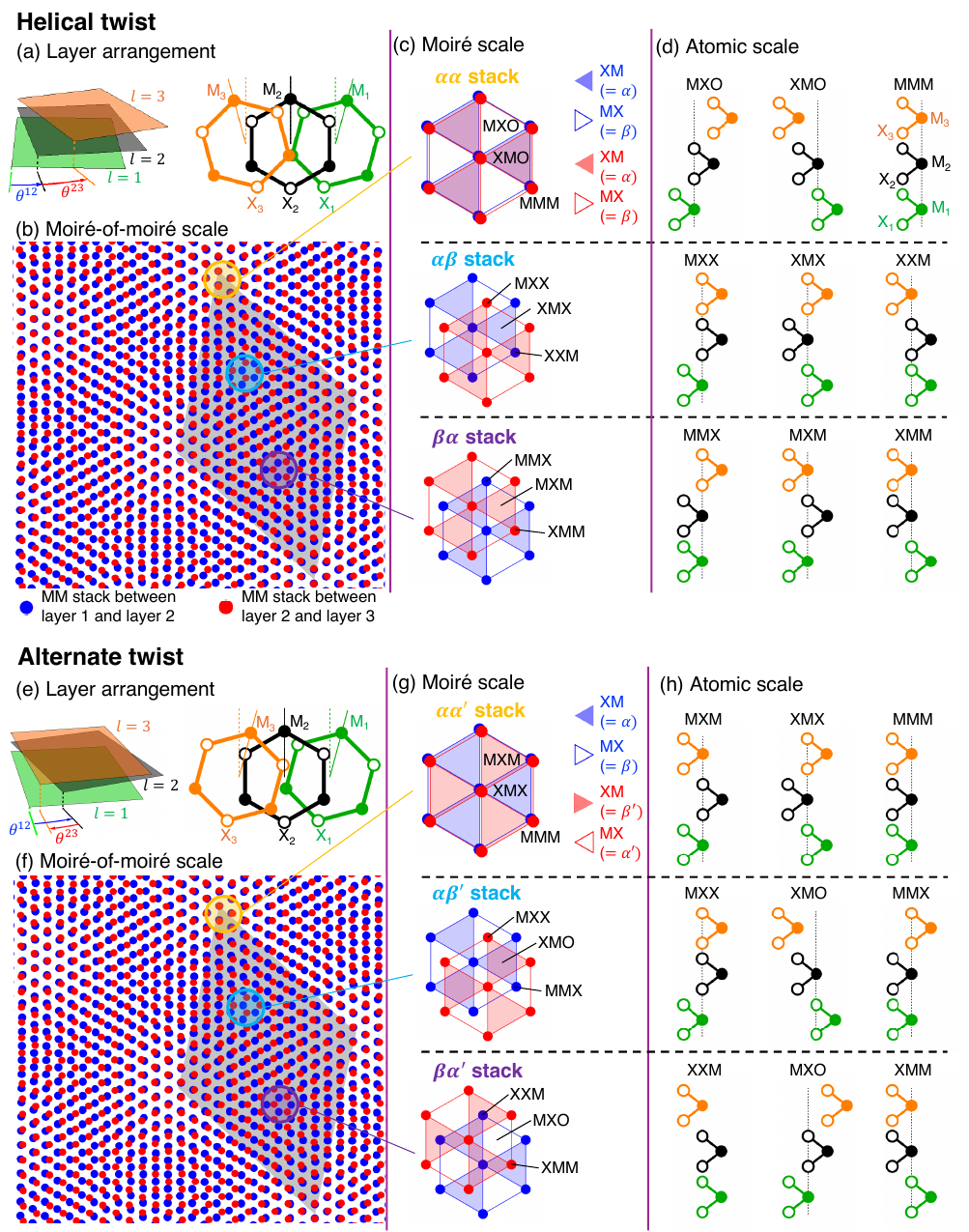}
\caption{
(a) Stacking geometry of a twisted homo-trilayer TMD with a helical twist. Green, black, and orange sheets represent layers 1, 2, and 3, respectively.
(b) Moir\'e pattern arrangement in the helical twist trilayer. Blue and red dots indicate the MM points of moir\'e 12 (formed between layers 1 and 2) and moir\'e 23 (formed between layers 2 and 3), respectively. The gray rhombus marks a moir\'e-of-moir\'e unit cell.
(c) Representative local moir\'e structures in (b), where empty and filled triangles denote MX and XM domains, respectively.
(d) Side view of the local atomic configurations at specific points in (c).
(e–h) Corresponding structures and atomic configurations for the alternate twist case.
}

        \label{atomic_lattice_structure}
        \end{center}
    \end{figure*}

    Figure \ref{atomic_lattice_structure} illustrates the geometry of a non-relaxed twisted trilayer TMD of helical (upper panels) and alternate stackings (lower panels).
    Here, $\theta^{12}$ denotes the twist angles from layer 1 to 2, and $\theta^{23}$ is from layer 2 to 3.
    The helical and alternate twists correspond to structures with
    $\theta^{12} \theta^{23} >0$ and $\theta^{12} \theta^{23}<0$, respectively.       
    In Figs.~\ref{atomic_lattice_structure}(b) and (f), we show 
    a schematic figure to illustrate an interference of the two moir\'e patterns for these structures, %for helical and alternate twists, respectively, 
    where blue and red dots represent the MM stacking point of the moir\'{e} 12 (the interference pattern from layer 1 and 2) and the moir\'{e} 23 (layer 2 and 3).
     We define the symbol M, X and O as the transition-metal atom site, and the chalcogen atom site and the vacant site (the center of hexagon), respectively, in a unit cell of monolayer TMD.
    The MM stacking stands for the arrangement of neighboring layers with M sites vertically aligned.

    The interference of the two moir\'{e} pattens give rise to a moir\'{e}-of-moir\'{e} superstructure with a period indicated by a gray region.
   The local structure approximates a stack of two commensurate moir\'{e} lattices with a specific lateral shift.
    For the helical case,
    we define three local arrangements $\alpha\alpha$, $\alpha\beta$ and $\beta\alpha$,
    as shown in Fig.~\ref{atomic_lattice_structure}(c).
    Here the filled and open triangles represent the XM and MX domain, respectively, in the individual moir\'e patterns.
    %\mage{, where we note that the filled triangle in this figure represents the XM stacking not the high intralayer potential region like Fig.~\ref{schematic_figure}.
    %In Figs 1 and 2, the filled triangles of moire23 are inverted because, as we show later, the layer 2 feels the intralayer potential of the opposite sign when the two moire structures have locally identical structures.}
    In $\alpha\alpha$ structure, the MM stacking points of moir\'{e} 12 and those of 23 are aligned, while the $\alpha\beta$ and $\beta\alpha$ stacking take structures with MM stacking points repelled from each other. %Specifically, in $\alpha\beta$ ($\beta\alpha$), the MX point of moir\'e 12 (23) overlaps with the MM point of moir\'e 23 (12). 
    %The $\alpha\beta$ takes the lattice structure overlapped the MX stacking of moir\'e 12 and the MM stacking of moir\'e 23, while on $\beta\alpha$, the MM stacking of moir\'e 12 is placed on the MX stacking of moir\'e 12 as shown in the middle and bottom figures of Fig.~\ref{atomic_lattice_structure}(c).
    Figure \ref{atomic_lattice_structure}(d) shows the local atomic configurations in $\alpha\alpha/\alpha\beta/\beta\alpha$ structures.
    For instance, MXO represents a stacking where the M site of layer 1, the X site of layer 2, and the O site of layer 3 are vertically aligned.
    Note that, unlike trilayer graphene, the $\alpha\beta$ and $\beta\alpha$ structures are not related by a $C_{2z}$ operation, as a monolayer TMDs lacks the $C_{2z}$ symmetry. 
    %We note that the $\alpha\beta$ and $\beta\alpha$ stacking are not connected by the $C_{2z}$ symmetry unlike to TTG.
    %This is because the monolayer TMD constructed by metal and chalcogen atoms does not have the $C_{2z}$ symmetry, and the $\alpha\beta$ and $\beta\alpha$ stacking locally has the different atomic arrangements as shown in Fig.~\ref{atomic_lattice_structure}(c).
    For the alternate twist, the system hosts three local moir\'e structure; $\alpha\alpha'$, $\alpha\beta'$ and $\beta\alpha'$ as shown in Figs.~\ref{atomic_lattice_structure}(g) and \ref{atomic_lattice_structure}(h).

    %Compared to the helical case, the moir\'e 23 in the alternate case has a mirror reflected structure with respect to $y$ axis, resulting in completely different local atomic arrangements, as shown in  Fig.~\ref{atomic_lattice_structure}(h).
   
   % Since $\theta_{12}$ and $\theta_{23}$ have opposite signs, the moir\'e 23 has the atomic structure rotated moire 12 to 180 degrees along the y-axis.
%    As the result of this, the atomic lattice configurations are totaly different even the relative position of the MM stacking of each moir\'e are similar to helical twist.
%    As shown in Fig.~\ref{atomic_lattice_structure}(g), $\alpha\alpha'$ stacking aligns the MM stacking of two moir\'e patterns, and for $\alpha\beta'$ ($\beta\alpha'$) stacking, the MX (MM) stacking of moir\'e 12 is placed on the MM (XM) stacking of moir\'e 23.

In the following, we present a precise description of the geometric configuration for trilayer TMD systems. 
For TMD, we consider WSe$_2$ throughout the paper.
The primitive lattice vectors a monolayer are given by $\bm{a}_{1}=a(1,0)$ and $\bm{a}_{2}=a(1/2,\sqrt{3}/2)$,
where $a\approx 0.33$ nm is a lattice constant \cite{mounet2018two}.
The lattice vectors of layer $\ell(=1,2,3)$ in the twisted trilayer are written as  $\bm{a}^{(\ell)}_{j}=R(\theta^{(\ell)})\bm{a}_{j}\, (j=1,2)$, where $R(\theta)$ is a two-dimensional rotation matrix, and $\theta^{(\ell)}$ is the absolute twist angle of layer $l$ specified by $\theta^{(1)}=-\theta^{12}$, $\theta^{(2)}=0$ and $\theta^{(3)}=\theta^{23}$.
    Accordingly, the reciprocal lattice vector of layer $l$ is given by $\bm{b}^{(\ell)}_{j}=R(\theta^{(\ell)})\bm{b}_{j}$, where $\bm{b}_{1}=b(\sqrt{3}/2,-1/2)$ and $\bm{b}_{2}=b(0,1)$ with $b=4\pi/(\sqrt{3}a)$. 
    The Brillouin zone (BZ) corners are located at $\bm{K}^{(\ell)}_{\xi} = -\xi (2\bm{b}^{(\ell)}_{1} + \bm{b}^{(\ell)}_{2}) / 3$ where $\xi = \pm 1$ is the valley index.
%In this study, we focus on monolayer TMDs with a valence band maximum located at the BZ corners, such as MoTe$_2$ and WSe$_2$.

    The reciprocal lattice vectors of the moir\'e pattern 12 and 23 are defined by $\bm{G}_j^{\ell\ell'}=\bm{b}^{(\ell')}_{j}-\bm{b}^{(\ell)}_{j}$ where $(\ell,\ell')=(1,2)$ and $(2,3)$, respectively. Accordingly, the moir\'e lattice vectors in the real space are given by $\bm{G}^{\ell\ell'}_{i}\cdot\bm{L}^{\ell\ell'}_{j}=2\pi\delta_{ij}$,
    and the corresponding moir\'e lattice period is written as $|\bm{L}^{\ell\ell'}_{j}|= 4\pi/(\sqrt{3}|\bm{G}^{\ell\ell'}_{j}|) = a/2\sin(|\theta^{\ell\ell'}|/2)$.  
    When $\bm{G}^{12}_j$ and $\bm{G}^{23}_j$ are close to each other, the interference of the two moir\'e patterns gives rise to a moir\'e-of-moir\'e pattern with reciprocal lattice vectors $\bm{\mathcal{G}}_j=\bm{G}^{12}_j-\bm{G}^{23}_j$.
    We can define the associate periods $\bm{\mathcal{L}}_j$ in the real space.

    In this paper, we consider trilayer systems where the two moir\'e patterns have a common periodicity for computational purposes.
    In this case, the period of the whole system can be expressed in terms of the individual moir\'e periods as \cite{PhysRevX.13.041007}
    \begin{align} \label{eq_com_condition}
        \bm{L}_{1} &= n \bm{L}^{12}_{1} + m \bm{L}^{12}_{2} = n' \bm{L}^{23}_{1} + m' \bm{L}^{23}_{2}, \nn \\
        \bm{L}_{2} &= R(60^\circ)\bm{L}_1,
    \end{align}
    with intergers $n, m, n',$ and $m'$.
    Solving the above equation for the $\theta^{12}$ and $\theta^{23}$, we obtain 
    \begin{align}\label{eq_angle_formulas}
       \theta^{12} = \theta(n,m,n',m'), \quad 
       \theta^{23} = - \theta(n',m',n,m),
    \end{align}
    where 
    \begin{align}\label{eq_angle_formulas_2}
     &\theta(n,m,n',m') =\notag\\
     & \quad 2 \tan^{-1}\frac{\sqrt{3}\left\{m \left(2n'+m'\right)-\left(2n+m\right)m'\right\}}{\left(2n+m\right)\left(2n'+m'\right)+3mm'+\left(2n'+m'\right)^{2}+3m'^{2}}.
    \end{align}
%    The spatial period is given by 
%    $L = L^{12}\sqrt{n^{2}+m^{2}+nm}= L^{23}\sqrt{n'^{2}+m'^{2}+n'm'}$.
    Using the condition $\bm{G}_{i}\cdot\bm{L}_{j}=2\pi\delta_{ij}$, the associated reciprocal lattice vectors are written as
    \begin{align}\label{eq_Gs}
        \begin{pmatrix}
            \bm{G}_{1}\\
            \bm{G}_{2}
        \end{pmatrix}
        &=
        \frac{1}{n^2+nm+m^2}\left(
        \begin{array} {cc}
            n+m & m \\
            -m & n
	\end{array}
        \right)
        \begin{pmatrix}
           \bm{G}_{1}^{12}\\
           \bm{G}_{2}^{12}
        \end{pmatrix}
        \nn  \\
        &=
        \frac{1}{n'^2+n'm'+m'^2}
        \left(
        \begin{array} {cc}
            n'+m' & m' \\
            -m' & n'
	\end{array}
        \right)
        \begin{pmatrix}
            \bm{G}_{1}^{23}\\
            \bm{G}_{2}^{23}
        \end{pmatrix}.
    \end{align}
     The lattice vectors $\bm{L}_{j}$ of a commensurate double moir\'e lattice can always be written as an integer linear combination of the moir\'e-of-moir\'e interference periods, $\bm{\mathcal{L}}_j$ \cite{moon2013opticalabsorption}.
    In the reciprocal space, accordingly, $\bm{\mathcal{G}}_j$
    can be written as an integer linear combination of $\bm{G}_{i}$'s. 
In all the systems considered in the following,
we have $\bm{L}_{j} = \bm{\mathcal{L}}_j$ and
 $\bm{G}_{j} = \bm{\mathcal{G}}_j$.

\begin{figure}
        \begin{center}
        \leavevmode\includegraphics[width=.9 \hsize]{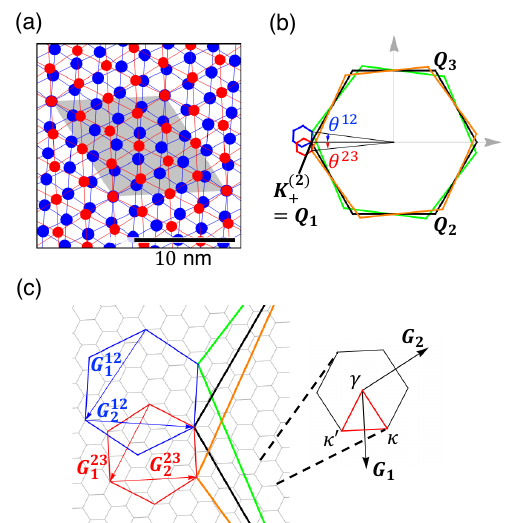}
\caption{
A double-moir\'e system with a common periodicity, which is given by
$(n,m,n',m')=(2,3,2,2)$, corresponding to $(\theta^{12},\theta^{23})=(7.3^\circ,5.8^\circ)$.
(a) Real space structure of the non-relaxed moir\'e lattices 12 (blue) and 23 (red), where the vertexes represent MM stacking point.  A gray rhombus is a common period (the moir\'e-of-moir\'e period). 
(b) Momentum -space sturcture. Green, black, and orange hexagons represent the monolayer BZ of layer 1, 2, and 3 respectively,
and blue and red hexagons are the BZs for the moir\'e 12 and 23.
(c) A magnified view of the region around the $K_{+}$-valley, where gray hexagons represent the BZ of the moir\'e-f-moir\'e pattern.
}
       % The real- and the k-space picture of twisted tirlayer TMD for the helical twist with $(\theta^{12},\theta^{23})=(7.3^\circ,5.8^\circ)$.
       % (a) is the non-relaxed moir\'e-of-moir\'e pattern in real space, where blue and red represent MM stacking of moir\'e 12 and 23, gray rhombus is moir\'e-of-moir\'e unit cell, and scale bar is $10$ nm.
       % Figure(b) shows the relation of moir\'e Brillouin zone (B.Z.) and the monolayer TMD's B.Z..
       % Green, black, and orange hexagonals are the monolayer TMD's B.Z. of layers 1, 2, and 3, and blue and red hexagonals represent the moir\'e B.Z. of moir\'e between layers 1 and 2, and between layers 2 and 3.
      %  (b) The enlarged picture around $K_{+}$-valley in Fig.~(b).
      %  Blue and red arrows show the reciprocal lattice vector of moir\'e 12 and moir\'e 23.
      %  The right figure represents the moir\'e-of-moir\'e B.Z. arranged in the k-space as shown in the left figure as gray hexagons.
        \label{kspcace_bz}
        \end{center}
    \end{figure}

Figure ~\ref{kspcace_bz} shows  an example of such a double-moir\'e system with a common periodicity, which is given by
$(n,m,n',m')=(2,3,2,2)$, or 
$(\theta^{12},\theta^{23})=(7.3^\circ,5.8^\circ)$.
Figure \ref{atomic_lattice_structure}(a) shows the real space structure of the moir\'e lattices 12 (blue) and 23 (red), with the common period is indicated by a gray rhombus.
Figure \ref{atomic_lattice_structure}(b) shows the corresponding $k$-space structure. Green, black, and orange hexagons represent the monolayer BZ of layer 1, 2, and 3 respectively,
and blue and red hexagons are the BZs for the moir\'e 12 and 23.
Figure \ref{atomic_lattice_structure}(c) provides a magnified view of the region around the $K_{+}$-valley, where gray hexagons represent the BZ of the common period, i.e., the moir\'e-of-moir\'e pattern.

For alternate twist cases with small twist angles, two moir\'e lattice vectors are nearly parallel, requiring a huge unit cell for commensuration of the two moir\'e periods.
In this case, we neglect the tiny misorientation of the moir\'e lattice vectors $\bm{L}_{j}^{12}$ and $\bm{L}_{j}^{23}$, while retaining their norms\cite{PhysRevX.13.041007}.
Under this approximation, the moir\'e-of-moir\'e commensurate period is expressed as
    \begin{align}
    \label{eq_comensurate_condition_approx} 
    \bm{L}_{1} = n\bm{L}^{12}_{1} = n'\bm{L}^{23}_{1},
    \quad \bm{L}_{2} = R(60^\circ)\bm{L}_{1},
    \end{align}
instead of Eqs.~\eqref{eq_com_condition}and Eq.~\eqref{eq_angle_formulas}.

%    In the later section, we show that the lattice relaxation constructs the domain structure for both helical and alternate twist, and we calculate the electronic properties of the domains by considering the system with its local structure in the whole region.
 %   These domains have two twist angles which are approximately equal $|\theta^{12}|=|\theta^{23}|$, and moir\'e lattice are parallel.
  %  Thus the moir\'e patterns can ve defined by $(n,m,n',m')=(1,0,1,0)$ and $(1,0,1,0)$ for the domains of alternate and helical twist respectively.
%This treatment is good even for the helical twist case $\theta=\theta^{12}=\theta^{23}$ which generally has the quasicrystal-like structure because two moir\'e lattice vectors are aligned by lattice relaxation in the domain.
%For the alternate twist, $\theta=\theta^{12}=-\theta^{23}$ gives the completely same moir\'e pattern for moir\'e 12 and 23 similar to the symmetric twisted trilayer graphene.

%%%%%%%%%%%%%%%%%%%%%%%%%%%%%%%%%%%%%%%%%%%%%%%%%%%%%%%%%%%%%%%%%%%%%%%%%%%%%%
%%%%%%%%%%%%%%%%%%%%%%%%%%%%%%%%%%%%%%%%%%%%%%%%%%%%%%%%%%%%%%%%%%%%%%%%%%%%%%
\subsection{Continuum method for lattice relaxation}
\label{subsec_relax_model}

We compute the relaxed lattice structure of the trilayer TMD using a continuum method similar to that used for twisted graphene systems \cite{nam2017lattice, koshino2020effective,PhysRevX.13.041007}.
In this model, the lattice distortion is described by a displacement field $\bm{s}^{(\ell)}(\bm{r})$, which represents the shift of an atom on layer $l$ at position $\bm{r}$.
We treat $\bm{s}^{(\ell)}(\bm{r})$ as a smooth function in $\bm{r}$, assuming a long wavelength displacement where its spatial scale is much larger than the atomic lattice constant.
%We introduce $\bm{s}(\bm{R}^{(\ell)}_{X})$ as the displacement vector of $X$ sublattice atom of layer $l$ at position $\bm{R}_{X}$.
%Here we assume a long wavelength displacement where the spatial scale of $\bm{s}(\bm{R}^{(\ell)}_{X})$ is much larger than the atomic lattice constant of TMD.
%Under this assumption, the displacement vector can be written as the continuum function of the space $\bm{s}(\bm{R}^{(\ell)}_{X})\to\bm{s}^{(\ell)}(\bm{r})$, where we ignore the sublattice dependence. 
We assume that the total structural energy is given by $U=U_{E}+U_{B}^{12}+U_{B}^{23}$
as a functional of $\bm{s}^{(\ell)}(\bm{r})$,
where $U_{E}$ is the elastic energy and $U_{B}^{\ell\ell'}$ is the interlayer binding energy of layers $l$ and $l'$.
Here $U_{E}$ is written as \cite{PhysRevB.65.235412,PhysRevB.90.115152} 
 \begin{align}\label{eq:elastic}
        U_E=\sum_{l=1}^{3}\frac{1}{2}\int&\left[\left(\mu+\lambda\right)\left(s_{xx}^{(\ell)}+s_{yy}^{(\ell)}\right)^{2} \right.\notag \\
           &\left. +\mu\left\{\left(s_{xx}^{(\ell)}-s_{yy}^{(\ell)}\right)^{2}+4\left(s_{xy}^{(\ell)}\right)^{2}\right\}\right]\diff^2\bm{r},
    \end{align}
where $\lambda$ and $\mu$ are Lam\'e factors of monolayer TMD, and $s_{ij}^{(\ell)}=(\partial_{i}s_{j}^{(\ell)}+\partial_{j}s_{i}^{(\ell)})/2$ is the strain tensor.
The interlayer binding energy of adjacent layers $(l,l')=(1,2), (2,3)$
is given by \cite{nam2017lattice}
 \begin{align}\label{eq:binding}
     U_{B}^{\ell\ell'}&=\int\diff^{2}\bm{r} \sum_{j=1}^{3}2U_{0}\cos\left[\bm{G}_{j}^{\ell\ell'}\cdot\bm{r}+\bm{b}_{j}\cdot\left(\bm{s}^{(\ell')}-\bm{s}^{(\ell)}\right)\right], 
\end{align}
where $\bm{b}_{3}=-\bm{b}_{1}-\bm{b}_{2}$, $\bm{G}_{3}^{\ell\ell'}=-\bm{G}_{1}^{\ell\ell'}-\bm{G}_{2}^{\ell\ell'}$. 
We ignore the binding energy of layers 1 and 3, which is known to be relevant in small twist angles of the order of $0.1^\circ$ \cite{park2024tunable,yananose2025metamorphicquantumdotarrays}.

%%%

    The parameters for WSe$_2$ are listed as $\lambda=185.2$~eV/nm$^{2}$, $\mu=302.2$~eV/nm$^{2}$\cite{PhysRevLett.124.206101}, and an interlayer potential strength of $U_{0}=0.07889$~eV/nm$^{2}$ \cite{PhysRevLett.124.206101}.
    The optimized lattice structure is numerically obtained by minimizing the lattice energy,  $U = U_{E} + U_{B}^{12} + U_{B}^{23}$, with respect to $\bm{s}^{(\ell)}(\bm{r})$.  
    The details of the numerical calculation used to obtain the optimized structure are presented in Appendix \ref{app_sec_num_relax}.

  %  \mage{
  %  We note that, in this model, we ignore the out-plane component of the displacement vector, it does not much contribute to the domain formation, but gives the building of the MM stacking region\cite{PhysRevLett.124.206101,PhysRevB.104.125440}.
  %  For the electronic calculation, we describe its effect later in Sec.~\ref{subsec_cont_ham}.
  %  }

     %In this paper, we assume that all TMD have similar parameters of the continuum model, Lam\'e parameters $\lambda,\mu$ and binding energy $V_0$. 
    %Under this assumption, we calculate the optimized lattice structure of trilayer MoS$_{2}$, and apply this result to calculate the electronic properties of trilayer WSe$_{2}$ including the effect of the lattice relaxation.
    %For MoS$_{2}$, these parameters are taken as $\lambda=423$~eV/$nm^{2}$, $\mu=423$~eV/$nm^{2}$, and $V_{0}=0.0889$~eV/nm$^{2}$ for MoS$_2$\cite{PhysRevLett.124.206101,weston2020atomic,krisna2024low}.
    
%%%%%%%%%%%%%%%%%%%%%%%%%%%%%%%%%%%%%%%%%%%%%%%%%%%%%%%%%%%%%%%%%%%%%%%%%%%%%%
%%%%%%%%%%%%%%%%%%%%%%%%%%%%%%%%%%%%%%%%%%%%%%%%%%%%%%%%%%%%%%%%%%%%%%%%%%%%%%
\subsection{Continuum Hamiltonian with lattice relaxation}
\label{subsec_cont_ham}

   We construct a continuum Hamiltonian of twisted trilayer TMD
   by extending that for twisted bilayers \cite{PhysRevLett.122.086402,devakul2021magic,PhysRevB.108.085117,PhysRevResearch.5.L032022,PhysRevLett.132.036501,doi:10.1073/pnas.2316749121} and including the effect of the lattice relaxation.
    By using the basis of the valence band around $K_\xi$ point, it is give by a $3\times 3$ matrix,
    \begin{align} \label{eq_Hamiltonian}
        {H}^{(\xi)} =
	    \left(
			\begin{array} {ccc}
		      H_{1}\left(\bm{k}\right) + \Delta^{12}_{+} & \Delta_{T}^{12,*} & \\
		      \Delta_{T}^{12} & H_{2}\left(\bm{k}\right)+ \Delta^{12}_{-}+ \Delta^{23}_{+} & \Delta_{T}^{23,*}\\
                 & \Delta_{T}^{23} & H_{3}\left(\bm{k}\right)+ \Delta^{23}_{-},
			\end{array}
		\right),
    \end{align}
    where the $\ell$-th component corresponds to layer $\ell$.
    We neglect the remote hopping between layer 1 and layer 3.
    % the basis $(\psi^{(1)},\psi^{(2)},\psi^{(3)})$.
    %Here $H_{\ell}$ is the valence band Hamiltonian for monolayer TMD of the layer $\ell$,  $\Delta^{\ell\ell'}_{\pm}(\bm{r})$ is the intralayer potential by the moir\'e pattern between layers $l$ and $l'$, and $\Delta_{T}^{\ell\ell'}(\bm{r})$ is the interlayer hopping between layers $l$ and $l'$.
Here $H_{\ell}$ is the valence band Hamiltonian for monolayer TMD of the layer $\ell$, given by
\begin{align}
\label{eq_effective_strained_TMD}
        H_{\ell} =&  - \frac{\hbar^2}{2m^{*}}\left(\bm{k}+\frac{e}{\hbar}\bm{A}^{(\ell)}\right)^{2} 
        +\delta (s_{xx}^{(\ell)}+s_{yy}^{(\ell)})
        \nn \\ 
        &+ \frac{\hbar^2}{2m^{*}}\alpha\left(k_{+}+\frac{e}{\hbar}A_{+}^{(\ell)}\right)(s_{xx}^{(\ell)}+s_{yy}^{(\ell)})\left(k_{-}+\frac{e}{\hbar}A_{-}^{(\ell)}\right), 
\end{align}
where $\bm{k} = (k_x,k_y)$ is a momentum vector,
$s^{(\ell)}_{\mu\nu}=(\partial_{\mu}s^{(\ell)}_{\nu}+\partial_{\nu}s^{(\ell)}_{\mu})/2$ is the strain tensor,
    $\bm{A}^{(\ell)}$ is the effective vector potential due to the distortion, given by
    \begin{equation} \label{eq_eff_A}
        \bm{A}^{(\ell)} = \xi\frac{\hbar}{e a}\beta\left(s_{xx}^{(\ell)}-s_{yy}^{(\ell)},-2s_{xy}^{(\ell)}\right),
    \end{equation}
and we used the notation $k_{\pm}=\xi k_{x} \pm i k_{y}$ and $A_{\pm}=\xi A_{x} \pm i A_{y}$.
For the parameters, we take $\alpha=-1.83$~eV, $\beta=1.99$~eV and $\delta=-2.24$~eV \cite{PhysRevB.98.075106}, and
$m^{*} = 0.43 m_{\rm e}$ \cite{PhysRevLett.116.086601,rasmussen2015computational,devakul2021magic} for WSe$_2$, where $m_{\rm e}$ is the bare electron mass.
The Hamiltonian of monolayer TMD including the strain effect [Eq.~\eqref{eq_effective_strained_TMD}] is derived by projecting
the $2\times2$ version \cite{PhysRevB.98.075106}
to the valence band by the second perturbation [See Appendix~\ref{app_sec_eff_ham}].

The $\Delta^{\ell\ell'}_{\pm}(\bm{r})$ is the intralayer potential by the moir\'e pattern $\ell\ell'$, and it is given by
\begin{align}
\label{eq_intra_pot} \Delta_{\pm}^{\ell\ell'}\left(\bm{r}\right) = 2V\sum_{j=1}^{3}\cos\left[\bm{G}_{j}^{\ell\ell'}\cdot\bm{r}+\bm{b}_{j}\cdot\left(\bm{s}^{(\ell')}-\bm{s}^{(\ell)}\right)\pm\psi\right],
    \end{align}
where $\pm$ correspond to the layer $\ell$ and $\ell'$.
%, and $\Delta_{T}^{\ell\ell'}(\bm{r})$ is the interlayer hopping between layers $l$ and $l'$.
The $\Delta_{T}^{\ell\ell'}(\bm{r})$ is the interlayer hopping term between layers $\ell$ and $\ell'$, and is written as
    \begin{align}
    \label{eq_inter_hop}
        \Delta_{T}^{\ell\ell'}\left(\bm{r}\right) = w\sum_{j=1}^{3}\e^{\i \bm{q}_{j}^{\ell\ell'}\cdot\bm{r}+\i \bm{Q}_{j}\cdot\left(\bm{s}^{(\ell')}-\bm{s}^{(\ell)}\right)},
    \end{align}
    where 
    \begin{align} \label{eq_dk}
    \bm{q}_{1}^{\ell\ell'}&=\bm{K}_{\xi}^{(\ell)}-\bm{K}_{\xi}^{(\ell')}, \nn \\
    \bm{q}_{2}^{\ell\ell'}&=\bm{q}_{1}^{\ell\ell'}+\xi\bm{G}_{1}^{\ell\ell'}, \nn \\
    \bm{q}_{3}^{\ell\ell'}&=\bm{q}_{1}^{\ell\ell'}+\xi\left(\bm{G}_{1}^{\ell\ell'}+\bm{G}_{2}^{\ell\ell'}\right),
    \end{align}
    and
    \begin{align}\label{eq_qj} &\bm{Q}_{1}=\bm{K}_{\xi},~~~\bm{Q}_{2}=\bm{Q}_{1}+\xi\bm{b}_{1},~~~\bm{Q}_{3}=\bm{Q}_{1}+\xi\left(\bm{b}_{1}+\bm{b}_{2}\right).
    \end{align}
    The parameters $V$, $w$ and $\psi$ in  
    Eqs.~(\ref{eq_intra_pot}) and (\ref{eq_inter_hop}) are determined from the DFT calculation for the non-twist bilayer WSe$_2$ with different lateral shifts, as detailed in Appendix.~\ref{app_sec_para_det}.
    The obtained values are listed as
     $V=5.966~\mathrm{meV}, w=14.32~\mathrm{meV}, \psi=77.02^\circ$.
     
%    In this paper, we deterine $V$ and $w$ from full lattice relaxed DFT calculation for the non-twist bilayer TMD having different interlayer distance.
%    So the effect of the out-plane relaxation on $\Delta_{T}$ and $\Delta_{b/t}$ might be included through these parameters.

%    The parameters $V$ and $w$ in the Eq.~(\ref{eq_intra_pot}) and (\ref{eq_inter_hop}) are determined from the DFT calculation for the non-twist bilayer TMD with lateral shift as shown in Appenddix.~***.
%    In the DFT calculation, we consider the relaxation of the distance of layers for each lateral shift, ensuring that the effect of the out-plane lattice relaxation is included in this continuum model.
%    We note that the effect of the lateral shift of each layer can be included in this continuum Hamiltonian by the $\bm{G}=\bm{0}$ component of the displacement vector $\bm{s}^{(\ell)}_{\bm{G}=0}$, similar to the continuum method \ref{subsec_relax_model}.

For the intralayer potential
Eq.~\eqref{eq_intra_pot},
the relationship between $\Delta_{+}^{\ell\ell'}$ and $\Delta_{-}^{\ell\ell'}$  depends on the parameter $\psi$:
we have $\Delta_{+}^{\ell\ell'} = \Delta_{-}^{\ell\ell'}$ for $\psi=0$,
and $\Delta_{+}^{\ell\ell'} = -
\Delta_{-}^{\ell\ell'}$ for $\psi=90^\circ$.
For $\psi=77^\circ$ in the current model, we have an approximate relationship $\Delta_{+}^{\ell\ell'} \approx -
\Delta_{-}^{\ell\ell'}$
For the later convenience, we also define $V_\ell(\bm{r})$ as the intralayer potential term of layer $l$ in
the diagonal blocks of Eq.~\eqref{eq_Hamiltonian}, i.e.,
\begin{equation}\label{eq_V_l}
    V_1(\bm{r}) = \Delta^{12}_{+},\quad
    V_2(\bm{r}) = \Delta^{12}_{-}+ \Delta^{23}_{+},\quad
    V_3(\bm{r}) = \Delta^{23}_{-}.
\end{equation}
For the current choice of $\psi$,
$V_2$ is nearly equal to 
$-V_1-V_3$, because $\Delta^{\ell\ell'}_{-} \approx -\Delta^{\ell\ell'}_{+}$.

The numerical calculation goes as follows.
The Hamiltonian of Eq.~\eqref{eq_Hamiltonian} hybridizes a set of wavenumbers $\bm{k} = \bm{k}_0 + \bm{G}$, where $\bm{G} = n_1 \bm{G}_1 + n_2 \bm{G}_2$ ($n_1,n_2$: integer) are reciprocal vectors of the moir\'e-of-moir\'e period [Eq.~\eqref{eq_Gs}], and $\bm{k}_0$ is a wavenumber within the moir\'e-of-moir\'e BZ.
To obtain the low-energy spectrum, we truncate the high-momentum components and retain only the wavenumbers with
$|\bm{G}|\leq 4\max(|\bm{G}_{j}^{12}|,|\bm{G}_{j}^{23}|)$.
We construct the Hamiltonian matrix within the finite set of 
wavenumbers, and obtain energy bands by plotting its eigenvalue as a function of $\bm{k}_0$.

    %In our numerical calculation, we only take the second order of $\bm{s}^{(\ell)}$, which is enough to obtain the result.    
%    We also take a finite number of the Fourier components as the basis of the Hamiltonian included in $|\bm{G}|\leq 4*\max(|\bm{G}_{j}^{12}|,|\bm{G}_{j}^{23}|)$.
%    We neglect the remote hopping between layer 1 and layer 3.
%    In the calculation of moir\'e-of-moir\'e scale band structure, we ignore the effective vector potential by the distortion since it does not have much influence on the band structure near the Fermi surface [See Appendix].

While out-of-plane lattice distortion is neglected in the structural relaxation, real moiré TMD systems exhibit corrugated structures, where the interlayer spacing is larger in the MM regions than in the MX/XM regions \cite{PhysRevLett.124.206101,PhysRevB.104.125440}. In the electronic band calculations, the effect of corrugation is partially incorporated into the DFT-based band parameters $v$, $w$, and $\psi$, as the structural optimization (including interlayer distance) was performed for non-twisted bilayers with MM, MX, and XM stackings [Appendix~\ref{app_sec_para_det}].

%\mage{
%We note that, in this model, we ignore the out-plane component of the displacement vector, it does not much contribute to the domain formation, but gives the building of the MM stacking region\cite{PhysRevLett.124.206101,PhysRevB.104.125440}.
%}

%    \mage{
%    In the electronic calculation, we include the out-plane distortion effect on the interlayer hopping term, but ignore on the hamiltonian of the monolayer TMD.
%    For TBG, it is known that the distortion in the out-plane gives the difference between the interlayer hopping parameter of AA and AB/BA stacking regions, leading to the opening of the gap between the flat band and the secondary band\cite{doi:10.1073/pnas.2316749121}, however, the out-plane distortion is smooth enough\cite{PhysRevB.106.115410,PhysRevB.101.125409} to ignore the effect of the monolayer level Hamiltonian.
%    }
    
%    \mage{
%    To inclued the out-plane distortion effect on the interlayer hopping, we perform the structural optimization including interlayer distance in the calculation of each non-twisted bilayer.
%    In the band calculation of the twisted trilayer, the effect of the spatial modulation of the interlayer  (wider in MM stack and narrower in MX stack) distance is included through the obtained band parameters.
%    }

%%%%%%%%%%%%%%
\section{Helical trilayers}
\label{sec_helical}

\begin{figure*}
\begin{center}
\leavevmode\includegraphics[width=.8 \hsize]{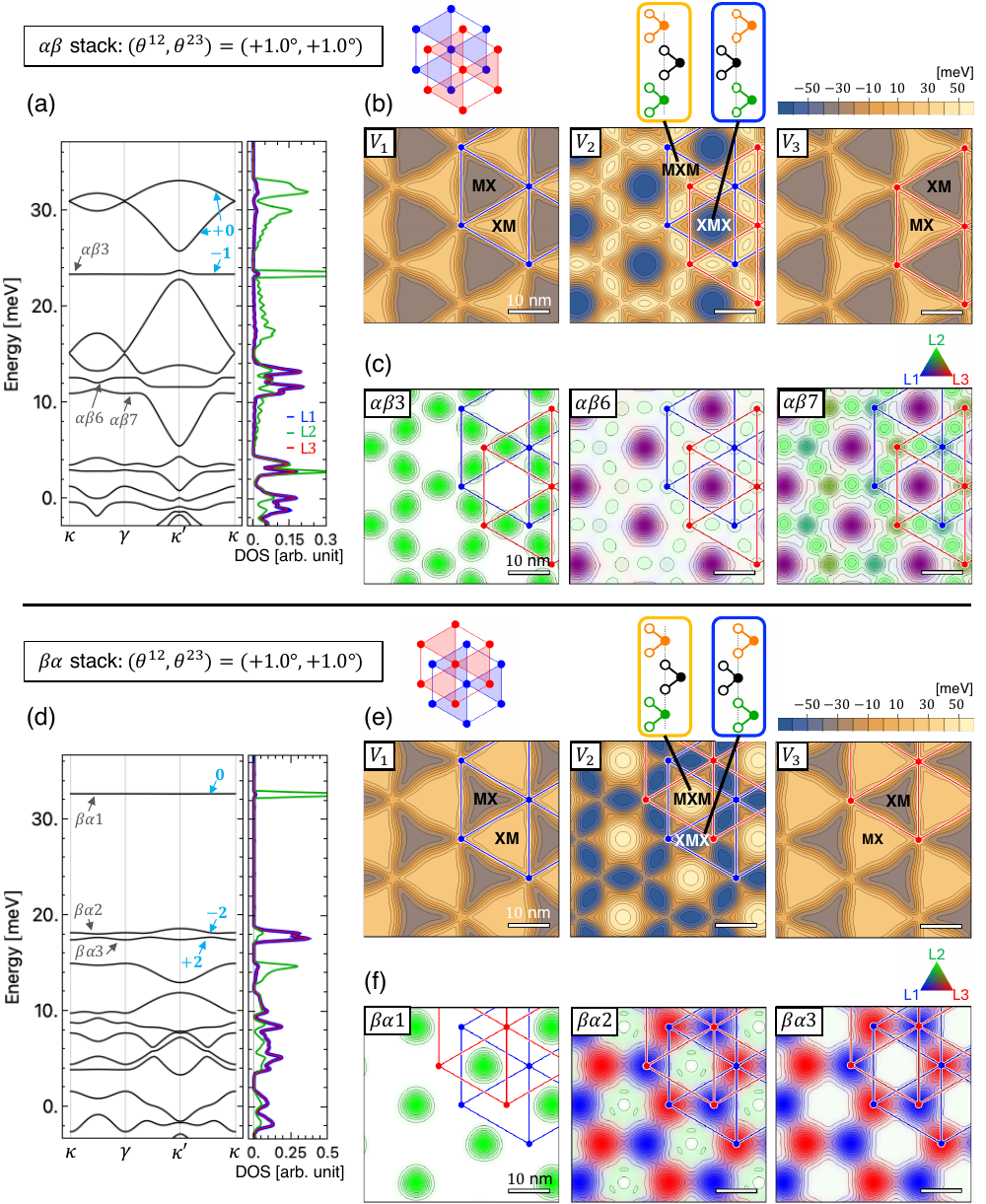}
\caption{
(a) Band structure and the layer-projected DOS near the valence band edge for a relaxed $\alpha\beta$-stacked twist trilayer WSe$_{2}$  with $(\theta^{12}, \theta^{23})=(1.0^{\circ}, 1.0^{\circ})$.
Blue numbers by the bands indicate the Chern number of the $K_+$ valley.
(b) Contour maps of the intralayer potential $V_1, V_2, V_3$.
Blue and red lattices represent moir\'e lattices of 12 and 23, respectively.
(c) Real-space distribution of wavefunctions for representative bands, averaged over the Brillouin zone. The colors indicate the layer distribution, as shown in the triangular diagram in the inset.
(d-f) Corresponding figures for a commensurate $\beta\alpha$-stacked trilayer with $(\theta^{12}, \theta^{23})=(1.0^{\circ}, 1.0^{\circ})$.
       }
        \label{local_band_BBA}
        \end{center}
    \end{figure*}

\begin{figure*}
\begin{center}
\leavevmode\includegraphics[width=1. \hsize]{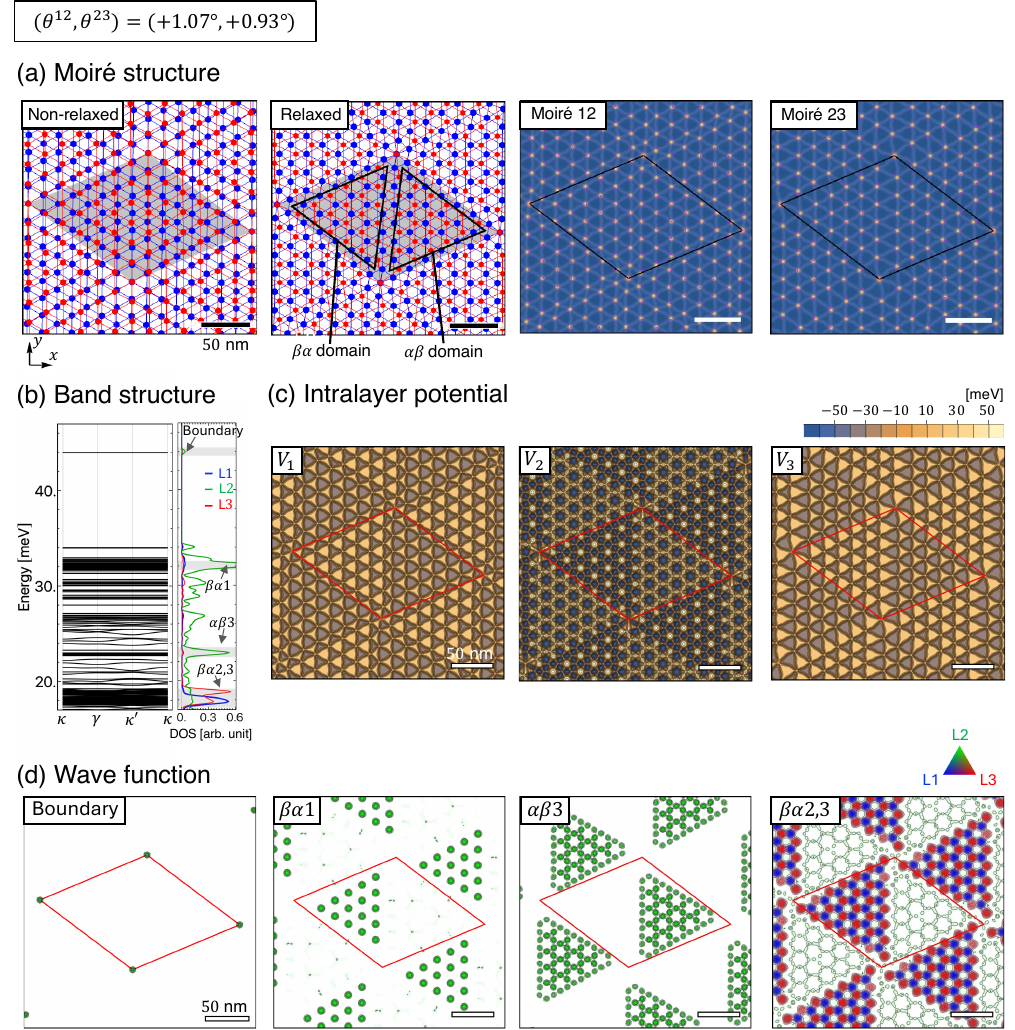}
\caption{
(a) Structural relaxation in a helical twist trilayer WSe$_2$ with $(\theta^{12},\theta^{23})=(1.07^\circ,0.93^\circ)$.
The left two panels present schematic illustrations of the double moir\'e lattice for non-relaxed and relaxed cases, where blue and red dots represents MM stacking points of moir\'e 12 and moir\'e 23.
A gray rhombus represents a moir\'e-of-moir\'e unit cell.
The right two panels plot the interlayer binding energy $U_{B}^{12}$ and $U_{B}^{23}$
in the relaxed structure,
where bright and dark regions correspond to MM and MX/XM stacking, respectively. Small magenta dots indicate the MM stacking without lattice relaxation.
(b) Band structure and layer-resolved density of states (DOS) of the same system.
(c) Contour maps of the intralayer potentials 
$V_1, V_2, V_3$.
(d) Real-space distribution of representative wavefunctions, averaged over the shaded energy regions in the DOS plot of (b).
The colors indicate the layer distribution.
}
        \label{mom_band_helical}
        \end{center}
    \end{figure*}

\subsection{Commensurate helical trilayers}
\label{subsec_helical_local_elec}

First, we consider a commensurate helical trilayer TMD,  
where the two moir\'e patterns share the exact same periodicity,  
as seen in the $\alpha\alpha$, $\alpha\beta$, and $\beta\alpha$ stackings shown in Fig.~\ref{atomic_lattice_structure}(c).  
Such a structure can be obtained by starting from an equal-angle trilayer with  
$\theta^{12} = \theta^{23}$ and slightly expanding the atomic period of the middle layer 
to align the orientations of moir\'e 12 and moir\'e 23 (i.e., $\bm{G}_{j}^{12}=\bm{G}_{j}^{23}$) \cite{PhysRevLett.128.026402,doi:10.1126/sciadv.adi6063,PhysRevX.13.041007}.

The interlayer stacking energy of the commensurate trilayer depends on the relative shift between the two moir\'e lattices.  
Similar to twisted trilayer graphene \cite{PhysRevX.13.041007},
the lowest energy state is achieved in the $\alpha\beta$ and $\beta\alpha$ structures,
where the MM points of moir\'e 12 and moir\'e 23 repel each other. 
The stability of the $\alpha\beta$ and $\beta\alpha$ domains can be understood as a result of frustration in the local relaxation of the middle layer \cite{PhysRevX.13.041007}. %\red{as summarized in Appendix *.}
Here, we perform the band calculations for relaxed commensurate trilayers of $\alpha\beta$ and $\beta\alpha$ stacking,  
with $\theta^{12} = \theta^{23} = 1^{\circ}$.  
In Fig.~\ref{local_band_BBA}, the upper panels present the results for the $\alpha\beta$ case, showing (a) the energy bands near the valence band top, along with the layer-projected density of states (DOS), (b) the intralayer potential maps $V_1$, $V_2$, and $V_3$ [Eq.~\eqref{eq_V_l}], and (c) the real-space distribution of wavefunctions in representative bands, averaged over the Brillouin zone.
Here we labeled the bands as $\alpha\beta1, \alpha\beta2, \alpha\beta3, \dots$ in descending order from the top.  
The lower panels [Figs.~\ref{local_band_BBA} (d), (e) and (f)] show the corresponding figures for the $\beta\alpha$ case.

In this system, the interlayer hopping terms $\Delta_T^{12},\Delta_T^{23}$
are significantly suppressed by the lattice relaxation [see, Appendix \ref{app_non-relax_vs_relax}], so the electronic states are primarily governed by the intralayer potential terms, $V_1, V_2, V_3$.
The potential of the outer layers, $V_1$ and $V_3$, are similar to that of a moir\'e bilayer, where $V_1(V_3)$ take the maxima and minima 
at the XM (MX) and MX (XM) domains, respectively \cite{doi:10.1126/science.154.3751.895,suzuki2014valley,jiang2014valley,PhysRevLett.124.206101}.
These domains are not perfect triangles but exhibit convex and concave shapes, and this deformation can be attributed to the interaction between moir\'e 12 and moir\'e 23 through the middle layer\cite{PhysRevX.13.041007}. %\red{(see Appendix~*)}.  
The second-layer potential $V_2$ is approximately given by $-V_1-V_3$ as argued in the previous section,
and therefore that $V_2$ has an energy range approximately twice as large as that of $V_1$ and $V_3$.

In the $\alpha\beta$ structure, the potential maxima of $V_1$ at XM and those of $V_3$ at MX coincide in position (XMX),  
and hence the minima of $V_2\approx-V_1-V_3$ form a triangular lattice.
In contrast, the minima of $V_1$ (MX) and $V_3$ (XM) are located at different positions within the moir\'e unit cell,  
occupying different sublattices of a honeycomb lattice. As a result, the maxima of $V_2$ appear at the midpoints of these sublattices (MXM),  
forming a Kagome lattice.  
The situation is also illustrated in a schematic diagram in Fig.~\ref{schematic_figure}.
The potential map of $\beta\alpha$ is nearly identical to that of $\alpha\beta$ but inverted along the energy axis.  
Thus, in $\beta\alpha$, the roles of the Kagome and triangular lattices for $V_2$ are reversed,  
with its minima forming a Kagome lattice and its maxima forming a triangular lattice.
The inverted profiles in $\alpha\beta$ and $\beta\alpha$ can be explained using Eq.~\eqref{eq_intra_pot},  
considering that $\bm{G}_{j}^{\ell\ell'}$ and $\bm{s}^{(\ell)}$ have opposite signs,  
and that $\psi = 77^\circ$ is close to $\pi/2$.

The energy bands strongly reflect the structure of the potential maps of $V_1, V_2, V_3$. 
In the energy spectrum, states in the upper energy region ($E \gtrsim 20$ meV)  
are predominantly localized on layer 2 due to the high potential energy of $V_2$. 
In the $\alpha\beta$ structure,  
the energy bands in this region correspond to states bound to the maxima of $V_2$, which form a Kagome lattice.  
As a result, a characteristic Kagome band structure emerges,  
including a %\mage{$C_K=-1$ topological} 
flat band ($\alpha\beta3$) with a wavefunction  
following the Kagome pattern.
The band $\alpha\beta6$ and $\alpha\beta7$ are bound states at the maxima of $V_1$ and $V_3$, which are arranged in a shared trigonal lattice.
While the overall band structures are primarily determined by the intralayer potentials $V_{\ell}$, additional features arise from the strain-induced effective vector potentials $\mathbf{A}^{(\ell)}$ [Eq.\eqref{eq_eff_A}] and the interlayer hopping terms $\Delta_T^{(\ell\ell')}$ [Eq.\eqref{eq_inter_hop}].
In the absence of both contributions, the flat band $\alpha\beta3$ is attached to the topmost graphene-like band cluster.
Including these terms shifts the flat band downward toward the lower dispersive bands.
They are also responsible for opening a small gap between the flat and dispersive bands and for generating a non-zero valley Chern number.

%&\mage{The flat band nearly sticks to the second band cluster, not the first, due to the interlayer hopping and the effective vector potential.}
%{\bf \red{[@ Mention the flat band nearly sticks to the second band cluster, not the first. Mention the Chern number $C=-1$.]}}

In the $\beta\alpha$ structure,  
the inverted potential results in a completely distinct energy band structure.  
The maxima of $V_2$ form a triangular pattern, giving rise to the $\beta\alpha1$ state.  
The maxima of $V_1$ and $V_3$ do not share the center position but instead form a hexagonal lattice.  
The bands $\beta\alpha2$ and $\beta\alpha3$ are bound states at these potential maxima,  
where layers 1 and 3 serve as different sublattices of a honeycomb lattice, 
as seen in Fig.~\ref{local_band_BBA}(f).
The corresponding energy bands in Fig.~\ref{local_band_BBA}(d) have small dispersions and Chern numbers of $\pm2$,
which can presumably be captured by a Haldane-type model based on a honeycomb lattice \cite{PhysRevLett.61.2015,PhysRevB.87.115402,Chang_2022}.

%\mage{
%@@ non-relax vs relax @@
%We note that lattice relaxation is quite important for the electronic properties due to the increasing of the intralayer potential and decreasing of interlayer hopping defectively [See Appendix].
%Moreover, specially for the $\alpha\beta$ stacking, lattice relaxation changes the real space distribution of the intralayer potential from honycomb lattice to kagome lattice as we shown in Appendix\ref{app_non-relax_vs_relax}.
%}

\subsection{General helical trilayers}

Now we consider a general helical trilayer TMD, where the two moir\'e patterns have different periodicities.
Here we calculate the relaxed structure and the electronic band structure of the helical twist trilayer WSe$_2$ with $(\theta^{12}, \theta^{23})=(1.07^\circ, 0.93^\circ)$,
which corresponds to
$(n,m,n',m') = (1,7,1,6)$
in Eq.~\eqref{eq_angle_formulas}.
Figure \ref{mom_band_helical}(a) summarizes the optimized lattice structure.
The left two panels show the moir\'e pattern before (left) and after (right) lattice relaxation, where blue and red dots indicate the MM stacking of moir\'e 12 and moir\'e 23, respectively.
The right two panels plot the local binding energy $U_{B}^{12}$ and $U_{B}^{23}$ in the relaxed structure, where
the bright and dark regions correspond to the MM and MX/XM stacking.
In each of moir\'e 12 and 23, the MM stacking points shrink, while the MX and XM regions expand to form a triangular domains as in moir\'e bilayer.
In the moir\'e-of-moir\'e scale, MM stacking points of moir\'e 12 and those of 23 avoid each other, leading to the emergence of commensurate moir\'e domains of $\alpha\beta$ and $\beta\alpha$ argued above.

Figure \ref{mom_band_helical}(b) shows the band structure and the layer-projected DOS near the valence band top of the relaxed trilayer.
Figure \ref{mom_band_helical}(c) presents the intralayer potential $V_1, V_2$ and $V_3$. Figure \ref{mom_band_helical}(d) illustrates the real-space distribution of the wave functions corresponding to representative peaks in (b). Here, we take the average of the squared wave amplitude over the states within the shaded region in (b), with the color representing the layer distribution as defined in the inset.

Notably, the DOS in the original moir\'e-of-moir\'e trilayer in Fig.~\ref{mom_band_helical}(b) shows reasonable agreement with the sum of DOS for $\alpha\beta$ and $\beta\alpha$
commensurate trilayers
[Figs.~\ref{local_band_BBA} (a) and (c)].
In particular, the peaks labeled by 
$\beta\alpha1, \,\beta\alpha2,\,\beta\alpha3$ and $\alpha\beta3$
in Fig.~\ref{mom_band_helical}(b) correspond to the flat bands with the same labels in the commensurate band structure.
We also see that the corresponding wavefunctions shown in Fig.~\ref{mom_band_helical}(d) 
share the same spatial distributions of the local-band wavefunctions in Figs.~\ref{local_band_BBA}(b) and (d).
These states are mainly distributed in either of $\alpha\beta$ or $\beta\alpha$ domains with a small inter-domain connection, leading to nearly flat dispersions in the moir\'e-of-moir\'e energy bands.

\begin{figure}
        \begin{center}
        \leavevmode\includegraphics[
        width=.95 \hsize]{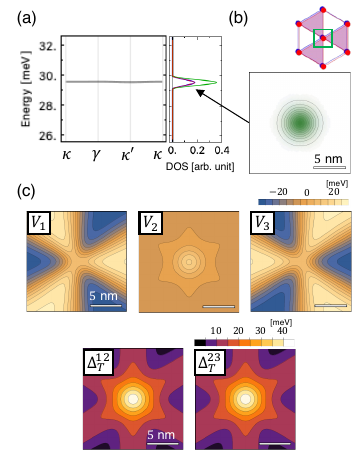}
        \caption{
(a) The topmost valence band and its projected density of states (DOS) in a relaxed helical twist trilayer WSe$_2$ with local $\alpha\alpha$ stacking for $(\theta^{12},\theta^{23})=(1.0^\circ,1.0^\circ)$.%, similar to Fig.~\ref{local_band_BBA}(a,d).
(b) Real-space distribution of the topmost band around the MMM stacking region, corresponding to the area enclosed by a green rectangle in the schematic figure above. The color scheme follows Fig.~\ref{local_band_BBA}.
(c) Contour maps of intralayer potentials $V_1, V_2, V_3$ and the absolute values of the interlayer hopping terms $|\Delta_{T}^{12}|, |\Delta_{T}^{23}|$.
        }
        \label{local_AAA_helical}
        \end{center}
    \end{figure}

On the other hand, there are also bands that do not correspond to the $\alpha\beta/\beta\alpha$ domain states. Specifically, the topmost band is located at a higher energy ($\sim 42$ meV) than the highest bands of $\alpha\beta/\beta\alpha$, separated by the rest of the spectrum by an energy gap. This band is associated with boundary states, which are sharply localized at the corners
(MMM positions) of the moir\'e-of-moir\'e domains, as shown in Figure \ref{mom_band_helical}(d).
These modes can be understood by considering the electronic structure of a commensurate  $\alpha\alpha$ trilayer, which has a MMM stack at the corners.
Figure \ref{local_AAA_helical} shows the result of the band calculation for $\alpha\alpha$ trilayer with $(\theta^{12},\theta^{23})=(1.0^\circ,1.0^\circ)$.
Here we show (a) the energy band of the MMM corner mode,
(b) its wave function,
and (c) the intralayer potential $V_1,V_2,V_3$ and the interlayer hopping amlitudes $|\Delta_T^{12}|,|\Delta_T^{23}|$ in the vicinity of the MMM corner.
%We see that $V_1$ and $V_3$ are nearly sign-reversed, resulting in a small component of $V_2=-V_1-V_3$.  
%On the other hand, 
We see that the interlayer hopping $|\Delta_T^{12}|,|\Delta_T^{23}|$  peaks at the corner with a greater magnitude than the $V_i$ terms, and consequently,
the MMM corner hosts a localized state bound by the dominant interlayer hopping.
%which is mirror-symmetric with respect the middle layer.
This MMM bound state qualitatively explains the corner state observed in Fig.~\ref{mom_band_helical}(b), although its energy does not match precisely due to discrepancies in the strain-induced terms in the Hamiltonian.

We note that these boundary states are not topologically-protected corner states, as they are fragile under a lateral shift between the moir\'e patterns, which remove the MMM stacking. 
In real twist trilayers, the two moir\'e patterns do not form exact commensuration within the moir\'e-of-moir\'e cell unlike the assumed structures in this calculation. 
As a result, the domain corners can adopt different local moir\'e configurations depending on the lateral shift, leading to corner-dependent variations in the boundary states.
%In such cases, we expect that the structure at domain boundaries—and consequently, the boundary states—should depend sensitively on the relative periodicities of moir\'e 12 and 23. 
%On the other hand, the formation of stable $\alpha\beta$ and $\beta\alpha$ domains is insensitive to these variations and should always occur when the two moir\'e periodicities are close to each other.

The size of the $\alpha\beta$ and $\beta\alpha$ domains increases as the two twist angles approach each other. In the case of equal angles, $\theta^{12} = \theta^{23} = \theta$, the moir\'e-of-moir\'e period is given by $\mathcal{L} = (4\pi/\sqrt{3})|\bm{G}_{j}^{12} - \bm{G}_{j}^{23}| = a / 4 \sin^{2}(\theta/2)$. The period becomes 0.8 $\mu$m at $\theta = 1.0^\circ$ and decreases further as $\theta$ becomes smaller. Therefore, it is in principle possible to fabricate a periodic Kagome potential over a scale on the order of micrometers.

 \begin{figure*}
        \begin{center}
        \leavevmode\includegraphics[width=.8 \hsize]{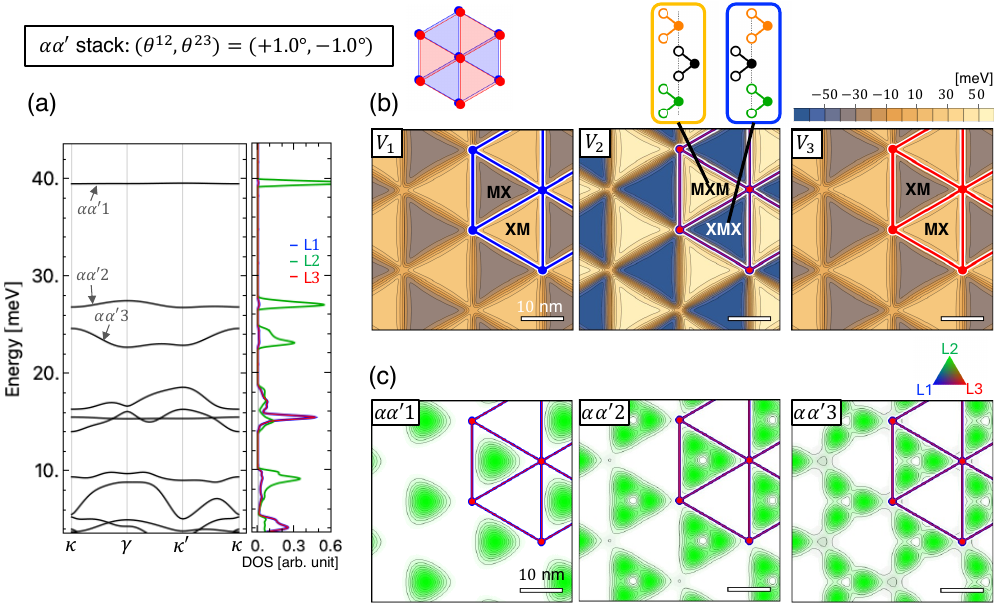}
        \caption{
(a) Band structure near the valence band edge,
(b) contour maps of the intralayer potential, and
(c) wavefunctions of representative bands,
calculated for a relaxed $\alpha\alpha'$ twisted trilayer WSe$_{2}$ with $(\theta^{12}, \theta^{23})=(1.0^{\circ}, -1.0^{\circ})$.
The results are plotted in a similar manner to Fig.~\ref{local_band_BBA}.
%        The result of the local $\alpha\alpha'$ stacking twisted trilayer WSe$_{2}$ with $(\theta^{12}, \theta^{23})=(1.0^{\circ}, -1.0^{\circ})$, corresponding to Fig.~\ref{local_band_BBA}.
        }
        \label{local_band_AAA}
        \end{center}
    \end{figure*}

\begin{figure*}
        \begin{center}
        \leavevmode\includegraphics[width=1.\hsize]{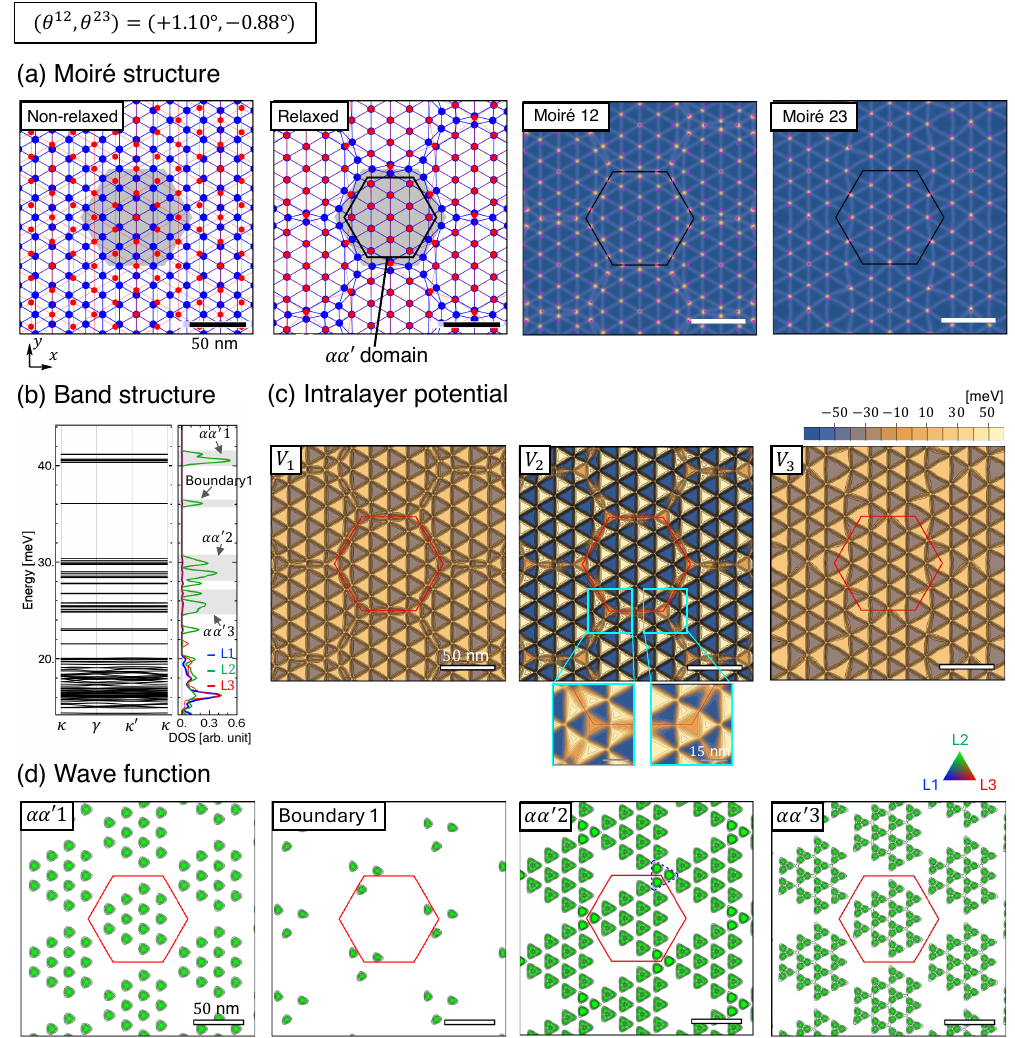}
        \caption{
(a) Structural relaxation,
(b) band structure,
(c) contour maps of the intralayer potentials, and
(d) wavefunctions averaged over representative energy regions, calculated for an alternate twist trilayer WSe$_{2}$ with $(\theta^{12},\theta^{23})=(1.10^\circ,-0.88^\circ)$.
The results are plotted in a similar manner to Fig.~\ref{mom_band_helical}.
%        The result for alternate twist trilayer WSe$_{2}$ with $(\theta^{12},\theta^{23})=(1.10^\circ,-0.88^\circ)$, corresponding to Fig.~\ref{mom_band_helical}.
        }
        \label{mom_band_alternate}
        \end{center}
    \end{figure*}

%%%%%%%%%
\section{Alternate twist trilayers}
\label{sec_alternate}

\subsection{Commensurate alternate trilayers}

Alternate twist trilayers ($\theta_{12}\cdot \theta_{23} < 0$) have fundamentally distinct electronic properties.
First, we consider a commensurate alternate trilayer with $\theta^{12}=-\theta^{23}$, where the moir\'e 12 and 23 have the exactly same period.
The total structural energy depends on the lateral shift of the two  moir\'e lattices,
the most stable registry is $\alpha\alpha'$ structure [Fig.~\ref{atomic_lattice_structure}(g)] where the MM stacking points of moir\'e 12 and those of 23 are overlapped.
The underlying mechanism of domain formation is essentially the same as in alternate twisted trilayer graphene \cite{PhysRevX.13.041007}.  %as outlined in Appendix~*.
Here we calculate the energy bands for a relaxed twisted trilayer WSe$_2$ with $(\theta^{12},\theta^{23}) = (1^{\circ},-1^{\circ})$ with $\alpha\alpha'$ stacking.  
Figure \ref{local_band_AAA} shows the energy bands, the intralayer potential map and the wave functions.
The $V_1$ and $V_3$ plotted in Fig.~\ref{local_band_AAA}(b) are equivalent to the intralayer potential of twisted bilayer,
where $V_1(V_3)$ take the maxima and minima 
in triangular XM (MX) and MX (XM) domains, respectively.
In contrast to the helical case,
$V_1$ and $V_3$ are exactly identical due to the mirror symmetry with respect to the middle layer.
Therefore, the middle layer potential, $V_2 \approx -V_1-V_3$, is simply twice as deep as $V_1$ and $V_3$, leading to an array of trigonal potential wells with a depth about 100 meV.

The topmost bands $\alpha\alpha'1, 2, 3$ 
originate from bound states at the maxima of the triangular potential well of $V_2$.
The $\alpha\alpha'1$ corresponds to an $s$-like orbital,  which is compactly localized in the well, resulting in an exponentially small band dispersion.
The $\alpha\alpha'2$ and $\alpha\alpha'3$ states originate from $p$-like orbitals with broader spatial distributions, resulting in bandwidths of a few meV. These two bands correspond to bound states with opposite angular momenta $l_z=\pm 1$, 
which correspond to the eigenstate $e^{2\pi i l_z /3}$ in $C_3$ rotation.
The energy splitting between $l_z=\pm 1$ arises from a strain-induced effective magnetic field.
As the twist angle decreases,  
the potential well widens in space,  
causing $\alpha\alpha'2$ and $\alpha\alpha'3$ to become more localized. 
Simultaneously, higher bound states $\alpha\alpha'4, \alpha\alpha'5, \cdots$ emerge.  
Note that in twisted TMD bilayers, the potential wells have only half the depth, leading to fewer bound states compared to a trilayer with the same twist angle.

\subsection{General alternate trilayers}

As a representative example of general alternate trilayers, we consider a trilayer WSe$_2$  with $(\theta^{12}, \theta^{23})=(1.10^\circ, -0.88^\circ)$, which corresponds to
$(n,n') = (5,-4)$ in Eq.~\eqref{eq_comensurate_condition_approx}.
Figure \ref{mom_band_alternate}(a) shows the optimized lattice structure presented with the same figure arrangement as in Fig.~\ref{mom_band_helical}(a) for the helical case.
By comparing the non-relaxed and relaxed cases, we observe that lattice relaxation induces  
locally commensurate $\alpha\alpha'$ domains arranged in a hexagonal pattern, along with domain boundaries.
The energy band, the intralayer potential, and representative wavefunctions are presented in
Fig.~\ref{mom_band_alternate}(b), (c) and (d), respectively.
In the potential map [Fig.~\ref{mom_band_alternate}(c)], 
$\alpha\alpha'$ domains accommodate an array of triangular quantum well discussed above.
%The electronic bands are basically bound states of these wells.
%The higher energy region $E>25$meV is dominated by the layer 2, which has the highest potential.
%Here $\alpha\alpha'1,2,3$ correspond to the bound states of wells in $\alpha\alpha'$ domain.
%We can also see a bunch of boundary modes localized near the domain corners, where the the intralayer potential has different structures from the domain.

In the energy spectrum  [Fig.~\ref{mom_band_alternate}(b)],
$\alpha\alpha'1,2,3$ bands appear as peaks
at the corresponding energy,
with an identical wave distribution
as shown in Fig.~\ref{mom_band_alternate}(d).
We can also see a bunch of boundary modes which do not correspond to $\alpha\alpha'$ domain states.
The peaks between $\alpha\alpha'1$ and 
$\alpha\alpha'2,3$ are the modes
localized around the corners of the
$\alpha\alpha'$ domain, where the the intralayer potential has different well structure  [Fig.~\ref{mom_band_alternate}(c)] from the $\alpha\alpha'$ domain.
As previously noted in the helical case, the moir\'e configurations near domain boundaries are highly sensitive to the relative periodicities of the moir\'e patterns, resulting in variations in the boundary states.

%\mage{The real-space distribution and the energy of states at the corner can be modulated by the lateral shift, altering the corner's configuration of $V_2$.
%In the real sample of twisted trilayer WSe$_2$, its lattice structure is generally incommensurate and locally takes a moir\'e-of-moir\'e pattern with different lateral shift.
%As a result, we expect that the various corner state locally appear in the real sample.
%}

%\textbf{\red{ 
%[@ Mention the potential structure at the boundary is sensistive the commensruation, as explainged in helical case. Really need the following explanation in this length?? ]}}
%As seen in the enlarged images of $V_2$,  
%the two inequivalent corners of the hexagonal domain exhibit distinct potential profiles: one consists of six maxima (denoted as corner A), while the other consists of three (corner B).  
%For the corner A, the height of $V_2$ at the six maxima is identical.
%However, we also have an additional on-site potential (not included in $V_i$'s) from the strain term $\delta$ in Eq.~\eqref{eq_effective_strained_TMD}.
%This gives a potential difference between 
%points marked by red and blue circles in the inset, leading to splitting of the corner states in energy [Boundary 1, 2 in Fig.~\ref{mom_band_alternate}(b)].
%For the corner B,
%three maxima have the same height due to $C_3$ symmetry, resulting in bound states at a single energy,
%which happens to be close to lower states 
%of the corner A.

%%%%%%%%%%%
%(垂直電場依存)
\section{Effects of perpendicular elecrtric field}
\label{sec_effect_electric_field}

   \begin{figure*}
        \begin{center}      
        \leavevmode\includegraphics[width=1. \hsize]{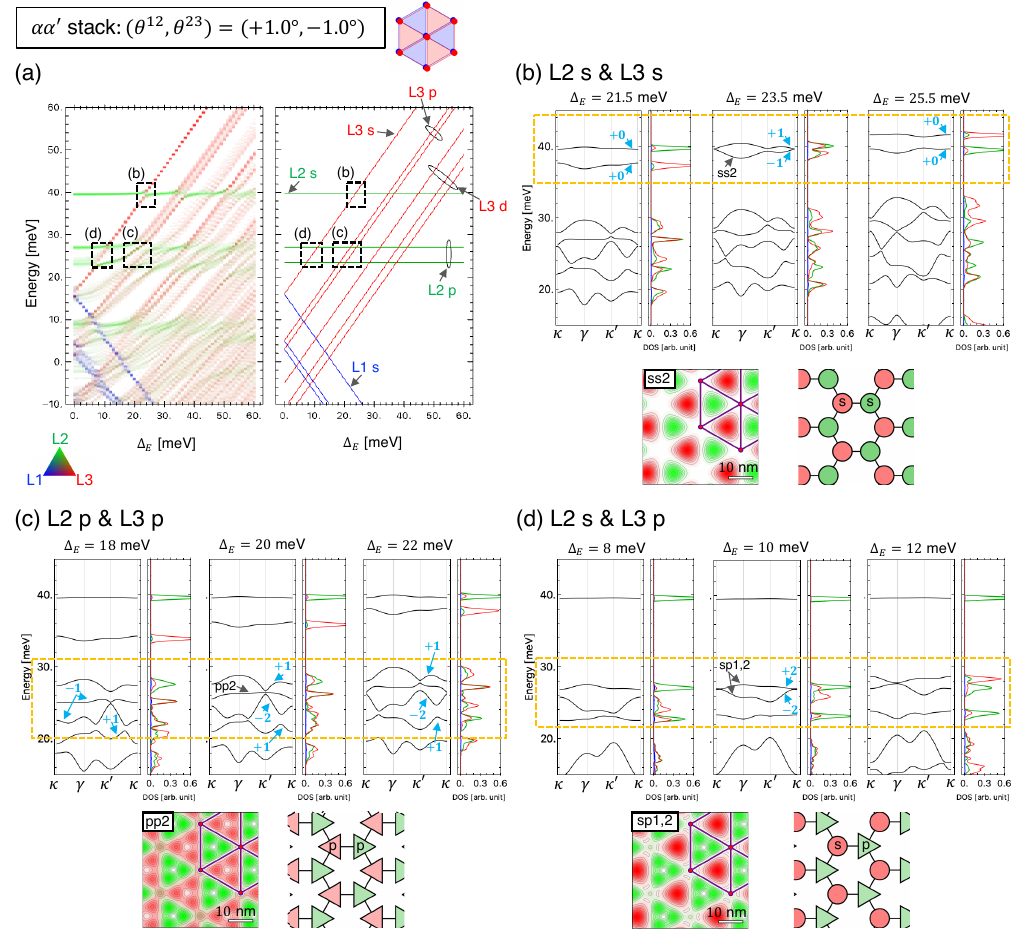}
        \caption{
        (a) Density map of the layer-projected density of states (DOS) for 
 the $\alpha\alpha'$-stacked twisted trilayer WSe$_2$ with $(\theta^{12}, \theta^{23}) = (+1.0^\circ, -1.0^\circ)$. The horizontal and vertical axes represent the electric field $\Delta_E$ and energy, respectively. The right panel presents a schematic diagram illustrating the energies of localized orbitals.
(b) Energy band structure in the region labeled (b) in Fig.\ref{fig_local_AAA_elec}(a). The bottom panels show the BZ-averaged wavefunction of a representative band, labeled $ss2$ (left), and the corresponding effective lattice (right).
(c,d) Band structures and representative wavefunctions for the crossing points labeled (c) and (d) in Fig.\ref{fig_local_AAA_elec}(a). 
The color schemes follow those in Fig.~\ref{local_band_BBA}.
        }
        \label{fig_local_AAA_elec}
        \end{center}
    \end{figure*}

    \begin{figure}
        \begin{center}        \leavevmode\includegraphics[width=1. \hsize]{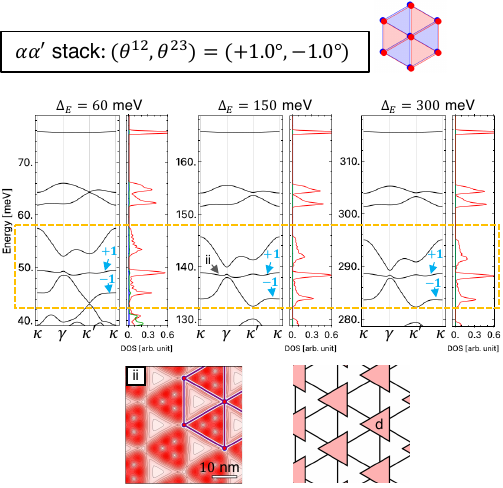}
        \caption{
        (Top) Band structures near the valence band top of the $\alpha\alpha'$-stacked twisted trilayer WSe$_2$ with $(\theta^{12}, \theta^{23}) = (+1.0^\circ, -1.0^\circ)$, under electric fields $\Delta_E = 60$, $150$, and $300$ meV.
(Bottom left) Density plot of the wavefunction for the flat band labeled ii at $\Delta_E = 150$ meV, averaged over the Brillouin zone.
(Bottom right) Schematic illustration of the corresponding effective triangular lattice model.
        }
        \label{fig_local_AAA_d}
        \end{center}
    \end{figure}

%We can control the layer polarization of the electronic states
%\mage{
%Twisted trilayer WSe$_2$ has the s-, p- and d-like states polarized on each layer, and we can controll the hibridization of these electron states of different layer by applying a perpendicular electric field.}

 By applying a perpendicular electric field, we can control the layer polarization by modulating the electrostatic potential of each layer. This also allows us to tune the hybridization of energy bands across layers, enabling the formation of diverse electronic structures composed of $s$- and $p$-like orbitals localized in triangular dots.
 
The effect of a vertical electric field can be introduced by adding a term to the Hamiltonian  [Eq.~\eqref{eq_Hamiltonian}],
     \begin{align} \label{eq_u_mat}
    	H_{E}=\left(
    		\begin{array}{ccc}
    		  -\Delta_{E} & & \\
    		   & 0 &  \\
                 & & \Delta_{E} 
    		\end{array}
    	\right),
        \end{align}
where $\Delta_{E}$ is the electrostatic potential difference between adjacent layers.

Figure~\ref{fig_local_AAA_elec} summarizes the electric field dependence of the electronic structure of the $\alpha\alpha'$-stacked twisted trilayer WSe$_2$ with $(\theta^{12}, \theta^{23}) = (+1.0^\circ, -1.0^\circ)$.
Figure~\ref{fig_local_AAA_elec}(a) presents a density map of the layer-projected DOS, plotted as a function of the electric field ($\Delta_E$) and energy. The $\Delta_E=0$ corresponds to the energy bands in Fig.~\ref{local_band_AAA}.
The right panel shows a schematic diagram illustrating the energies of localized orbitals; for example, $L2s$ denotes the $s$-like bound state localized in a triangular quantum well in layer 2.
With increasing $\Delta_E$, we clearly observe that the bands originating from layer 3 (red) shift upward in energy and eventually surpass those of layer 2 (green), while the bands from layer 1 (blue) move downward.
%    Figure \ref{fig_local_AAA_elec}(b-d) represent the valence edge energy band  and wave function under the perpendicular electric field.
%    Here we label the characteristic bands hybridized $\mu-$orbit of layer 2 and $\nu-$orbit of layer 3 as the $\mu\nu1, 2,...$ in the Fig.~\ref{local_AAA_elec}(b-e).
 %   All plot follow the similar color scheme to Fig.~\ref{fig_local_band_BBA}.

  %  \mage{
  %  Figure \ref{fig_local_AAA_elec} sumarizes the electric filed dependence of the electronic properties of twisted trilayer WSe$_2$ with $(\theta^{12},\theta^{23})=(+1.0^\circ,-1.0^\circ)$ with $\alpha\alpha'$ stacking.
   % Figure~\ref{fig_local_AAA_elec}(a) is the layer-projected DOS, taking horizontal and vertical axis as the electric field $\Delta_E$ and energy, where left is the actual calculation and right figure shows only specific line of the orbits.
   % Figure \ref{fig_local_AAA_elec}(b-d) represent the valence edge energy band  and wave function under the perpendicular electric field.
   % Here we label the characteristic bands hybridized $\mu-$orbit of layer 2 and $\nu-$orbit of layer 3 as the $\mu\nu1, 2,...$ in the Fig.~\ref{local_AAA_elec}(b-e).
   % All plot follow the similar color scheme to Fig.~\ref{local_band_BBA}.
   % }

At the crossing points of orbital energies from different layers, we observe significant band hybridization and anticrossing behavior.
Figures~\ref{fig_local_AAA_elec}(b–d) show the band structure and representative wavefunctions near the crossings labeled (b), (c), and (d) in Fig.~\ref{fig_local_AAA_elec}(a).
Panel (b) corresponds to the intersection of $L2s$ and $L3s$, where the $s$-like states from layers 2 and 3 form a honeycomb lattice, resulting in a graphene-like band structure.
In this system, $\Delta_E$ effectively acts as the Dirac mass.
Likewise, the crossing point (c) gives rise to a honeycomb lattice formed by $p$ orbitals.
At $\Delta_E = 20$ meV, in particular, we observe a characteristic structure where a Dirac cone and a flat band coexist, resembling the $S = 1$ Dirac band \cite{PhysRevLett.119.206401,PhysRevLett.119.206402,PhysRevLett.122.076402,yang2019topological,sanchez2019topological,rao2019observation}.
At point (d), an unusual situation arises where $s$ and $p$ orbitals occupy different sublattices of a honeycomb lattice.
In this case, we observe quadratic band touching at $\Delta_E = 10$ meV and 12 meV.
These characteristic features can be captured by simple tight-binding models of $s$ and/or $p$ orbitals on a honeycomb lattice, as discussed in Appendix~\ref{app_effec_model}.
We also observe that the subbands often carry non-zero valley Chern numbers $C_K = -C_{K'}$.
In the effective $s/p$ tight-binding model, this behavior can be attributed to the effect of the strain-induced effective magnetic field.

At large $\Delta_E$, the electronic states near the valence band top are fully polarized on layer 3, where interlayer hybridization is absent [Fig.~\ref{fig_local_AAA_elec}(a)].
Nonetheless, non-trivial band structures still emerge due to the interplay between different orbitals within the same layer.
Figure~\ref{fig_local_AAA_d} shows the band structure near the valence band top for $\Delta_E=60$, 150 and 300 meV.
The topmost band originates from an 
$s$-orbital, and the second and third bands from $p$-orbitals (with $l_z=\pm 1$), with all orbitals arranged on a triangular lattice.
The three lower bands, enclosed by the dashed yellow rectangle, stem from the orbitals which we refer to as $d$-orbitals for convenience.
In an isolated triangular quantum well, these correspond to a single orbital with angular momentum $l_z = 0$ and a doubly-degenerate set with $l_z = \pm 1$, which happen to be nearly degenerate in energy.
The $d$-bands exhibit a characteristic structure, where a flat band is sandwiched between massive Dirac bands.
At large $\Delta_E$, where interlayer hybridization is negligible, the Dirac mass becomes smaller, leading to an approximate triple degeneracy with the flat band at the band touching point.
This characteristic band structure is well reproduced by an effective tight-binding model for $d$-orbitals, as discussed in Appendix~\ref{app_effec_model}.
For lower twist angles,
we can find energy bands related with even higher orbitals of triangular well.

 %   \mage{
 %   For the more than $\Delta_{E} >50$ meV, we can observe that the electron states around fermi energy are poralized at layer 3 as shown in Fig.~\ref{fig_local_AAA_elec}(a).
 %   In such a large electric region, d-like bands poralized at layer 3 appears and it constructs the trigonal lattice with 3 orbits characterized by the eigenvalue of $C_{3z}$ symmetry.
 %   Top figures of Fig.~\ref{fig_local_AAA_d} shows the varence band edge energy band, layer-projected DOS under the perpendicular electric field with $\Delta_E=60$, $150$ and $300$ meV.
 %   Bottom is the wave function of bands indicated as (i), (ii) and (iii) in the energy band for $\Delta_E=150$ meV.
 %   For larger electric field like $\Delta_E=300$ meV, these bands gives the Dirac and flat bands, and for smaller electric field, the gap between Dirac and flat band bigger due to the interlayer hopping.
 %   }

 %   \mage{
 %   Here we consider only $\Delta_E>0$ since the band structure for $\Delta_E<0$ corresponding to the $xy-$plane symmetry pair of $\Delta_E>0$.
 %   We can find the hybridizing states related the higher electronic orbit state like d, f for the lower twist angle.
 %   }

The band structures in the $\alpha\beta$ and $\beta\alpha$ stackings can also be modulated by applying a perpendicular electric field.
In the $\alpha\beta$ case, for example, the Kagome band localized on layer 2 can hybridize with $s$- or $p$-like orbitals on the triangular lattices of layers 1 and 3, depending on the field strength.
For the $\beta\alpha$ stacking, the electric field allows for hybridization among $s/p$ orbitals residing on three distinct triangular lattices associated with layers 1, 2, and 3.
Further details of these calculations are presented in Appendix~\ref{app_sec_elec_dependence_ab_n_ba}

Although we performed the band calculation above
using the lattice structure optimized at $\Delta_E=0$,
the vertical electric field generally affects the relaxed lattice structure.
%The vertical electric field generally modifies the relaxed lattice structure,
%while we ignored these effects in the band calculation above.
In twisted bilayer TMD, the MX and XM regions have
opposite electric polarizations along  $z$ direction \cite{li2017binary,tong2020interferences,ferreira2021weak,wang2022interfacial},
and consequently, a prependicular electric field modifies the relative sizes of these domains.\cite{enaldiev2022scalable}
In a twist angle around $\sim 1^\circ$,   
such a modulation becomes significant in an electric field range
of $\Delta_E$ exceeding a few eV \cite{enaldiev2022scalable}, which is much larger than the values considered above.
We expect relevant effects to occur in moderate electric fields
in marginal twisted trilayers with $\theta \sim \mathcal{O}(0.1^\circ)$,
which we leave for future studies.

\section{Conclusion}

We have conducted a systematic theoretical study on the structural and electronic properties of twisted trilayer WSe$_2$, revealing distinct behaviors arising from the interplay of two independent moir\'e patterns. As in twisted trilayer graphene, lattice relaxation plays a crucial role in domain formation: helical trilayers develop $\alpha\beta$ and $\beta\alpha$ domains where the two moir\'e lattices shift to minimize overlap, while alternating trilayers stabilize $\alpha\alpha'$ domains where the two moir\'e lattices align.
A key feature unique to trilayer TMD systems is the summation of moir\'e potentials from the outer layers onto the middle layer, effectively doubling its potential depth. This mechanism leads to qualitatively distinct electronic structures compared to bilayer systems. In helical trilayers, a Kagome lattice potential naturally emerges in $\alpha\beta$ domains, producing localized flat bands that could host strongly correlated states. In alternating trilayers, the deeper triangular potential wells provide stronger confinement than in twisted bilayers. The global electronic spectrum reflects a combination of bulk-like states from commensurate domains and isolated boundary modes localized at domain intersections.

%Furthermore, we have demonstrated that the layer polarization of valence states can be controlled by a vertical electric field, introducing an additional degree of tunability to trilayer moir\'e systems. 
Furthermore, we have demonstrated that the layer polarization of valence states can be controlled by a vertical electric field, introducing an additional degree of tunability to trilayer moir\'e systems. In particular, electric field modulation enables hybridization between orbitals localized on different layers, allowing for the engineering of novel electronic band structures, including Dirac bands, flat bands, and their coexistence.

Our findings highlight the significance of moir\'e potential summation as a design principle in twisted multilayer structures, offering an expanded platform for engineering moir\'e-based quantum phenomena beyond bilayer TMDs. 
Future research directions include investigating the effects of external perturbations, many-body interactions, and optical responses in these trilayer moir\'e systems.

\label{sec_con}

\section*{Acknowledgments}

We thank Kazunari Matsuda, Young-Woo Son, and Manato Fujimoto for the discussions and suggestion about this work.
This work was supported in part by 
JSPS KAKENHI Grant Number JP20H01840, JP21H05236, JP21H05232, JP25K00938 and by JST CREST Grant Number JPMJCR20T3, Japan.

%%%%%%%%%%%%%%%%%%%%%%%%%%%%%%%%%%%%%%%%%%%%%%%%%%%%%%%%%%%%%%%%%%%%%%%%%%%%%%%%%%%%%%%%%%%%%%%%%%%%%%%%%%%%%%%%%%%%%%
\appendix

%\section{Lateral shift dependence}
%figure only the corner(main)
%origin of the corner state(main)
%however, it depends on the lateral shift(discuss in the lateral shift dependence)

\section{Numerical method for lattice relaxation}
\label{app_sec_num_relax}

    Here, we derive the self-consistent equation for the structural relaxation from the Euler-Lagrange equation for the total energy $U=U_{E}+U_{B}^{12}+U_{B}^{23}$ 
    [See, Sec.~\ref{subsec_relax_model}], following the method adopted in the previous paper \cite{PhysRevX.13.041007}.
    %[See Eqs.~(\ref{eq:elastic}) and (\ref{eq:binding})].
    First, we introduce 
    \begin{align}\label{eq:transform}
        \bm{w} &= \bm{s}^{(1)}+\bm{s}^{(2)}+\bm{s}^{(3)} \nn \\ 
        \bm{u} &= \bm{s}^{(1)}-2\bm{s}^{(2)}+\bm{s}^{(3)} \nn \\
        \bm{v} &= \bm{s}^{(1)}-\bm{s}^{(3)},
    \end{align}
    and rewrite $U$ as a functional of $\bm{w}, \bm{u}$ and $\bm{v}$.
    Here $\bm{w}$ represents an overall translation of three layers, while $\bm{u}$ and $\bm{v}$ are relative sliding
    which are mirror-even and odd, respectively, with respect to the layer 2.
    In the subsequent analysis, we set $\bm{w}=\bm{0}$ and focus solely on $\bm{u}$ and $\bm{v}$, as $\bm{w}$ does not alter the interlayer registration and therefore does not impact the formation of moir\'e domains.

    We assume $\bm{s}^{(\ell)}$ (so $\bm{u}$ and $\bm{v}$) are periodic in the original moir\'e-of-moir\'e period, and define the Fourier components as
     \begin{align}\label{eq:Fourier_transform_uv}
      \bm{u}\left(\bm{r}\right) =   \sum_{\bm{G}}\bm{u}_{\bm{G}}\e^{\i\bm{G}\cdot\bm{r}},
    \quad
    \bm{v}\left(\bm{r}\right) =   \sum_{\bm{G}}\bm{v}_{\bm{G}}\e^{\i\bm{G}\cdot\bm{r}}, 
    \end{align}
    where $\bm{G}=m_{1}\bm{G}_{1}+m_{2}\bm{G}_{2}$ are the moir\'e-of-moir\'e reciprocal lattice vectors.

    The Euler-Lagrange equations %$\partial U/\partial u_{\mu}$ and $\partial U/\partial v_{\mu}$ 
    are written as
    \begin{align}\label{eq:static_sc}
        \bm{u}_{\bm{G}} &= -6V_{0}\sum_{j=1}^{3}\left(f_{\bm{G},j}^{12}+f_{\bm{G},j}^{23}\right)\hat{K}_{\bm{G}}^{-1}\bm{b}_{j}, \notag \\
        \bm{v}_{\bm{G}} &= -2V_{0}\sum_{j=1}^{3}\left(f_{\bm{G},j}^{12}-f_{\bm{G},j}^{23}\right)\hat{K}_{\bm{G}}^{-1}\bm{b}_{j},
    \end{align}
    where
    \begin{align} \label{eq:def_K}
        \hat{K}_{\bm{G}} =
	    \left(
			\begin{array} {cc}
			\left(\lambda+2\mu\right)G_{x}^{2}+\mu G_{y}^{2} & \left(\lambda+\mu\right)G_{x}G_{y} \\
			\left(\lambda+\mu\right)G_{x}G_{y} & \left(\lambda+2\mu\right)G_{y}^{2}+\mu G_{x}^{2}
			\end{array}
		\right),
    \end{align}
    and
    \begin{align}
    &\sin\left[\bm{G}_{j}^{12}\cdot\bm{r}-\bm{b}_{j}\cdot(\bm{u}+\bm{v})/2\right]
    =  \sum_{\bm{G}}f_{\bm{G},j}^{12}\e^{\i\bm{G}\cdot\bm{r}},
    \notag\\
    &\sin\left[\bm{G}_{j}^{23}\cdot\bm{r}+
    \bm{b}_{j}\cdot(\bm{u}-\bm{v})/2\right]
    =  \sum_{\bm{G}}f_{\bm{G},j}^{23}\e^{\i\bm{G}\cdot\bm{r}}.
    \label{eq:Fourier_transform}
    \end{align}

    We obtain the optimized $\bm{u}_{\bm{G}}$ and $\bm{v}_{\bm{G}}$ by solving Eq.~\eqref{eq:Fourier_transform_uv} and \eqref{eq:static_sc} self-consistently.
    We assume that the relaxation preserves the symmetries of the model $C_{3z}$.
    %Under this assumption, we take the Fourier components in $0^\circ \leq \phi(\bm{G}) < 60^\circ$, where $\phi(\bm{G})$ is the angle of the vector $\bm{G}$ respected with x-axes.
    %By the requirement of the displacement vector is real $\operatorname{Im}[\bm{s}(\bm{r})]=\bm{0}$, others Fourier components are giving by $\bm{s}_{R(2\pi/3)\bm{G}}=R(2\pi/3)s_{\bm{G}}$ and $\bm{s}_{R(\pi/3)\bm{G}}=R(-2\pi/3)s_{\bm{G}}^{*}$.
    Furthermore, We only take a finite number of the Fourier components in $|\bm{G}|<4\max \left(|n|,|m|,|n'|,|m'|\right)|G_{j}|$, which is sufficient to consider the lattice relaxation of the systems considered in this work.

    % We obtain the optimized $\bm{u}_{\bm{G}}$ and $\bm{v}_{\bm{G}}$ by solving Eq.~\eqref{eq:Fourier_transform_uv} and \eqref{eq:static_sc} self-consistently.
    % We assume that the relaxation preserves the symmetries of the model $C_{3z}$ and $C_{2z}$.
    % Under this assumption, we take the Fourier components in $0^\circ \leq \phi(\bm{G}) < 60^\circ$, where $\phi(\bm{G})$ is the angle of the vector $\bm{G}$ respected with x-axes.
    % Others Fourier components are giving by $s_{R(\pi/3)\bm{G}}=R(\pi/3)s_{\bm{G}}$.
    % Here, we should note that the continuum method for twisted TMD has the $C_{2z}$ symmetry even the original system breaks this symmetry due to the atomic alignments of MX and XM stacks. 
    % This additional $C2_z$ symmetry of the continuum method arises because the model ignores the atomic structure, and the local binding energies of MX and XM stacks are identical.
    % Furthermore, We only take a finite number of the Fourier components in $|\bm{G}|<3\max \left(|n|,|m|,|n'|,|m'|\right)$, which is enough to consider the lattice relaxation of the system.

In this scheme, the $\bm{G} = 0$ component of $\bm{s}^{(\ell)}$, which corresponds to a uniform translation of layer $\ell$, cannot be optimized directly due to $\hat{K}_{\bm{G}=0} = 0$ in Eq.~\eqref{eq:static_sc}. 
These components should instead be treated as parameters,
and the optimized structure can then be determined by comparing the total energies of self-consistent solutions obtained for different choices of $\bm{s}^{(\ell)}_{\bm{G}=0}$.
For commensurate trilayers where the moir\'e 12 and 23 share the same periodicity (i.e., $\bm{G}_j^{12}=\bm{G}_j^{23}$), the total energy significantly depends on $\bm{s}^{(\ell)}_{\bm{G}=0}$.
In the case of helical trilayers, the energetically favored structures are the $\alpha\beta$ and $\beta\alpha$ stackings, which are realized by setting
$\bm{s}^{(3)}_{\bm{G}=0} = \pm (\bm{a}_{1}^{(3)} + \bm{a}_{2}^{(3)})/3$ and $\bm{s}^{(1)}_{\bm{G}=0} = \bm{s}^{(2)}_{\bm{G}=0} = \bm{0}$.
For the alternate trilayers, the lowest-energy configuration is the $\alpha\alpha'$ stacking, achieved when all layers are unshifted $\bm{s}^{(\ell)}_{\bm{G}=0} = \bm{0}$ for all $\ell = 1, 2, 3$.
For general trilayers $\bm{G}_j^{12}\neq\bm{G}_j^{23}$,
the total energy does not sensitively depend on the $\bm{G}=0$ components because of a large moir\'e-of-moir\'e period, so we primarily show the case with no lateral shift $\bm{s}^{(\ell)}_{\bm{G}=0}=0$.

%    It is important to note that the components of $\bm{G}=0$, which corresponds to the global lateral shift of the layer, cannot be determined by this scheme because of $\hat{K}_{\bm{G}} = 0$ in Eq.~\eqref{eq:static_sc}.
%    We can include the effect of the lateral shift by treating $\bm{s}^{(\ell)}_{\bm{G}=0}$ as a parameter and solve the above self-consistent equation for each $\bm{s}^{(\ell)}_{\bm{G}=0}$.
 %   In this paper, for the moir\'e-of-moir\'e scale calculation, we primarily show the case with no lateral shift $\bm{s}^{(\ell)}_{\bm{G}=0}=0$ for all layer.
 %   The effect of lateral shift is more dominant in the case in which two twist angles are not close, in short, the system has a quasi-crystalline lattice structure.
 %   However, when two moir\'e patterns are close but slightly different as we mainly consider in this paper, the shift effectively redefines the unit cell by translating it, even though it alters the center of the domain constructed by lattice relaxation.
 %   Thus, we mainly focus on the case without lateral shift, and discuss the effect of the lateral shift on the electronic properties in the conclusion.

 %   For the local domain calculation, we take the lateral shift of layer 3 as $\bm{s}^{(3)}_{\bm{G}=0}=-(\bm{a}_{1}^{(3)}+\bm{a}_{2}^{(3)})/3$ and $+(\bm{a}_{1}^{(3)}+\bm{a}_{2}^{(3)})/3$ for the $\alpha\beta$ and $\beta\alpha$ stacking of the helical twist domains, and $\bm{s}^{(3)}_{\bm{G}=0}=\bm{0}$ for $\alpha\alpha'$ stacking of alternate twist domain.

%%%%%%%%%%%%%%%%%%%%%
%%%%%%%%%%%%%%%%%%%%%
\section{Effective Hamiltonian for strained monolayer TMD} \label{app_sec_eff_ham}

We derive an effective valence-band Hamiltonian for a monolayer TMD in the presence of lattice distortion.
%We set the x, y-axis on the in-plane direction of monolayer TMD and the z-axis on the out-plane.
%We define the lattice vector $a_{1}=a(1,0)$ and $a_{2}=a(\sqrt{3}/2,1/2)$ and the reciprocal lattice vector $b_{1}=(4\pi/\sqrt{3}/a)(-1/2,-\sqrt{3}/2)$ and $b_{2}=(4\pi/\sqrt{3}/a)(0,1)$ for the non-strained TMD.
%The corner of the Bulliuan zone is defined $\bm{K}_{\xi}=-\xi(2b_{1}+b_{2})/3$ where $\xi=\pm1$.
We assume the system has the only in-plane lattice distortion specified the displacement vector $\bm{s}(\bm{r})$.

In the basis of the conduction and valence band $(\psi_{c},\psi_{v})$, the  effective Hamiltonian at valley $\xi=\pm$ is given by \cite{PhysRevB.98.075106}
\begin{align}
\label{eq1}
H^{\xi} = F_{0}\sigma_{0} + F_{z}\sigma_{z} + \bm{F}\cdot\bm{\sigma},
\end{align}
where $\sigma_\mu$ is the Pauli matrix,
$\sigma_{0}$ is the $2\times2$ identity matrix. The $F_i$ are defined by
\begin{align}
\label{def:F_p}
 &   F_{0} = f_{0}+f_{3}\sum_{\mu=x,y}s_{\mu\mu}, \nn \\
&  F_{z} = f_{1}+f_{4}\sum_{\mu=x,y}s_{\mu\mu}, \nn \\
 &  \bm{F} = (F_x,F_y) = f_{2}a \left(\bm{k}+\frac{e}{\hbar}\bm{A}\right),
\end{align}
where $s_{\mu\nu}=\left(\partial_{\mu}s_{\nu}+\partial_{\nu}s_{\mu}\right)/2$ is the tensole strain, $\bm{A} = \xi(\hbar f_{5})/(e f_{2}a)\left(s_{xx}-s_{yy},-2s_{xy}\right)$ is the effective vector potential, and
$f_{j}$ are the material dependent parameters \cite{PhysRevB.98.075106}.
Here we ignore the second order terms in $s_{\mu\nu}$.

%Since TMD has a gap between the conduction and valence bands, we can apply the second perturbation theory to Hamiltonian Eq.~(\ref{eq1}) as follows.
%Assuming a sufficiently large gap between the conduction and valence bands, 
We derive the effective Hamiltonian projected onto the valence band 
by the second order perturbation as follows.
\begin{align}
&H^{\xi}= \left(F_{0}-F_{z}\right)
        + \left(
        \begin{array} {cc}
		2F_{z} & F_{-} \\
	    F_{+} & 0
		\end{array}
        \right), \nn \\
\to \, & H^{\xi}_{\rm{v}}\simeq \left(F_{0}-F_{z}\right) - F_{+} \left(2F_{z}\right)^{-1} F_{-},
\end{align}
where $F_{\pm}=F_{x}\pm\i F_{y}$.
By using Eq.~\eqref{def:F_p}, it is explicitly written as
\begin{align}
\label{eq:effective_strained_TMD}
    H^{\xi}_{\rm{v}} =& \left(f_{0}-f_{1}\right) +\left(f_{3}-f_{4}\right)\sum_{\mu=x,y}s_{\mu\mu} \, - \frac{\hbar^2}{2m^{*}}\left(\bm{k}+\frac{e}{\hbar}\bm{A}\right)^{2} \nn \\ 
    &+ \frac{\hbar^2}{2m^{*}}\left(\frac{f_{4}}{f_{1}}\right)\left(k_{+}+\frac{e}{\hbar}A_{+}\right)\left(\sum_{\mu=x,y}s_{\mu\mu}\right)\left(k_{-}+\frac{e}{\hbar}A_{-}\right),
\end{align}
where $m^{*} = (\hbar^{2}f_{1})/(f_{2}a)^2$ is the effective mass.

\section{Determining Parameters for the Continuum Model} \label{app_sec_para_det}

    \begin{figure}
        \begin{center}
        \leavevmode\includegraphics[width=1. \hsize]{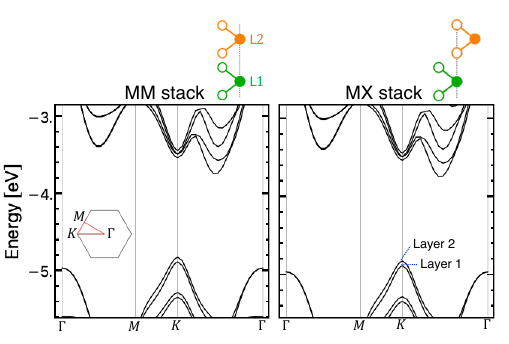}
        \caption{
        DFT band calculation of the non-twist bilayer WSe$_2$ for MM and MX stacking.
        Hexagon in the left panel shows the Brilloouin zone of bilayer WSe$_{2}$, where red line represents the k-space path we took in the DFT calculation.
        Right top figures of each panel are atomic arrangement of each stacking, where green and orange correspond to the layer 1 and 2 respectively.
        }
        \label{app_dft_band}
        \end{center}
    \end{figure}

    We determine the parameters $V,w$ and $\psi$ of continuum Hamiltonian for twisted bilayer TMDs based on DFT calculations \cite{PhysRevLett.122.086402,devakul2021magic,PhysRevB.108.085117,PhysRevResearch.5.L032022,PhysRevLett.132.036501,doi:10.1073/pnas.2316749121}.
     First, we perform the DFT band calculation for non-twist bilayer TMDs with MM,
     MX and XM stacking, which correspond to
    interlayer lateral shift vector $\bm{d}=\bm{0}$ and $\pm(\bm{a}_{1}+\bm{a}_{2})/3$, respectively.
 For each case, we obtain the $k\cdot p$ Hamiltonian at near $K$ point
 to fit the DFT band structure
 and then determine the parameters to reproduce the valence band structure.

    We performed density functional theory (DFT) calculations for non-twisted bilayer WSe2 by using Vienna Ab initio Simulation Package (VASP)\cite{KRESSE199615,PhysRevB.54.11169}, with
 the Perdew-Burke-Ernzerhof (PBE) generalized gradient approximation (GGA)\cite{PhysRevLett.77.3865} and DFT-D3 method of Grimme with zero-damping function\cite{10.1063/1.3382344}. 
    The spin-orbit-coupling effects are included self-consistently\cite{PhysRevB.93.224425}. 
    Plane wave cutoff energy was set to 335 eV. 
    We sampled the 11×11×1 grid of $k$ points for the Brillouin-zone by Gamma centered mesh. 
    In $z$ direction, a vacuum layer thicker than 25 \AA is introduced to eliminate interactions across supercells.
%    We firstly optimized the structure of AB stacking. 
    We shift the top layer in parallel relative to the bottom layer over 6×6 grid in the unit cell and calculated 21 symmetrically unique shifted bilayers. 
The interlayer distance of each laterally shifted bilayer configuration is optimized to the value that minimizes the total energy.

    Fig.~\ref{app_dft_band} is the DFT band structure of bilayer TMD with MM and MX stacking.
    To compare the calculation of the different stackings, we take the vacuum energy as the zero energy.
    Here, the vacuum energy can be calculated as a planar average potential at the
    center of the vaccuum layer.
    The first and second valence bands at the $K$-point are split by $\Delta E_{K}^{MM} = 85.91$ meV and $\Delta E_{K}^{MX} = 60.4$ meV for MM and MX stackings, respectively. The valence band tops of MM and MX stackings differ by $E^{MM}_{K,1st} - E^{MX}_{K,1st} = 25.20$ meV.
    The first and second valence band states in MX stacking are predominantly polarized to layer 2 and layer 1, respectively.

%    Fig.~\ref{app_dft_band} is the DFT band structure of bilayer TMD with MM and MX stacking.
%    To compare the calculation of the different stackings, we take the vacuum energy as the zero energy.
%    Here, we define the vacuum energy as the constant term of the palanar average potential along to $z$-axis, whre the palanar average potential is the average of the potential on the $xy$-plane at the specific $z$ value.
%    We can see the first and second valence bands at K-point are split with $\Delta E_{K}^{MM}=85.91$ meV and $\Delta E_{K}^{MX}=60.4$ meV for MM and MX stacking, and valence band top of MM and MX stacking are shifted with $\Delta E'_{K}=E^{MM}_{K,1st}-E^{MX}_{K,1st}=25.20$ meV.

    We define the effective valence-band Hamiltonian of TMD bilayer at $K$ point as \cite{PhysRevLett.122.086402,devakul2021magic,PhysRevB.108.085117,PhysRevResearch.5.L032022,PhysRevLett.132.036501,doi:10.1073/pnas.2316749121} 
    \begin{align}\label{eq_app_hamiltonian_non_twist}
        {H}_{K} =
	    \left(
			\begin{array} {cc}
		    \Delta_{+}(\bm{d}) & \Delta_{T}^{*}(\bm{d})\\
		      \Delta_{T}(\bm{d}) & \Delta_{-}(\bm{d})
			\end{array}
		\right),
    \end{align}
    where 
    \begin{equation} \Delta_{T}\left(\bm{d}\right)=w\left[1+\e^{\bm{b}_{1}\cdot\bm{d}}+\e^{\left(\bm{b}_{1}+\bm{b}_{2}\right)\cdot\bm{d}}\right],
    \end{equation}
    and
    \begin{equation} \label{app_intra_pot}
        \Delta_{\pm}\left(\bm{d}\right)=2V\sum_{j=1}^{3} \cos\left[\bm{b}_{j}\cdot\bm{d}\pm\psi\right].
    \end{equation}
    Here we take the valence band of layer 1 and 2 as the basis $(\psi^{(1)},\psi^{(2)})$.
    The parameters $(w, V, \psi)$ are determined by fitting to  $\Delta E_{K}^{MM}$, $\Delta E_{K}^{MX}$, $E^{MM}_{K,1st} - E^{MX}_{K,1st}$, and the layerwise polarization in MM and MX stackings in the DFT calculation. The resulting parameters are $(w, V, \psi) = (14.32~\mathrm{meV}, 5.966~\mathrm{meV}, 77.02^\circ)$.

\section{Impact of lattice relaxation on electronic structure}
%\section{Comparison of Non-relaxed and relaxed}
\label{app_non-relax_vs_relax}

\begin{figure*}
        \begin{center}
        \leavevmode\includegraphics[width=1. \hsize]{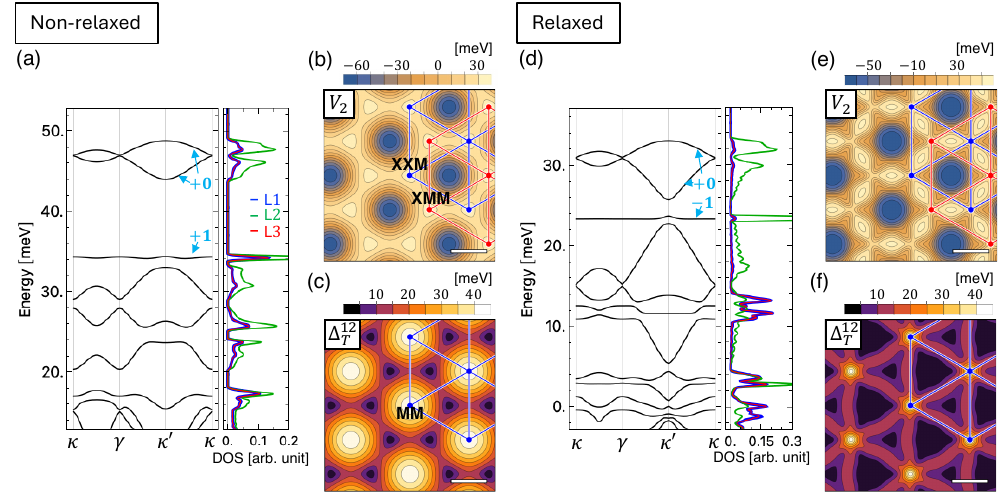}
        \caption{
        (a) Band structure and the layer-projected DOS near the valence band edge for non-relaxed $\alpha\beta$-stacked twist trilayer WSe$_2$ with helical twist angles, $(\theta^{12},\theta^{23})=(+1.0^\circ,+1.0^\circ)$. (b,c) Contour plot of (b) the intralayer potential of middle layer $V_2$ and (c) the absolute value of the interlayer hopping of moiré 12 $|\Delta_T^{12}|$ for non-relaxed case. (d-f) Similar plots for the relaxed twisted trilayer WSe$_2$, also shown in Fig.~\ref{local_band_BBA}(a) and (b). The white scale bar represents $10$ nm.
        %Similar plot of relaxed electric band structure and projected DOS of $(\theta^{12},\theta^{23})=(1.50^\circ,-1.20^\circ)$ under the perpendicular electric field (a) $V=-50$ meV and (c) $V=+50$ meV.
        %(b,d) Real space distribution of the squared amplitude of the wave function of the top energy band in Fig.~(a) and (c) respectively.
        %The white scale bar represents $10$ nm.
        }
        \label{app_BBA_non-relax_n_relax}
        \end{center}
    \end{figure*}

The lattice relaxation plays a singinicant role in the electronic structures of trilayer TMD
 Figure~\ref{app_BBA_non-relax_n_relax}
 compares the band structure and the moir\'e potentials in 
non-relaxed (left) and relaxed (right)
 $\alpha\beta$-stacked twist trilayer WSe$_{2}$  with $(\theta^{12}, \theta^{23})=(1.0^{\circ}, 1.0^{\circ})$.
 In each figure, the left panels shows the band structure and layer-resolved DOS, 
 and the right figures display the middle-layer intralayer potential $V_2$ and the  interlayer hopping  $\Delta_T^{12}$ (between layer 1 and 2).
We observe that, in the non-relaxed case, the potential $V_2$ exhibits a honeycomb-like pattern, in contrast to the kagome lattice structure seen in the relaxed case.
As discussed in the main text, the kagome potential arises from the overlap of triangular potentials $V_1$ and $V_3$, each featuring flat tops/bottoms and sharply defined boundaries.
Without lattice relaxation, however, these individual potentials become smoother and less distinct, causing the kagome structure to dissolve into a simpler hexagonal pattern.

   % \mage{
   % Figure~\ref{app_BBA_non-relax_n_relax}(a-c) shows the band structure, the intralayer potential of layer 2, and interlayer hopping of moir\'e 12 for the $\alpha\beta$ stacking without relaxation.
   % The corresponding plots for the relaxed case are shown Fig.~\ref{app_BBA_non-relax_n_relax} in (d-f).
   % As shown in Fig.~\ref{app_BBA_non-relax_n_relax}(b), the intralayer potential of layer 2 in the non-relaxed case exhibits a honeycomb lattice patter unlike relaxed case having kagome lattice.
   % For non-relaxed case, the local atomic structure near the unit cell boundary does not correspond to MX/XM stacking for both moir\'e patterns, and there is no exact MXM stacking.
   % As a result, the maxima of $V_2$ appear at MXX and XXM stacking, forming the honeycomb lattice.
   % }

The energy band structure is sensitively dependent on this change.
The flat band comprising a Kagome band in the relaxed case
is significantly moved downward in the non-relaxed case,
and it can be interpreted as a flat band in $p_x$-$p_y$ honeycomb lattice \cite{PhysRevLett.99.070401,PhysRevB.77.235107,PhysRevB.101.205311}.
We also observe that 
The projected DOS for the non-relaxed case shows that the states around Fermi energy are only weakly localized at layer 2 compared to the relaxed case.
This occurs because the interlayer hopping $\Delta_T^{12}$ is significantly suppressed by lattice relaxation as shown in Figure~\ref{app_BBA_non-relax_n_relax}.

 %   \mage{
 %   The energy band structure is strongly influenced by $V_2$, similar to relaxed case.
 %   The highest two bands in the spectrum form the s-orbit structure in the honeycomb lattice [See Fig.~\ref{app_BBA_non-relax_n_relax}(a)], and the next four bands correspond to the px- and py-orbit bands.
 %   The finite dispersion of flat bands and energy gaps are result from the effect of the interlayer hopping and weak electron confinement.
 %   }
    
%    Figure~\ref{app_BBA_non-relax_n_relax} (c,f) show the absolute value of the interlayer hopping of moir\'e 12 in real space.
%    From the comparison between Fig.~(c) and (f), we can see that relaxation reduce the MM stacking region which has the large interlayer hopping, leading the effective reduction of the interlayer hopping.

\begin{figure}
        \begin{center}
        \leavevmode\includegraphics[width=1. \hsize]{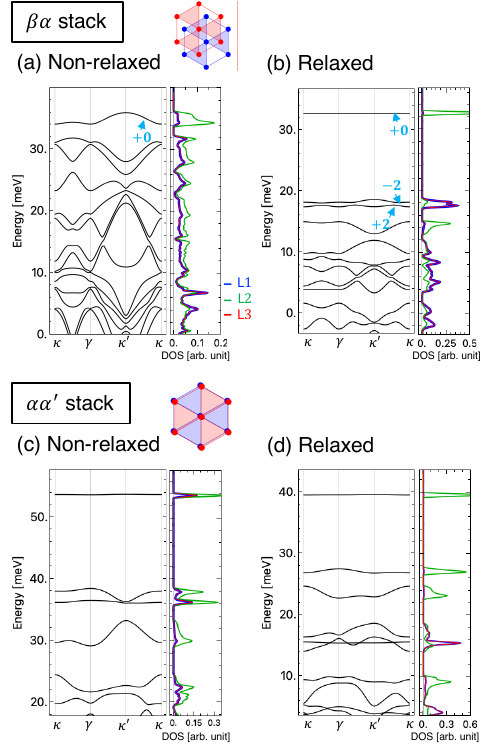}
        \caption{
        (Top) Energy band structure near the valence edge of (a) non-relaxed and (b) relaxed $\beta\alpha$-stacked twisted trilayer WSe$_2$ with $(\theta^{12},\theta^{23})=(+1.0^\circ,+1.0^\circ)$.
        (Bottom) The similar plot for the $\alpha\alpha'$ stacking.}
        \label{app_non-relax_n_relax}
        \end{center}
    \end{figure}

The energy band structures of other stackings, $\beta\alpha$ and $\alpha\alpha'$, are also affected by lattice relaxation. Figure~\ref{app_non-relax_n_relax} compares the band structures of these stackings with and without relaxation. Similar to the $\alpha\beta$ case, lattice relaxation greatly enhances the layer polarization, primarily due to the suppression of interlayer hopping. In the $\beta\alpha$ stacking, the band dispersion is significantly reduced as the triangular potential pockets deepen through relaxation.

%    \mage{
 %   The enegy band structure of others stackings, $\beta\alpha$ and $\alpha\alpha'$ are also affected by relaxation.
 %   Figure~\ref{app_non-relax_n_relax} shows the energy band structure of $\beta\alpha$ and $\alpha\alpha'$ stacking for non-relaxed and relaxed cases.
 %   In the $\beta\alpha$ stacking, the band dispersion are suppressed by relaxation because the maximum region of $V_{2}$(MXM stacking) expand, enhancing the electron confinement.
 %   For both stackings, the reduction of interlayer hopping leads the finite amplitude of layer 1 and 3 on the state near the fermi energy.
 %   }
    
    % \begin{figure*}
    %     \begin{center}        \leavevmode\includegraphics[width=.8 \hsize]{app_local_AAA_elec_depend.pdf}
    %     \caption{
    %     \mage{
    %     a
    %     }
    %     }
    %     \label{app_local_AAA_elec_depend}
    %     \end{center}
    % \end{figure*}

    % \begin{figure*}
    %     \begin{center}        \leavevmode\includegraphics[width=.8 \hsize]{app_local_BBA_elec_depend.pdf}
    %     \caption{
    %     \mage{
    %     a
    %     }
    %     }
    %     \label{app_local_BBA_elec_depend}
    %     \end{center}
    % \end{figure*}

    % \begin{figure*}
    %     \begin{center}        \leavevmode\includegraphics[width=.8 \hsize]{app_local_AAB_elec_depend.pdf}
    %     \caption{
    %     \mage{
    %     a
    %     }
    %     }
    %     \label{app_local_AAB_elec_depend}
    %     \end{center}
    % \end{figure*}

\section{The effective model for hybridized electron state under the electric field} \label{app_effec_model}

In this section, we introduce effective models that qualitatively capture the field-tunable $s/p/d$-hybridized electron states in the $\alpha\alpha'$ TMD trilayer   [Fig.~\ref{fig_local_AAA_elec}, Sec.~\ref{sec_effect_electric_field}].

\subsection{p-p honeycomb model}\label{app_subsec_pp_model}

\begin{figure}
        \begin{center}  \leavevmode\includegraphics[width=1. \hsize]{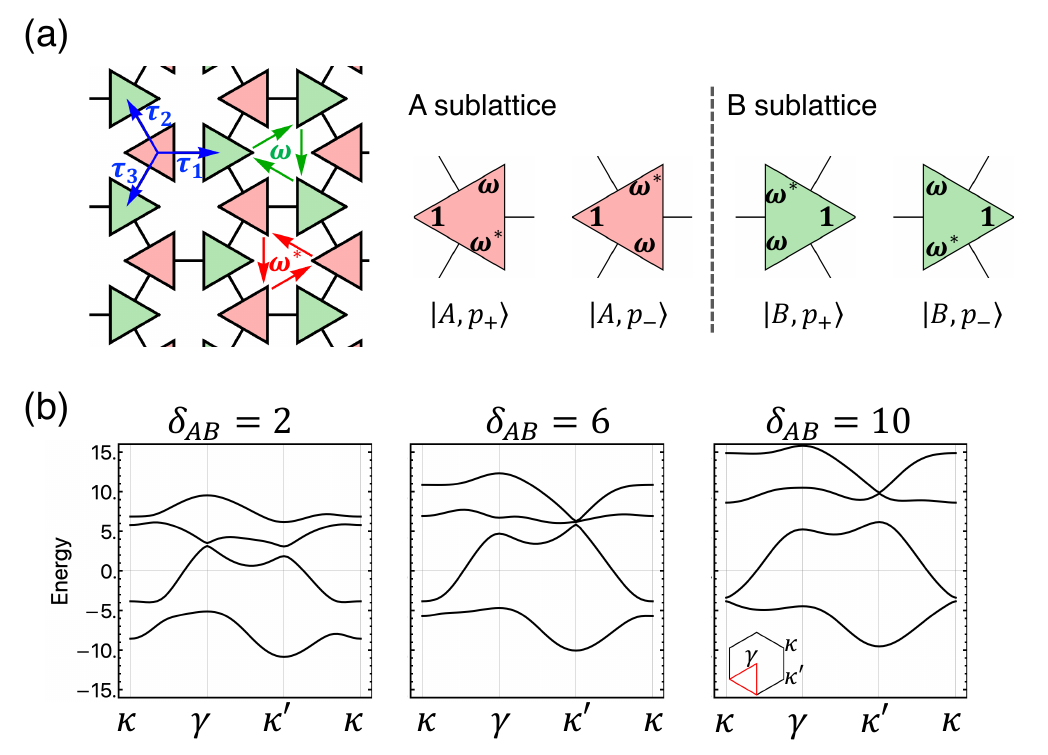}
        \caption{
(a) Schematic illustration of $p$-$p$ honeycomb model and the orbital $\Ket{X, p_{\pm}} (X=A,B)$.
The numbers $1,\omega,\omega^*$ represent the phase of the wavefunction at triangle corners.
(b) Energy band structures of the $p$-$p$ honeycomb model for different values of
$\delta_{AB} = E_A - E_B$, with parameters
$(\Delta h_{A},~\Delta h_{B},~t_{pp}^{AA},~t_{pp}^{BB},t_{pp}^{AB}) = (+5, +10, -0.8, -0.4, -1)$.
The model qualitatively captures the band structure shown in Fig.~\ref{fig_local_AAA_elec}(c).
%The inserted hexagonal of the right panel shows the B.Z. and red line is the pass we took in this calculation.
}
        \label{fig_pp_model}
        \end{center}
    \end{figure}

We first consider the energy bands highlighted in Fig.~\ref{fig_local_AAA_elec}(c). These can be effectively modeled by a honeycomb lattice composed of $p$-orbitals, denoted as $\ket{X, p_{\pm}}$, where $X = A, B$ indicates the sublattice, and $\pm$ labels the eigenstates of the $C_{3z}$ rotation symmetry. Specifically, the $C_{3z}$ operation acts as $C_{3z} \ket{X, p_{\pm}} = \omega^{\pm 1} \ket{X, p_{\pm}}$, with $\omega = \exp(i 2\pi/3)$.
Figure~\ref{fig_pp_model}(a) shows a schematic illustration of the lattice structure with orbital $\ket{X, p_{\pm}}$, where the numbers at each corner indicate the phase of the orbital wave function.
%Figure~\ref{fig_pp_model}(a) presents a schematic illustration of the $\ket{X, p_{\pm}}$ orbitals for sublattices $X = A, B$, where the numbers at each corner indicate the phase of the orbital wave function. Figure~\ref{fig_pp_model}(b) shows a schematic diagram of the lattice structure.
We define the primitive lattice vectors $\bm{a}_{j}=R(\pi (j-1)/3)(0,-1)$ for $(j=1,\cdots,6)$,
and also vectors connecting an $A$ site to three neighboring $B$ sites,
$\bm{\tau}_{1}=(1/\sqrt{3},0)$,
$\bm{\tau}_{2}=(-1/(2\sqrt{3}),1/2)$,
and
$\bm{\tau}_{3}=(-1/(2\sqrt{3}),-1/2)$.

 %   As shown in Fig.~\ref{fig_local_AAA_elec}(c), twisted trilayer WSe$_{2}$ hosts the honeycomb lattie electronic state constructed by p-like state localizing on the MXM and XMX trigonal domains.
 %   From the above result, here, we consider the honeycomb lattice with two sublattice, A and B, where both sublattice have the two orbits $\Ket{Xp_{+}}$ and $\Ket{Xp_{-}}$ for $X=A,B$, characterized by the eigenvalue of the $C_{3z}$ symmetry: $C_{3z}\Ket{Xp_{\pm}}=\omega^{\pm1}\Ket{Xp_{\pm}}$, where $\omega=\exp(\i 2\pi/3)$. 

%    Figure \ref{fig_pp_model}(a) shows the schematic figures of the $\Ket{Xp_{\pm}}$ orbits for $X=A,B$ sublattice, where values at the each corner shows the phase of the wave function of orbits, and Figure \ref{fig_pp_model}(b) indicates a schematic figure of the lattice structure.
%    As we show later, this model has the Haldane type hopping with imaginary phase $\omega$ and $\omega^{*}$ for each sublattice, representing by the red and green arrows.

The Hamiltonian is written in the basis $(\Ket{Ap_{+}}, \Ket{Ap_{-}}, \Ket{Bp_{+}}, \Ket{Bp_{-}})$ as 
    \begin{align} \label{eq_pp_ham}
        	H_{pp}=\left(
        		\begin{array}{cccc}
        		  E_{A+} + h_{A+} & h^{*}_{A-,A+} & u^{*}_{B+,A+} & u^{*}_{B-,A+} \\
        		  h_{A-,A+} & E_{A-} + h_{A-} & u^{*}_{B+,A-} & u^{*}_{B-,A-} \\
    		  u_{B+,A+} & u_{B+,A-} & E_{B+} + h_{B+} & h^{*}_{B-,B+} \\
                      u_{B-,A+} & u_{B-,A-} & h_{B-,B+} & E_{B-} + h_{B-}
        		\end{array}
        	\right).
    \end{align}
Here $u_{B\mu, A\nu} (\mu, \nu=\pm)$ is the nearest-neighbor hopping term from A to B sublattice, given by
    \begin{align} \label{eq_pp_hop_nn}
    	u_{B+,A+}(\bm{k})&=u_{B-,A-}(\bm{k})) \\ \nn
        &=t_{pp}^{AB}\left(-\e^{-\i \bm{k}\cdot\bm{\tau}_{1}}-\e^{-\i \bm{k}\cdot\bm{\tau}_{2}}-\e^{-\i \bm{k}\cdot\bm{\tau}_{3}}\right), \nn \\
    	u_{B-,A+}(\bm{k})&=t_{pp}^{AB} \sum_{j=1}^{3} 2 \omega^{j-1} \e^{-\i\bm{k}\cdot\bm{\tau}_{j}}, \nn \\
    	u_{B+,A-}(\bm{k})&=\left( \omega \leftrightarrow \omega^{*}~\text{in}~u_{B-,A+}\right),
    \end{align}
where we assume that the hopping integral between neighboring A and B sites is proportional to a summation of the phase differences at the two shared corners.
%For the hopping from $|A,p_+\rangle$ to $|B,p_+\rangle$ displaced by $\bm{\tau}_{1}$, for instance,
%the phases at the upper angle changed from $\omega$
%to $\omega^*$,
%and those at the lower angle changed from
%$\omega^*$ to $\omega$.
%Consequently, the hopping is given by
%$t_^{AB}_{pp} [(\omega^*)^* \omega + omega^* \omega^*] = $
The $h_{A\pm}$, $h_{B\pm}$, $h_{A+,A-}$ and $h_{B+,B-}$ , are the next nearest-neighbor hopping terms between the same sublattices, defined by 
    \begin{align} \label{app_pp_hop_nnn1}
    	&h_{A+}(\bm{k})=h_{B-}(\bm{k}) \nn \\
        &=t_{pp}^{AA/BB} \left[\sum_{j=1,3,5} \omega \e^{-\i \bm{k}\cdot\bm{a}_{j}}+\sum_{j=2,4,6} \omega^{*} \e^{-\i \bm{k}\cdot\bm{a}_{j}}\right], \nn \\
    	&h_{A-}(\bm{k})=h_{B+}(\bm{k}) \nn \\
        &=t_{pp}^{AA/BB} \left[\sum_{j=1,3,5} \omega^{*} \e^{-\i \bm{k}\cdot\bm{a}_{j}}+\sum_{j=2,4,6} \omega \e^{-\i \bm{k}\cdot\bm{a}_{j}}\right],
    \end{align}
    and
    \begin{align} \label{app_pp_hop_nnn2}
    	&h_{A-,A+}(\bm{k})=h_{B-,B+}(\bm{k})\nn \\
        &=t_{pp}^{AA/BB}\left[\left(\e^{-\i \bm{k}\cdot\bm{a}_{1}} + \e^{-\i \bm{k}\cdot\bm{a}_{4}}\right)\right. \nn \\
        &\left.\quad\quad +\omega^{*} \left(\e^{-\i \bm{k}\cdot\bm{a}_{2}} + \e^{-\i \bm{k}\cdot\bm{a}_{5}}\right)\nn + \omega\left(\e^{-\i \bm{k}\cdot\bm{a}_{3}} + \e^{-\i \bm{k}\cdot\bm{a}_{6}}\right) \right].
    \end{align}
%    where we introduce the primitive lattice vector as $\bm{a}_{1}=(0,-1)$ and $\bm{a}_{j}=R(\pi j/3)\bm{a}_{1}$ for $(j=2,3,4,5,6)$.

$E_{A\pm}$ and $E_{B\pm}$ are the constant on-site energy given by
\begin{align} %\label{app_pp_onsite}
        E_{A\pm} &= E_A \pm\Delta h_A/2 \nn \\
        E_{B\pm} &= E_B \pm\Delta h_B/2.
\end{align}
Noting that the A and B sublattices correspond to layer 3 and layer 2, respectively, in the original calculation of Sec.~\ref{sec_effect_electric_field},
$E_A$ and $E_B$ represent the on-site energies of the layers under a perpendicular electric field, where $E_A - E_B$ corresponds to $\Delta_E$.
$\Delta h_A$ and $\Delta h_B$ describe the orbital Zeeman splitting of $p_{\pm}$ orbitals due to the strain-induced effective magnetic field.
Here we assume $\Delta h_B = 2\Delta h_A$, reflecting the fact that the displacement vector from lattice relaxation on layer 2 is approximately twice as large as that on layer 3.

 Figure~\ref{fig_pp_model}(b) shows the the band structure in different values of $\delta_{AB} \equiv E_A-E_B$, with $(\Delta h_{A},~\Delta h_{B},~t_{pp}^{AA},~t_{pp}^{BB},~t_{pp}^{AB})=(5,10,-0.8,-0.4,-1)$.
At $\delta_{AB}=6$, we observe that the Dirac cones and flat bands are nearly degenerate at a single energy, showing a qualitative agreement with the band structure at $\Delta_E = 20$meV
in Fig.~\ref{fig_local_AAA_elec}(c).

  %  $E_{\mu}$ for $(\mu=A+,~A-,~B+,~B-)$ are the constant on-site energy for each orbit.
%    Here we introduce the constant energy: $\Delta E_{A}$ and $\Delta E_{B}$ as the energy shift between $p_{+}$ and $p_{-}$ for each sublattice coming from the time-reversal symmetry breaking, and energy difference between two sublattice by $\delta_{AB}$ corresponding to the energy difference between layer given by the electric field $\Delta_E$ in the real calculation.
%    With these values, on-site energy of each orbit and sublattice can be written by
%    \begin{align}\label{app_pp_ham}
%        E_{A\pm}&=\pm\Delta E_{A}/2 + \delta_{AB}, \nn \\
%        E_{B\pm}&=\pm\Delta E_{A}/2.
%    \end{align}
    
%    In the real calculation shown in Ch.~\ref{sec_effect_electric_field}, A and B sublattice correspond to the layer 3 and 2, and $p_{\pm}$ on each sublattice are split by the orbital Zeeman splitting due to the strain-induced magnetic field.
%    Since the displacement vector of relaxation on layer 2 is almost double than one on layer 3, we assume $\Delta E_{A}=2\Delta E_{B}$.
%    We also assume that all hopping parameter have the same value in this model.
    
 %   Figure~\ref{fig_pp_model}(c) shows the $\delta_{AB}$ dependence of the energy band structure of this model with $(\Delta E_{A},~\Delta E_{B},~t_{pp}^{AA},~t_{pp}^{BB},~t_{pp}^{AB})=(+5,+10,-0.8,-0.4,-1)$.
 %   At $\delta_{AB}=6$, we observe that the Dirac and flat bands, compared with Fig.~\ref{fig_local_AAA_elec}(c).

\subsection{s-p honeycomb model}
 
\begin{figure}
        \begin{center}  \leavevmode\includegraphics[width=1. \hsize]{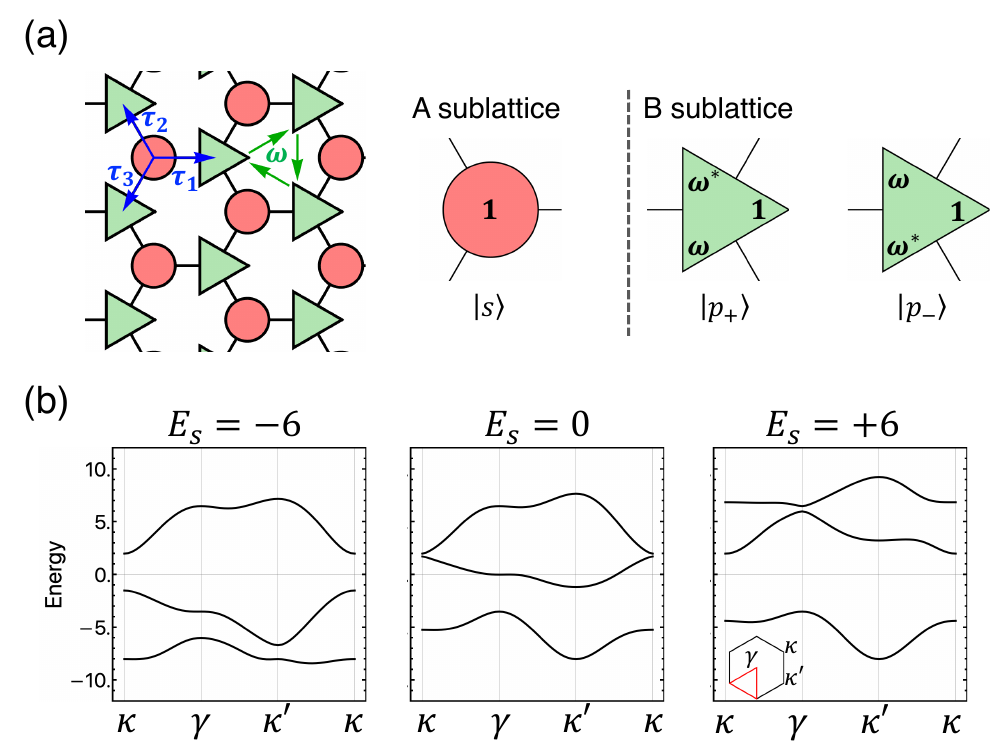}
        \caption{
(a) $s$-$p$ honeycomb model.
(b) Energy band structures of the $p$-$p$ honeycomb model for different values of
$E_s$, with parameters $(E_{p_{+}},~E_{p_{-}},~t_{ss},~t_{pp},~t_{sp})=(5,-5,0,-0.5,-1)$, to be compared with Fig.~\ref{fig_local_AAA_elec}(d).
        }
        \label{fig_sp_model}
        \end{center}
    \end{figure}

In a similar manner, the energy band structure in Fig.\ref{fig_local_AAA_elec}(d) can be described by an effective $s$–$p$ orbital model.
We consider a honeycomb lattice composed of an isotropic orbital $\Ket{s}$ at the A site, and two orbitals, $\Ket{p_{+}}$ and $\Ket{p_{-}}$, at the B site, as illustrated in Fig.\ref{fig_sp_model}(a).
%Similar to pp effective model given by \ref{app_subsec_pp_model}, p-like states has the Halden type hopping but s-like orbit does not.
%    Therefore this model might be considered as the alternative Haldane model.
The Hamiltonian in the basis $(\Ket{s}, \Ket{p_{+}}, \Ket{p_{-}})$ is given by,
    \begin{align} \label{eq_sp_ham}
        	H_{sp}=\left(
        		\begin{array}{ccc}
        		  E_{s} + h_{s} & u^{*}_{p_{+},s} & u^{*}_{p_{-},s} \\
        		  u_{p_{+},s}& E_{p_{+}} + h_{p_{+}} & h^{*}_{p_{-},p_{+}} \\
                    u_{p_{-},s}& h_{p_{-},p_{+}} & E_{p_{-}} + h_{p_{-}}
        		\end{array}
        	\right),
    \end{align}
where $E_{\mu}$ for $(\mu=s,p_{+},p_{-})$ are the constant on-site energy, and $u_{\mu\nu}$ is the nearest-neighbor hopping terms defined as 
\begin{align}\label{app_sp_hop_nn}
    	u_{p_{+},s}(\bm{k})&=t_{sp}\left[\left(\omega+\omega^{*}\right)\e^{-\i\bm{k}\cdot\bm{\tau}_{1}}\right. \nn \\
        &\left.~~~~~~~~ + \left(1+\omega^{*}\right)\e^{-\i\bm{k}\cdot\bm{\tau}_{2}} + \left(1+\omega\right)\e^{-\i\bm{k}\cdot\bm{\tau}_{3}}  \right], \nn \\
    	u_{p_{-},s}(\bm{k})&=\left( \omega \leftrightarrow \omega^{*}~\text{in}~u_{s,p_{+}}\right).
    \end{align}
The $h_{\mu}$ and and $h_{\mu\nu}$ are 
the next nearest-neighbor hopping at the same sublattice, which are given by
    \begin{align} \label{app_sp_hop_nnn1}
    	h_{s}(\bm{k})&=t_{ss}\sum_{j=1}^{6} \e^{-\i \bm{k}\cdot\bm{a}_{j}}, \nn \\
    	h_{p_{+}}(\bm{k})&=t_{pp} \left[\sum_{j=1,3,5} \omega^{*} \e^{-\i\bm{k}\cdot\bm{a}_{j}}+\sum_{j=2,4,6} \omega \e^{-\i \bm{k}\cdot\bm{a}_{j}}\right], \nn \\
        h_{p_{-}}(\bm{k})&=\left( \omega \leftrightarrow \omega^{*}~\text{in}~h_{p+}\right),
    \end{align}
    and
    \begin{align} \label{app_sp_hop_nnn2}
    	h_{p_{-},p_{+}}(\bm{k})&=t_{pp}\left[\left(\e^{-\i \bm{k}\cdot\bm{a}_{1}} + \e^{-\i \bm{k}\cdot\bm{a}_{4}}\right)\right. \nn \\
        &\left.~~~~~~~~~~~~~~~~~ +\omega^{*} \left(\e^{-\i \bm{k}\cdot\bm{a}_{2}} + \e^{-\i \bm{k}\cdot\bm{a}_{5}}\right)\right. \nn \\
        &\left.~~~~~~~~~~~~~~~~~~~~~~ + \omega\left(\e^{-\i \bm{k}\cdot\bm{a}_{3}} + \e^{-\i \bm{k}\cdot\bm{a}_{6}}\right) \right].
    \end{align}
In the above formula, 
we adopt similar definitions of $\bm{a}_{j}$ and $\bm{\tau}_{j}$ to Sec.~\ref{app_subsec_pp_model}.

 The band structure of this model shows a quantitative agreement with the result of Fig~\ref{fig_local_AAA_elec}(d), where the dependence of $E_{s}$ corresponds to the electric field on-site energy $\Delta_E$.
Here, we adopted the parameters of the effective model as $(E_{p_{+}},~E_{p_{-}},~t_{ss},~t_{pp},~t_{sp})=(+5,-5,+0,-0.5,-1)$.

\subsection{d triangular model}

The Dirac and flat bands shown in Fig.~\ref{fig_local_AAA_d} can be understood by a triangular lattice model with the 3 orbits 
$\Ket{j}=0, \pm1$ with
with $C_{3z}$ eigenvalue of $\omega^j$.
%C_{3z}\Ket{j}=\omega^{j}\Ket{j}$ for $j=0,1,-1$.
The Hamiltonian in the basis $(\Ket{0},\ket{+},\Ket{-}])$ is given
    \begin{align} \label{eq_d-tri_ham}
        	H_{d}=\left(
        		\begin{array}{ccc}
        		  E_{0} + h_{0} & h^{*}_{+,0} & h^{*}_{-,0} \\
        		  h_{+,0} & E_{+} + h_{+} & h^{*}_{-,+} \\
                    h_{-,0} & h_{-,+} & E_{-} + h_{-}
        		\end{array}
        	\right).
    \end{align}
     $E_{\mu}$ for $(\mu=0,+,-)$ are the constant on-site energy.
Here the intra-orbit hopping terms are defined by
    \begin{align} \label{app_intra-d_hop_nn}
        h_{0}(\bm{k})&=t\sum_{j=1}^{6} \e^{-\i \bm{k}\cdot\bm{a}_{j}}, \nn \\
        h_{+}(\bm{k})&=\tilde{t} \left[\sum_{j=1,3,5} \omega \e^{-\i\bm{k}\cdot\bm{a}_{j}}+\sum_{j=2,4,6} \omega^{*} \e^{-\i \bm{k}\cdot\bm{a}_{m}}\right], \nn \\
        h_{-}(\bm{k})&= \left( \omega \leftrightarrow \omega^{*}~\text{in}~h_{+}\right).
   \end{align}
The inter-orbits hopping terms are given by
   \begin{align} \label{app_inter-d_hop_nnn1}  
        h_{+,0}(\bm{k})&=t\left[\omega^{*} \left(\e^{-\i\bm{k}\cdot\bm{a}_{6}}+\e^{-\i\bm{k}\cdot\bm{a}_{1}}\right)\right. \nn \\
        &\left.~~~~~~~~ +\left(\e^{-\i\bm{k}\cdot\bm{a}_{2}}+\e^{-\i\bm{k}\cdot\bm{a}_{3}}\right)+ \omega \left(\e^{-\i\bm{k}\cdot\bm{a}_{4}}+\e^{-\i\bm{k}\cdot\bm{a}_{5}}\right)\right] \nn \\
        h_{-,0}(\bm{k})&= \left( \omega \leftrightarrow \omega^{*}~\text{in}~h_{+,0}\right),
    \end{align}
    and
    \begin{align} \label{app_inter-d_hop_nnn2}
        h_{-,+}(\bm{k})&=\tilde{t}\left[\left(\e^{-\i \bm{k}\cdot\bm{a}_{1}} + \e^{-\i \bm{k}\cdot\bm{a}_{4}}\right)\right. \nn \\
        &\left.~~~~~~~~~~ +\omega^{*} \left(\e^{-\i \bm{k}\cdot\bm{a}_{2}} + \e^{-\i \bm{k}\cdot\bm{a}_{5}}\right)\right. \nn \\
        &\left.~~~~~~~~~~~~  + \omega\left(\e^{-\i \bm{k}\cdot\bm{a}_{3}} + \e^{-\i \bm{k}\cdot\bm{a}_{6}}\right) \right].
    \end{align}

%The effect of the perpendicular electric field is emulated by the parameter $E_0$.
%Figure \ref{app_d-tri-model}(c) shows 
%the band structures in some different $E_0$'s
%with
%$(~E_{+},~E_{-},~t,~\tilde{t})=(-4,+3,-3,-1,-0.5)$.
%We observe the nearly degenerate
%Dirac and flat bands at $E_0=-4$,
%where 

\begin{figure}
        \begin{center}  \leavevmode\includegraphics[width=1. \hsize]{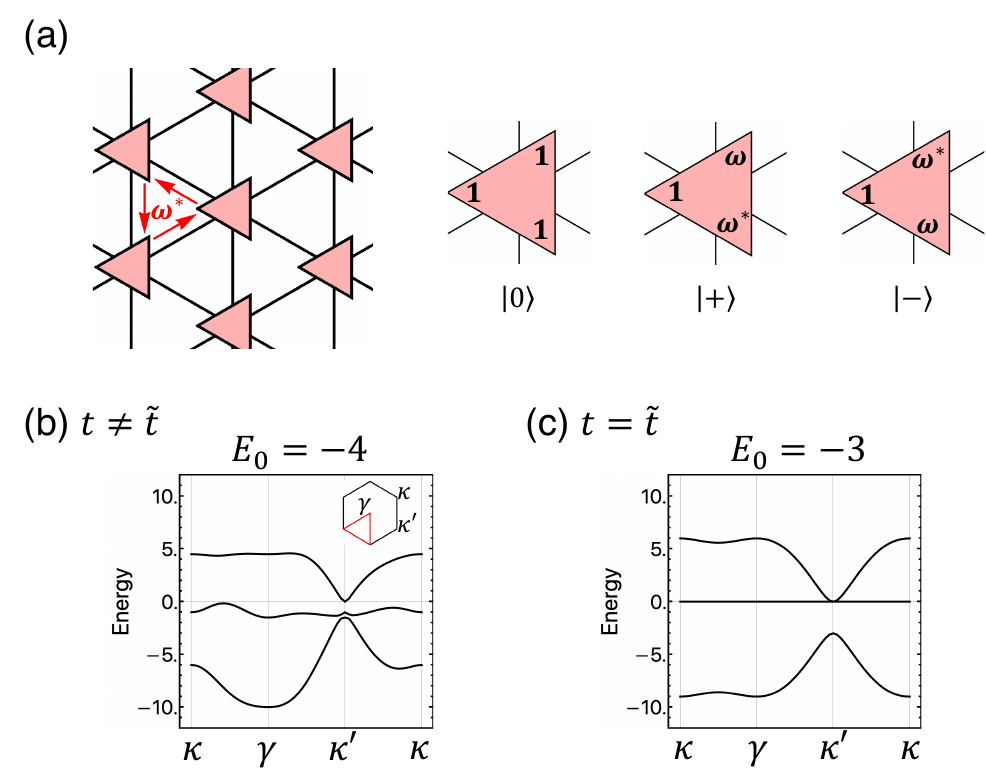}
        \caption{
        (a) $d$-orbital triangular model.
        (b) Band structure at
        $(E_{0},~E_{+},~E_{-},~t,~\tilde{t})=(-4,+3,-3,-1,-0.5)$, to be compared with Fig.~\ref{fig_local_AAA_d}.
        (c) Band structure with $(E_0, E_{+},~E_{-},~t,~\tilde{t})=(-3,+3,-3,-1,-1)$, where a perfect flat band appears.
        }
        \label{app_d-tri-model}
        \end{center}
    \end{figure}

Figure \ref{app_d-tri-model}(b) shows the band structure of the model with parameters $(E_0,E_{+},E_{-},t,\tilde{t}) = (-4, 3, -3, -1, -0.5)$, chosen to qualitatively reproduce the band structure in Fig.~\ref{fig_local_AAA_d}.
We observe a nearly degenerate Dirac band and flat band structure similar to Fig.~\ref{fig_local_AAA_d}.
Notably, this flat band is a remnant of a perfectly flat band realized in a special case with $(E_0,~E_{+},E_{-},t,\tilde{t}) = (-3, 3, -3, -1, -1)$, as shown in Fig.\ref{app_d-tri-model}(c).

%\mage{
%This flat band is came from the perfect flat, occurred for %$E_0=-3$ with $(E_{+},~E_{-},~t,~\tilde{t})=(+3,-3,-1,-1)$ of %this model [See Fig.~\ref{app_d-tri-model}(c)].
%}

%Also, this model has an exact flat band for $E_0=-3$ with $(E_{+},~E_{-},~t,~\tilde{t})=(+3,-3,-1,-1)$ [See Fig.~\ref{app_d-tri-model}(d)].

  %  As shown in Fig.~\ref{app_d-tri-model}(c), this model gives the Dirac and flat bands for $(E_{0},~E_{+},~E_{-},~t,~\tilde{t})=(-4,+3,-3,-1,-0.5)$, compared with Fig.~\ref{fig_local_AAA_d}.
  %  Also this model has the exact flat band for $E_0=-3$ and $3$ with $(E_{+},~E_{-},~t,~\tilde{t})=(+3,-3,-1,-1)$ [See Fig.~\ref{app_d-tri-model}(d)].

%%%%%%%%

\section{Effect of perpendicular electric field in $\alpha\beta$ and $\beta\alpha$ stacking} 

\label{app_sec_elec_dependence_ab_n_ba}

\begin{figure*}
         \begin{center}        \leavevmode\includegraphics[width=.95 \hsize]{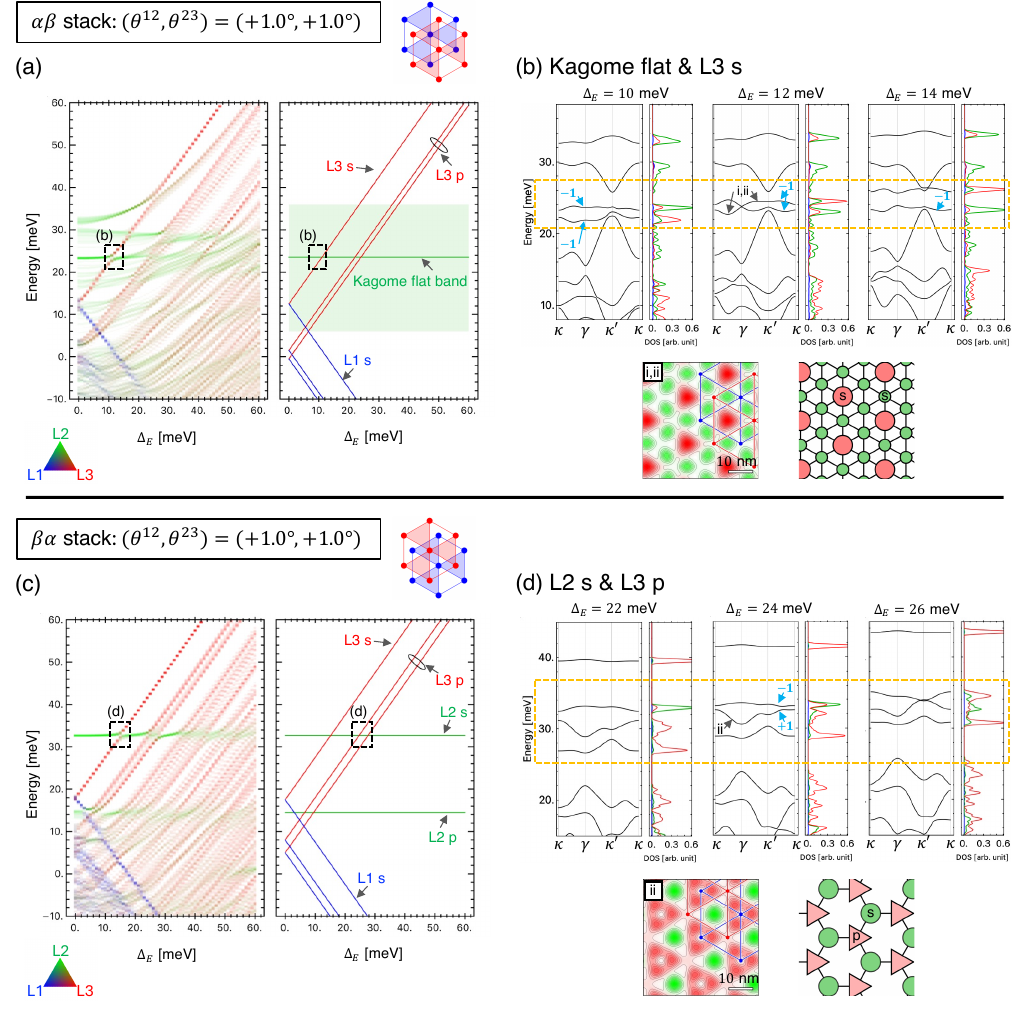}
         \caption{
Perpendicular electric field dependence of the energy band structure for $\alpha\beta$ and $\beta\alpha$ stacking twisted trilayer WSe$_{2}$ with helical twist with $(\theta^{12},\theta^{23})=(1.0,1.0)$, corresponding to Fig.~\ref{fig_local_AAA_elec}
for $\alpha\alpha'$ stacking.
         }
         \label{app_BBA_AAB_elec}
         \end{center}
     \end{figure*}

Figure~\ref{app_BBA_AAB_elec} shows the electric field dependence of the band structure for $\alpha\beta$- and $\beta\alpha$-stacked twisted trilayer WSe$_2$ with $(\theta^{12}, \theta^{23}) = (+1.0^\circ, +1.0^\circ)$, corresponding to Fig.~\ref{fig_local_AAA_elec} for the $\alpha\alpha'$ stacking.
In both cases, we observe the layer polarization switching by the electric field, similar to those seen in the $\alpha\alpha'$ trilayer.
The $\alpha\beta$ stacking hosts the kagome lattice state polarized on layer 2, and it can be hybridized with the triangular lattice state of layer 1 and 3 under the perpendicular electric field.
At $\Delta_E\approx 12$ meV, in particular, the kagome lattice and $s$-orbital trianglar lattice on layer 3 are hybridized as shown in Fig.~\ref{app_BBA_AAB_elec}(b).
For $\beta\alpha$ stacking, we find a hybrid band formed by the $s$ state of layer 2 and the $p$ state of layer 3, similar to the band observed in $\alpha\alpha'$ stacking [Fig.~\ref{fig_local_AAA_elec}(d)].
%The triangular lattice state of $\beta\alpha$ has the $C_{6z}$ symmetry not $C_{3z}$ like $\alpha\alpha'$, and it may give the different hopping term from the trianguler lattice of $\alpha\alpha'$ stacking.
    
%\section{The effective potential from the expand strain} %\label{app_sec_expand_pot}

\bibliography{reference}

\begin{thebibliography}{177}%
\makeatletter
\providecommand \@ifxundefined [1]{%
 \@ifx{#1\undefined}
}%
\providecommand \@ifnum [1]{%
 \ifnum #1\expandafter \@firstoftwo
 \else \expandafter \@secondoftwo
 \fi
}%
\providecommand \@ifx [1]{%
 \ifx #1\expandafter \@firstoftwo
 \else \expandafter \@secondoftwo
 \fi
}%
\providecommand \natexlab [1]{#1}%
\providecommand \enquote  [1]{``#1''}%
\providecommand \bibnamefont  [1]{#1}%
\providecommand \bibfnamefont [1]{#1}%
\providecommand \citenamefont [1]{#1}%
\providecommand \href@noop [0]{\@secondoftwo}%
\providecommand \href [0]{\begingroup \@sanitize@url \@href}%
\providecommand \@href[1]{\@@startlink{#1}\@@href}%
\providecommand \@@href[1]{\endgroup#1\@@endlink}%
\providecommand \@sanitize@url [0]{\catcode `\\12\catcode `\$12\catcode `\&12\catcode `\#12\catcode `\^12\catcode `\_12\catcode `\%12\relax}%
\providecommand \@@startlink[1]{}%
\providecommand \@@endlink[0]{}%
\providecommand \url  [0]{\begingroup\@sanitize@url \@url }%
\providecommand \@url [1]{\endgroup\@href {#1}{\urlprefix }}%
\providecommand \urlprefix  [0]{URL }%
\providecommand \Eprint [0]{\href }%
\providecommand \doibase [0]{http://dx.doi.org/}%
\providecommand \selectlanguage [0]{\@gobble}%
\providecommand \bibinfo  [0]{\@secondoftwo}%
\providecommand \bibfield  [0]{\@secondoftwo}%
\providecommand \translation [1]{[#1]}%
\providecommand \BibitemOpen [0]{}%
\providecommand \bibitemStop [0]{}%
\providecommand \bibitemNoStop [0]{.\EOS\space}%
\providecommand \EOS [0]{\spacefactor3000\relax}%
\providecommand \BibitemShut  [1]{\csname bibitem#1\endcsname}%
\let\auto@bib@innerbib\@empty
%</preamble>
\bibitem [{\citenamefont {Bistritzer}\ and\ \citenamefont {MacDonald}(2011{\natexlab{a}})}]{bistritzer2011moirepnas}%
  \BibitemOpen
  \bibfield  {author} {\bibinfo {author} {\bibfnamefont {R.}~\bibnamefont {Bistritzer}}\ and\ \bibinfo {author} {\bibfnamefont {A.}~\bibnamefont {MacDonald}},\ }\href@noop {} {\bibfield  {journal} {\bibinfo  {journal} {Proc. Natl. Acad. Sci.}\ }\textbf {\bibinfo {volume} {108}},\ \bibinfo {pages} {12233} (\bibinfo {year} {2011}{\natexlab{a}})}\BibitemShut {NoStop}%
\bibitem [{\citenamefont {Bistritzer}\ and\ \citenamefont {MacDonald}(2011{\natexlab{b}})}]{PhysRevB.84.035440}%
  \BibitemOpen
  \bibfield  {author} {\bibinfo {author} {\bibfnamefont {R.}~\bibnamefont {Bistritzer}}\ and\ \bibinfo {author} {\bibfnamefont {A.~H.}\ \bibnamefont {MacDonald}},\ }\href {\doibase 10.1103/PhysRevB.84.035440} {\bibfield  {journal} {\bibinfo  {journal} {Phys. Rev. B}\ }\textbf {\bibinfo {volume} {84}},\ \bibinfo {pages} {035440} (\bibinfo {year} {2011}{\natexlab{b}})}\BibitemShut {NoStop}%
\bibitem [{\citenamefont {Lopes~dos Santos}\ \emph {et~al.}(2012)\citenamefont {Lopes~dos Santos}, \citenamefont {Peres},\ and\ \citenamefont {Castro~Neto}}]{PhysRevB.86.155449}%
  \BibitemOpen
  \bibfield  {author} {\bibinfo {author} {\bibfnamefont {J.~M.~B.}\ \bibnamefont {Lopes~dos Santos}}, \bibinfo {author} {\bibfnamefont {N.~M.~R.}\ \bibnamefont {Peres}}, \ and\ \bibinfo {author} {\bibfnamefont {A.~H.}\ \bibnamefont {Castro~Neto}},\ }\href {\doibase 10.1103/PhysRevB.86.155449} {\bibfield  {journal} {\bibinfo  {journal} {Phys. Rev. B}\ }\textbf {\bibinfo {volume} {86}},\ \bibinfo {pages} {155449} (\bibinfo {year} {2012})}\BibitemShut {NoStop}%
\bibitem [{\citenamefont {Cao}\ \emph {et~al.}(2018{\natexlab{a}})\citenamefont {Cao}, \citenamefont {Fatemi}, \citenamefont {Demir}, \citenamefont {Fang}, \citenamefont {Tomarken}, \citenamefont {Luo}, \citenamefont {Sanchez-Yamagishi}, \citenamefont {Watanabe}, \citenamefont {Taniguchi}, \citenamefont {Kaxiras}, \citenamefont {Ashoori},\ and\ \citenamefont {Jarillo-Herrero}}]{cao2018_80}%
  \BibitemOpen
  \bibfield  {author} {\bibinfo {author} {\bibfnamefont {Y.}~\bibnamefont {Cao}}, \bibinfo {author} {\bibfnamefont {V.}~\bibnamefont {Fatemi}}, \bibinfo {author} {\bibfnamefont {A.}~\bibnamefont {Demir}}, \bibinfo {author} {\bibfnamefont {S.}~\bibnamefont {Fang}}, \bibinfo {author} {\bibfnamefont {S.~L.}\ \bibnamefont {Tomarken}}, \bibinfo {author} {\bibfnamefont {J.~Y.}\ \bibnamefont {Luo}}, \bibinfo {author} {\bibfnamefont {J.~D.}\ \bibnamefont {Sanchez-Yamagishi}}, \bibinfo {author} {\bibfnamefont {K.}~\bibnamefont {Watanabe}}, \bibinfo {author} {\bibfnamefont {T.}~\bibnamefont {Taniguchi}}, \bibinfo {author} {\bibfnamefont {E.}~\bibnamefont {Kaxiras}}, \bibinfo {author} {\bibfnamefont {R.~C.}\ \bibnamefont {Ashoori}}, \ and\ \bibinfo {author} {\bibfnamefont {P.}~\bibnamefont {Jarillo-Herrero}},\ }\href {\doibase 10.1038/nature26154} {\bibfield  {journal} {\bibinfo  {journal} {Nature}\ }\textbf {\bibinfo {volume} {556}},\ \bibinfo {pages} {80} (\bibinfo {year} {2018}{\natexlab{a}})}\BibitemShut {NoStop}%
\bibitem [{\citenamefont {Cao}\ \emph {et~al.}(2018{\natexlab{b}})\citenamefont {Cao}, \citenamefont {Fatemi}, \citenamefont {Fang}, \citenamefont {Watanabe}, \citenamefont {Taniguchi}, \citenamefont {Kaxiras},\ and\ \citenamefont {Jarillo-Herrero}}]{cao2018_43}%
  \BibitemOpen
  \bibfield  {author} {\bibinfo {author} {\bibfnamefont {Y.}~\bibnamefont {Cao}}, \bibinfo {author} {\bibfnamefont {V.}~\bibnamefont {Fatemi}}, \bibinfo {author} {\bibfnamefont {S.}~\bibnamefont {Fang}}, \bibinfo {author} {\bibfnamefont {K.}~\bibnamefont {Watanabe}}, \bibinfo {author} {\bibfnamefont {T.}~\bibnamefont {Taniguchi}}, \bibinfo {author} {\bibfnamefont {E.}~\bibnamefont {Kaxiras}}, \ and\ \bibinfo {author} {\bibfnamefont {P.}~\bibnamefont {Jarillo-Herrero}},\ }\href {\doibase 10.1038/nature26160} {\bibfield  {journal} {\bibinfo  {journal} {Nature}\ }\textbf {\bibinfo {volume} {556}},\ \bibinfo {pages} {43} (\bibinfo {year} {2018}{\natexlab{b}})}\BibitemShut {NoStop}%
\bibitem [{\citenamefont {Yankowitz}\ \emph {et~al.}(2019)\citenamefont {Yankowitz}, \citenamefont {Chen}, \citenamefont {Polshyn}, \citenamefont {Zhang}, \citenamefont {Watanabe}, \citenamefont {Taniguchi}, \citenamefont {Graf}, \citenamefont {Young},\ and\ \citenamefont {Dean}}]{doi:10.1126/science.aav1910}%
  \BibitemOpen
  \bibfield  {author} {\bibinfo {author} {\bibfnamefont {M.}~\bibnamefont {Yankowitz}}, \bibinfo {author} {\bibfnamefont {S.}~\bibnamefont {Chen}}, \bibinfo {author} {\bibfnamefont {H.}~\bibnamefont {Polshyn}}, \bibinfo {author} {\bibfnamefont {Y.}~\bibnamefont {Zhang}}, \bibinfo {author} {\bibfnamefont {K.}~\bibnamefont {Watanabe}}, \bibinfo {author} {\bibfnamefont {T.}~\bibnamefont {Taniguchi}}, \bibinfo {author} {\bibfnamefont {D.}~\bibnamefont {Graf}}, \bibinfo {author} {\bibfnamefont {A.~F.}\ \bibnamefont {Young}}, \ and\ \bibinfo {author} {\bibfnamefont {C.~R.}\ \bibnamefont {Dean}},\ }\href {\doibase 10.1126/science.aav1910} {\bibfield  {journal} {\bibinfo  {journal} {Science}\ }\textbf {\bibinfo {volume} {363}},\ \bibinfo {pages} {1059} (\bibinfo {year} {2019})},\ \Eprint {http://arxiv.org/abs/https://www.science.org/doi/pdf/10.1126/science.aav1910} {https://www.science.org/doi/pdf/10.1126/science.aav1910} \BibitemShut {NoStop}%
\bibitem [{\citenamefont {Kerelsky}\ \emph {et~al.}(2019)\citenamefont {Kerelsky}, \citenamefont {McGilly}, \citenamefont {Kennes}, \citenamefont {Xian}, \citenamefont {Yankowitz}, \citenamefont {Chen}, \citenamefont {Watanabe}, \citenamefont {Taniguchi}, \citenamefont {Hone}, \citenamefont {Dean}, \citenamefont {Rubio},\ and\ \citenamefont {Pasupathy}}]{Kerelsky2019}%
  \BibitemOpen
  \bibfield  {author} {\bibinfo {author} {\bibfnamefont {A.}~\bibnamefont {Kerelsky}}, \bibinfo {author} {\bibfnamefont {L.~J.}\ \bibnamefont {McGilly}}, \bibinfo {author} {\bibfnamefont {D.~M.}\ \bibnamefont {Kennes}}, \bibinfo {author} {\bibfnamefont {L.}~\bibnamefont {Xian}}, \bibinfo {author} {\bibfnamefont {M.}~\bibnamefont {Yankowitz}}, \bibinfo {author} {\bibfnamefont {S.}~\bibnamefont {Chen}}, \bibinfo {author} {\bibfnamefont {K.}~\bibnamefont {Watanabe}}, \bibinfo {author} {\bibfnamefont {T.}~\bibnamefont {Taniguchi}}, \bibinfo {author} {\bibfnamefont {J.}~\bibnamefont {Hone}}, \bibinfo {author} {\bibfnamefont {C.}~\bibnamefont {Dean}}, \bibinfo {author} {\bibfnamefont {A.}~\bibnamefont {Rubio}}, \ and\ \bibinfo {author} {\bibfnamefont {A.~N.}\ \bibnamefont {Pasupathy}},\ }\href {\doibase 10.1038/s41586-019-1431-9} {\bibfield  {journal} {\bibinfo  {journal} {Nature}\ }\textbf {\bibinfo {volume} {572}},\ \bibinfo {pages} {95} (\bibinfo {year} {2019})}\BibitemShut {NoStop}%
\bibitem [{\citenamefont {Xie}\ \emph {et~al.}(2019)\citenamefont {Xie}, \citenamefont {Lian}, \citenamefont {J{\"a}ck}, \citenamefont {Liu}, \citenamefont {Chiu}, \citenamefont {Watanabe}, \citenamefont {Taniguchi}, \citenamefont {Bernevig},\ and\ \citenamefont {Yazdani}}]{xie2019spectroscopic}%
  \BibitemOpen
  \bibfield  {author} {\bibinfo {author} {\bibfnamefont {Y.}~\bibnamefont {Xie}}, \bibinfo {author} {\bibfnamefont {B.}~\bibnamefont {Lian}}, \bibinfo {author} {\bibfnamefont {B.}~\bibnamefont {J{\"a}ck}}, \bibinfo {author} {\bibfnamefont {X.}~\bibnamefont {Liu}}, \bibinfo {author} {\bibfnamefont {C.-L.}\ \bibnamefont {Chiu}}, \bibinfo {author} {\bibfnamefont {K.}~\bibnamefont {Watanabe}}, \bibinfo {author} {\bibfnamefont {T.}~\bibnamefont {Taniguchi}}, \bibinfo {author} {\bibfnamefont {B.~A.}\ \bibnamefont {Bernevig}}, \ and\ \bibinfo {author} {\bibfnamefont {A.}~\bibnamefont {Yazdani}},\ }\href@noop {} {\bibfield  {journal} {\bibinfo  {journal} {Nature}\ }\textbf {\bibinfo {volume} {572}},\ \bibinfo {pages} {101} (\bibinfo {year} {2019})}\BibitemShut {NoStop}%
\bibitem [{\citenamefont {Jiang}\ \emph {et~al.}(2019)\citenamefont {Jiang}, \citenamefont {Lai}, \citenamefont {Watanabe}, \citenamefont {Taniguchi}, \citenamefont {Haule}, \citenamefont {Mao},\ and\ \citenamefont {Andrei}}]{jiang2019charge}%
  \BibitemOpen
  \bibfield  {author} {\bibinfo {author} {\bibfnamefont {Y.}~\bibnamefont {Jiang}}, \bibinfo {author} {\bibfnamefont {X.}~\bibnamefont {Lai}}, \bibinfo {author} {\bibfnamefont {K.}~\bibnamefont {Watanabe}}, \bibinfo {author} {\bibfnamefont {T.}~\bibnamefont {Taniguchi}}, \bibinfo {author} {\bibfnamefont {K.}~\bibnamefont {Haule}}, \bibinfo {author} {\bibfnamefont {J.}~\bibnamefont {Mao}}, \ and\ \bibinfo {author} {\bibfnamefont {E.~Y.}\ \bibnamefont {Andrei}},\ }\href@noop {} {\bibfield  {journal} {\bibinfo  {journal} {Nature}\ }\textbf {\bibinfo {volume} {573}},\ \bibinfo {pages} {91} (\bibinfo {year} {2019})}\BibitemShut {NoStop}%
\bibitem [{\citenamefont {Polshyn}\ \emph {et~al.}(2019)\citenamefont {Polshyn}, \citenamefont {Yankowitz}, \citenamefont {Chen}, \citenamefont {Zhang}, \citenamefont {Watanabe}, \citenamefont {Taniguchi}, \citenamefont {Dean},\ and\ \citenamefont {Young}}]{polshyn2019large}%
  \BibitemOpen
  \bibfield  {author} {\bibinfo {author} {\bibfnamefont {H.}~\bibnamefont {Polshyn}}, \bibinfo {author} {\bibfnamefont {M.}~\bibnamefont {Yankowitz}}, \bibinfo {author} {\bibfnamefont {S.}~\bibnamefont {Chen}}, \bibinfo {author} {\bibfnamefont {Y.}~\bibnamefont {Zhang}}, \bibinfo {author} {\bibfnamefont {K.}~\bibnamefont {Watanabe}}, \bibinfo {author} {\bibfnamefont {T.}~\bibnamefont {Taniguchi}}, \bibinfo {author} {\bibfnamefont {C.~R.}\ \bibnamefont {Dean}}, \ and\ \bibinfo {author} {\bibfnamefont {A.~F.}\ \bibnamefont {Young}},\ }\href@noop {} {\bibfield  {journal} {\bibinfo  {journal} {Nature Physics}\ }\textbf {\bibinfo {volume} {15}},\ \bibinfo {pages} {1011} (\bibinfo {year} {2019})}\BibitemShut {NoStop}%
\bibitem [{\citenamefont {Choi}\ \emph {et~al.}(2019)\citenamefont {Choi}, \citenamefont {Kemmer}, \citenamefont {Peng}, \citenamefont {Thomson}, \citenamefont {Arora}, \citenamefont {Polski}, \citenamefont {Zhang}, \citenamefont {Ren}, \citenamefont {Alicea}, \citenamefont {Refael}, \citenamefont {von Oppen}, \citenamefont {Watanabe}, \citenamefont {Taniguchi},\ and\ \citenamefont {Nadj-Perge}}]{Choi2019}%
  \BibitemOpen
  \bibfield  {author} {\bibinfo {author} {\bibfnamefont {Y.}~\bibnamefont {Choi}}, \bibinfo {author} {\bibfnamefont {J.}~\bibnamefont {Kemmer}}, \bibinfo {author} {\bibfnamefont {Y.}~\bibnamefont {Peng}}, \bibinfo {author} {\bibfnamefont {A.}~\bibnamefont {Thomson}}, \bibinfo {author} {\bibfnamefont {H.}~\bibnamefont {Arora}}, \bibinfo {author} {\bibfnamefont {R.}~\bibnamefont {Polski}}, \bibinfo {author} {\bibfnamefont {Y.}~\bibnamefont {Zhang}}, \bibinfo {author} {\bibfnamefont {H.}~\bibnamefont {Ren}}, \bibinfo {author} {\bibfnamefont {J.}~\bibnamefont {Alicea}}, \bibinfo {author} {\bibfnamefont {G.}~\bibnamefont {Refael}}, \bibinfo {author} {\bibfnamefont {F.}~\bibnamefont {von Oppen}}, \bibinfo {author} {\bibfnamefont {K.}~\bibnamefont {Watanabe}}, \bibinfo {author} {\bibfnamefont {T.}~\bibnamefont {Taniguchi}}, \ and\ \bibinfo {author} {\bibfnamefont {S.}~\bibnamefont {Nadj-Perge}},\ }\href {\doibase 10.1038/s41567-019-0606-5} {\bibfield  {journal} {\bibinfo  {journal} {Nature Physics}\ }\textbf
  {\bibinfo {volume} {15}},\ \bibinfo {pages} {1174} (\bibinfo {year} {2019})}\BibitemShut {NoStop}%
\bibitem [{\citenamefont {Sharpe}\ \emph {et~al.}(2019)\citenamefont {Sharpe}, \citenamefont {Fox}, \citenamefont {Barnard}, \citenamefont {Finney}, \citenamefont {Watanabe}, \citenamefont {Taniguchi}, \citenamefont {Kastner},\ and\ \citenamefont {Goldhaber-Gordon}}]{doi:10.1126/science.aaw3780}%
  \BibitemOpen
  \bibfield  {author} {\bibinfo {author} {\bibfnamefont {A.~L.}\ \bibnamefont {Sharpe}}, \bibinfo {author} {\bibfnamefont {E.~J.}\ \bibnamefont {Fox}}, \bibinfo {author} {\bibfnamefont {A.~W.}\ \bibnamefont {Barnard}}, \bibinfo {author} {\bibfnamefont {J.}~\bibnamefont {Finney}}, \bibinfo {author} {\bibfnamefont {K.}~\bibnamefont {Watanabe}}, \bibinfo {author} {\bibfnamefont {T.}~\bibnamefont {Taniguchi}}, \bibinfo {author} {\bibfnamefont {M.~A.}\ \bibnamefont {Kastner}}, \ and\ \bibinfo {author} {\bibfnamefont {D.}~\bibnamefont {Goldhaber-Gordon}},\ }\href {\doibase 10.1126/science.aaw3780} {\bibfield  {journal} {\bibinfo  {journal} {Science}\ }\textbf {\bibinfo {volume} {365}},\ \bibinfo {pages} {605} (\bibinfo {year} {2019})},\ \Eprint {http://arxiv.org/abs/https://www.science.org/doi/pdf/10.1126/science.aaw3780} {https://www.science.org/doi/pdf/10.1126/science.aaw3780} \BibitemShut {NoStop}%
\bibitem [{\citenamefont {Lu}\ \emph {et~al.}(2019)\citenamefont {Lu}, \citenamefont {Stepanov}, \citenamefont {Yang}, \citenamefont {Xie}, \citenamefont {Aamir}, \citenamefont {Das}, \citenamefont {Urgell}, \citenamefont {Watanabe}, \citenamefont {Taniguchi}, \citenamefont {Zhang}, \citenamefont {Bachtold}, \citenamefont {MacDonald},\ and\ \citenamefont {Efetov}}]{lu2019superconductors}%
  \BibitemOpen
  \bibfield  {author} {\bibinfo {author} {\bibfnamefont {X.}~\bibnamefont {Lu}}, \bibinfo {author} {\bibfnamefont {P.}~\bibnamefont {Stepanov}}, \bibinfo {author} {\bibfnamefont {W.}~\bibnamefont {Yang}}, \bibinfo {author} {\bibfnamefont {M.}~\bibnamefont {Xie}}, \bibinfo {author} {\bibfnamefont {M.~A.}\ \bibnamefont {Aamir}}, \bibinfo {author} {\bibfnamefont {I.}~\bibnamefont {Das}}, \bibinfo {author} {\bibfnamefont {C.}~\bibnamefont {Urgell}}, \bibinfo {author} {\bibfnamefont {K.}~\bibnamefont {Watanabe}}, \bibinfo {author} {\bibfnamefont {T.}~\bibnamefont {Taniguchi}}, \bibinfo {author} {\bibfnamefont {G.}~\bibnamefont {Zhang}}, \bibinfo {author} {\bibfnamefont {A.}~\bibnamefont {Bachtold}}, \bibinfo {author} {\bibfnamefont {A.~H.}\ \bibnamefont {MacDonald}}, \ and\ \bibinfo {author} {\bibfnamefont {D.~K.}\ \bibnamefont {Efetov}},\ }\href@noop {} {\bibfield  {journal} {\bibinfo  {journal} {Nature}\ }\textbf {\bibinfo {volume} {574}},\ \bibinfo {pages} {653} (\bibinfo {year} {2019})}\BibitemShut {NoStop}%
\bibitem [{\citenamefont {Cao}\ \emph {et~al.}(2020)\citenamefont {Cao}, \citenamefont {Chowdhury}, \citenamefont {Rodan-Legrain}, \citenamefont {Rubies-Bigorda}, \citenamefont {Watanabe}, \citenamefont {Taniguchi}, \citenamefont {Senthil},\ and\ \citenamefont {Jarillo-Herrero}}]{PhysRevLett.124.076801}%
  \BibitemOpen
  \bibfield  {author} {\bibinfo {author} {\bibfnamefont {Y.}~\bibnamefont {Cao}}, \bibinfo {author} {\bibfnamefont {D.}~\bibnamefont {Chowdhury}}, \bibinfo {author} {\bibfnamefont {D.}~\bibnamefont {Rodan-Legrain}}, \bibinfo {author} {\bibfnamefont {O.}~\bibnamefont {Rubies-Bigorda}}, \bibinfo {author} {\bibfnamefont {K.}~\bibnamefont {Watanabe}}, \bibinfo {author} {\bibfnamefont {T.}~\bibnamefont {Taniguchi}}, \bibinfo {author} {\bibfnamefont {T.}~\bibnamefont {Senthil}}, \ and\ \bibinfo {author} {\bibfnamefont {P.}~\bibnamefont {Jarillo-Herrero}},\ }\href {\doibase 10.1103/PhysRevLett.124.076801} {\bibfield  {journal} {\bibinfo  {journal} {Phys. Rev. Lett.}\ }\textbf {\bibinfo {volume} {124}},\ \bibinfo {pages} {076801} (\bibinfo {year} {2020})}\BibitemShut {NoStop}%
\bibitem [{\citenamefont {Serlin}\ \emph {et~al.}(2020)\citenamefont {Serlin}, \citenamefont {Tschirhart}, \citenamefont {Polshyn}, \citenamefont {Zhang}, \citenamefont {Zhu}, \citenamefont {Watanabe}, \citenamefont {Taniguchi}, \citenamefont {Balents},\ and\ \citenamefont {Young}}]{doi:10.1126/science.aay5533}%
  \BibitemOpen
  \bibfield  {author} {\bibinfo {author} {\bibfnamefont {M.}~\bibnamefont {Serlin}}, \bibinfo {author} {\bibfnamefont {C.~L.}\ \bibnamefont {Tschirhart}}, \bibinfo {author} {\bibfnamefont {H.}~\bibnamefont {Polshyn}}, \bibinfo {author} {\bibfnamefont {Y.}~\bibnamefont {Zhang}}, \bibinfo {author} {\bibfnamefont {J.}~\bibnamefont {Zhu}}, \bibinfo {author} {\bibfnamefont {K.}~\bibnamefont {Watanabe}}, \bibinfo {author} {\bibfnamefont {T.}~\bibnamefont {Taniguchi}}, \bibinfo {author} {\bibfnamefont {L.}~\bibnamefont {Balents}}, \ and\ \bibinfo {author} {\bibfnamefont {A.~F.}\ \bibnamefont {Young}},\ }\href {\doibase 10.1126/science.aay5533} {\bibfield  {journal} {\bibinfo  {journal} {Science}\ }\textbf {\bibinfo {volume} {367}},\ \bibinfo {pages} {900} (\bibinfo {year} {2020})},\ \Eprint {http://arxiv.org/abs/https://www.science.org/doi/pdf/10.1126/science.aay5533} {https://www.science.org/doi/pdf/10.1126/science.aay5533} \BibitemShut {NoStop}%
\bibitem [{\citenamefont {Chen}\ \emph {et~al.}(2020)\citenamefont {Chen}, \citenamefont {Sharpe}, \citenamefont {Fox}, \citenamefont {Zhang}, \citenamefont {Wang}, \citenamefont {Jiang}, \citenamefont {Lyu}, \citenamefont {Li}, \citenamefont {Watanabe}, \citenamefont {Taniguchi} \emph {et~al.}}]{chen2020tunable}%
  \BibitemOpen
  \bibfield  {author} {\bibinfo {author} {\bibfnamefont {G.}~\bibnamefont {Chen}}, \bibinfo {author} {\bibfnamefont {A.~L.}\ \bibnamefont {Sharpe}}, \bibinfo {author} {\bibfnamefont {E.~J.}\ \bibnamefont {Fox}}, \bibinfo {author} {\bibfnamefont {Y.-H.}\ \bibnamefont {Zhang}}, \bibinfo {author} {\bibfnamefont {S.}~\bibnamefont {Wang}}, \bibinfo {author} {\bibfnamefont {L.}~\bibnamefont {Jiang}}, \bibinfo {author} {\bibfnamefont {B.}~\bibnamefont {Lyu}}, \bibinfo {author} {\bibfnamefont {H.}~\bibnamefont {Li}}, \bibinfo {author} {\bibfnamefont {K.}~\bibnamefont {Watanabe}}, \bibinfo {author} {\bibfnamefont {T.}~\bibnamefont {Taniguchi}},  \emph {et~al.},\ }\href@noop {} {\bibfield  {journal} {\bibinfo  {journal} {Nature}\ }\textbf {\bibinfo {volume} {579}},\ \bibinfo {pages} {56} (\bibinfo {year} {2020})}\BibitemShut {NoStop}%
\bibitem [{\citenamefont {Saito}\ \emph {et~al.}(2020)\citenamefont {Saito}, \citenamefont {Ge}, \citenamefont {Watanabe}, \citenamefont {Taniguchi},\ and\ \citenamefont {Young}}]{saito2020independent}%
  \BibitemOpen
  \bibfield  {author} {\bibinfo {author} {\bibfnamefont {Y.}~\bibnamefont {Saito}}, \bibinfo {author} {\bibfnamefont {J.}~\bibnamefont {Ge}}, \bibinfo {author} {\bibfnamefont {K.}~\bibnamefont {Watanabe}}, \bibinfo {author} {\bibfnamefont {T.}~\bibnamefont {Taniguchi}}, \ and\ \bibinfo {author} {\bibfnamefont {A.~F.}\ \bibnamefont {Young}},\ }\href@noop {} {\bibfield  {journal} {\bibinfo  {journal} {Nature Physics}\ }\textbf {\bibinfo {volume} {16}},\ \bibinfo {pages} {926} (\bibinfo {year} {2020})}\BibitemShut {NoStop}%
\bibitem [{\citenamefont {Zondiner}\ \emph {et~al.}(2020)\citenamefont {Zondiner}, \citenamefont {Rozen}, \citenamefont {Rodan-Legrain}, \citenamefont {Cao}, \citenamefont {Queiroz}, \citenamefont {Taniguchi}, \citenamefont {Watanabe}, \citenamefont {Oreg}, \citenamefont {von Oppen}, \citenamefont {Stern} \emph {et~al.}}]{zondiner2020cascade}%
  \BibitemOpen
  \bibfield  {author} {\bibinfo {author} {\bibfnamefont {U.}~\bibnamefont {Zondiner}}, \bibinfo {author} {\bibfnamefont {A.}~\bibnamefont {Rozen}}, \bibinfo {author} {\bibfnamefont {D.}~\bibnamefont {Rodan-Legrain}}, \bibinfo {author} {\bibfnamefont {Y.}~\bibnamefont {Cao}}, \bibinfo {author} {\bibfnamefont {R.}~\bibnamefont {Queiroz}}, \bibinfo {author} {\bibfnamefont {T.}~\bibnamefont {Taniguchi}}, \bibinfo {author} {\bibfnamefont {K.}~\bibnamefont {Watanabe}}, \bibinfo {author} {\bibfnamefont {Y.}~\bibnamefont {Oreg}}, \bibinfo {author} {\bibfnamefont {F.}~\bibnamefont {von Oppen}}, \bibinfo {author} {\bibfnamefont {A.}~\bibnamefont {Stern}},  \emph {et~al.},\ }\href@noop {} {\bibfield  {journal} {\bibinfo  {journal} {Nature}\ }\textbf {\bibinfo {volume} {582}},\ \bibinfo {pages} {203} (\bibinfo {year} {2020})}\BibitemShut {NoStop}%
\bibitem [{\citenamefont {Wong}\ \emph {et~al.}(2020)\citenamefont {Wong}, \citenamefont {Nuckolls}, \citenamefont {Oh}, \citenamefont {Lian}, \citenamefont {Xie}, \citenamefont {Jeon}, \citenamefont {Watanabe}, \citenamefont {Taniguchi}, \citenamefont {Bernevig},\ and\ \citenamefont {Yazdani}}]{wong2020cascade}%
  \BibitemOpen
  \bibfield  {author} {\bibinfo {author} {\bibfnamefont {D.}~\bibnamefont {Wong}}, \bibinfo {author} {\bibfnamefont {K.~P.}\ \bibnamefont {Nuckolls}}, \bibinfo {author} {\bibfnamefont {M.}~\bibnamefont {Oh}}, \bibinfo {author} {\bibfnamefont {B.}~\bibnamefont {Lian}}, \bibinfo {author} {\bibfnamefont {Y.}~\bibnamefont {Xie}}, \bibinfo {author} {\bibfnamefont {S.}~\bibnamefont {Jeon}}, \bibinfo {author} {\bibfnamefont {K.}~\bibnamefont {Watanabe}}, \bibinfo {author} {\bibfnamefont {T.}~\bibnamefont {Taniguchi}}, \bibinfo {author} {\bibfnamefont {B.~A.}\ \bibnamefont {Bernevig}}, \ and\ \bibinfo {author} {\bibfnamefont {A.}~\bibnamefont {Yazdani}},\ }\href@noop {} {\bibfield  {journal} {\bibinfo  {journal} {Nature}\ }\textbf {\bibinfo {volume} {582}},\ \bibinfo {pages} {198} (\bibinfo {year} {2020})}\BibitemShut {NoStop}%
\bibitem [{\citenamefont {Stepanov}\ \emph {et~al.}(2020)\citenamefont {Stepanov}, \citenamefont {Das}, \citenamefont {Lu}, \citenamefont {Fahimniya}, \citenamefont {Watanabe}, \citenamefont {Taniguchi}, \citenamefont {Koppens}, \citenamefont {Lischner}, \citenamefont {Levitov},\ and\ \citenamefont {Efetov}}]{stepanov2020untying}%
  \BibitemOpen
  \bibfield  {author} {\bibinfo {author} {\bibfnamefont {P.}~\bibnamefont {Stepanov}}, \bibinfo {author} {\bibfnamefont {I.}~\bibnamefont {Das}}, \bibinfo {author} {\bibfnamefont {X.}~\bibnamefont {Lu}}, \bibinfo {author} {\bibfnamefont {A.}~\bibnamefont {Fahimniya}}, \bibinfo {author} {\bibfnamefont {K.}~\bibnamefont {Watanabe}}, \bibinfo {author} {\bibfnamefont {T.}~\bibnamefont {Taniguchi}}, \bibinfo {author} {\bibfnamefont {F.~H.}\ \bibnamefont {Koppens}}, \bibinfo {author} {\bibfnamefont {J.}~\bibnamefont {Lischner}}, \bibinfo {author} {\bibfnamefont {L.}~\bibnamefont {Levitov}}, \ and\ \bibinfo {author} {\bibfnamefont {D.~K.}\ \bibnamefont {Efetov}},\ }\href@noop {} {\bibfield  {journal} {\bibinfo  {journal} {Nature}\ }\textbf {\bibinfo {volume} {583}},\ \bibinfo {pages} {375} (\bibinfo {year} {2020})}\BibitemShut {NoStop}%
\bibitem [{\citenamefont {Arora}\ \emph {et~al.}(2020)\citenamefont {Arora}, \citenamefont {Polski}, \citenamefont {Zhang}, \citenamefont {Thomson}, \citenamefont {Choi}, \citenamefont {Kim}, \citenamefont {Lin}, \citenamefont {Wilson}, \citenamefont {Xu}, \citenamefont {Chu} \emph {et~al.}}]{arora2020superconductivity}%
  \BibitemOpen
  \bibfield  {author} {\bibinfo {author} {\bibfnamefont {H.~S.}\ \bibnamefont {Arora}}, \bibinfo {author} {\bibfnamefont {R.}~\bibnamefont {Polski}}, \bibinfo {author} {\bibfnamefont {Y.}~\bibnamefont {Zhang}}, \bibinfo {author} {\bibfnamefont {A.}~\bibnamefont {Thomson}}, \bibinfo {author} {\bibfnamefont {Y.}~\bibnamefont {Choi}}, \bibinfo {author} {\bibfnamefont {H.}~\bibnamefont {Kim}}, \bibinfo {author} {\bibfnamefont {Z.}~\bibnamefont {Lin}}, \bibinfo {author} {\bibfnamefont {I.~Z.}\ \bibnamefont {Wilson}}, \bibinfo {author} {\bibfnamefont {X.}~\bibnamefont {Xu}}, \bibinfo {author} {\bibfnamefont {J.-H.}\ \bibnamefont {Chu}},  \emph {et~al.},\ }\href@noop {} {\bibfield  {journal} {\bibinfo  {journal} {Nature}\ }\textbf {\bibinfo {volume} {583}},\ \bibinfo {pages} {379} (\bibinfo {year} {2020})}\BibitemShut {NoStop}%
\bibitem [{\citenamefont {Stepanov}\ \emph {et~al.}(2021)\citenamefont {Stepanov}, \citenamefont {Xie}, \citenamefont {Taniguchi}, \citenamefont {Watanabe}, \citenamefont {Lu}, \citenamefont {MacDonald}, \citenamefont {Bernevig},\ and\ \citenamefont {Efetov}}]{PhysRevLett.127.197701}%
  \BibitemOpen
  \bibfield  {author} {\bibinfo {author} {\bibfnamefont {P.}~\bibnamefont {Stepanov}}, \bibinfo {author} {\bibfnamefont {M.}~\bibnamefont {Xie}}, \bibinfo {author} {\bibfnamefont {T.}~\bibnamefont {Taniguchi}}, \bibinfo {author} {\bibfnamefont {K.}~\bibnamefont {Watanabe}}, \bibinfo {author} {\bibfnamefont {X.}~\bibnamefont {Lu}}, \bibinfo {author} {\bibfnamefont {A.~H.}\ \bibnamefont {MacDonald}}, \bibinfo {author} {\bibfnamefont {B.~A.}\ \bibnamefont {Bernevig}}, \ and\ \bibinfo {author} {\bibfnamefont {D.~K.}\ \bibnamefont {Efetov}},\ }\href {\doibase 10.1103/PhysRevLett.127.197701} {\bibfield  {journal} {\bibinfo  {journal} {Phys. Rev. Lett.}\ }\textbf {\bibinfo {volume} {127}},\ \bibinfo {pages} {197701} (\bibinfo {year} {2021})}\BibitemShut {NoStop}%
\bibitem [{\citenamefont {Wu}\ \emph {et~al.}(2018{\natexlab{a}})\citenamefont {Wu}, \citenamefont {Lovorn}, \citenamefont {Tutuc},\ and\ \citenamefont {MacDonald}}]{PhysRevLett.121.026402}%
  \BibitemOpen
  \bibfield  {author} {\bibinfo {author} {\bibfnamefont {F.}~\bibnamefont {Wu}}, \bibinfo {author} {\bibfnamefont {T.}~\bibnamefont {Lovorn}}, \bibinfo {author} {\bibfnamefont {E.}~\bibnamefont {Tutuc}}, \ and\ \bibinfo {author} {\bibfnamefont {A.~H.}\ \bibnamefont {MacDonald}},\ }\href {\doibase 10.1103/PhysRevLett.121.026402} {\bibfield  {journal} {\bibinfo  {journal} {Phys. Rev. Lett.}\ }\textbf {\bibinfo {volume} {121}},\ \bibinfo {pages} {026402} (\bibinfo {year} {2018}{\natexlab{a}})}\BibitemShut {NoStop}%
\bibitem [{\citenamefont {Pan}\ \emph {et~al.}(2020{\natexlab{a}})\citenamefont {Pan}, \citenamefont {Wu},\ and\ \citenamefont {Das~Sarma}}]{PhysRevResearch.2.033087}%
  \BibitemOpen
  \bibfield  {author} {\bibinfo {author} {\bibfnamefont {H.}~\bibnamefont {Pan}}, \bibinfo {author} {\bibfnamefont {F.}~\bibnamefont {Wu}}, \ and\ \bibinfo {author} {\bibfnamefont {S.}~\bibnamefont {Das~Sarma}},\ }\href {\doibase 10.1103/PhysRevResearch.2.033087} {\bibfield  {journal} {\bibinfo  {journal} {Phys. Rev. Res.}\ }\textbf {\bibinfo {volume} {2}},\ \bibinfo {pages} {033087} (\bibinfo {year} {2020}{\natexlab{a}})}\BibitemShut {NoStop}%
\bibitem [{\citenamefont {Pan}\ \emph {et~al.}(2020{\natexlab{b}})\citenamefont {Pan}, \citenamefont {Wu},\ and\ \citenamefont {Das~Sarma}}]{PhysRevB.102.201104}%
  \BibitemOpen
  \bibfield  {author} {\bibinfo {author} {\bibfnamefont {H.}~\bibnamefont {Pan}}, \bibinfo {author} {\bibfnamefont {F.}~\bibnamefont {Wu}}, \ and\ \bibinfo {author} {\bibfnamefont {S.}~\bibnamefont {Das~Sarma}},\ }\href {\doibase 10.1103/PhysRevB.102.201104} {\bibfield  {journal} {\bibinfo  {journal} {Phys. Rev. B}\ }\textbf {\bibinfo {volume} {102}},\ \bibinfo {pages} {201104} (\bibinfo {year} {2020}{\natexlab{b}})}\BibitemShut {NoStop}%
\bibitem [{\citenamefont {Pan}\ and\ \citenamefont {Das~Sarma}(2021)}]{PhysRevLett.127.096802}%
  \BibitemOpen
  \bibfield  {author} {\bibinfo {author} {\bibfnamefont {H.}~\bibnamefont {Pan}}\ and\ \bibinfo {author} {\bibfnamefont {S.}~\bibnamefont {Das~Sarma}},\ }\href {\doibase 10.1103/PhysRevLett.127.096802} {\bibfield  {journal} {\bibinfo  {journal} {Phys. Rev. Lett.}\ }\textbf {\bibinfo {volume} {127}},\ \bibinfo {pages} {096802} (\bibinfo {year} {2021})}\BibitemShut {NoStop}%
\bibitem [{\citenamefont {Zang}\ \emph {et~al.}(2021)\citenamefont {Zang}, \citenamefont {Wang}, \citenamefont {Cano},\ and\ \citenamefont {Millis}}]{PhysRevB.104.075150}%
  \BibitemOpen
  \bibfield  {author} {\bibinfo {author} {\bibfnamefont {J.}~\bibnamefont {Zang}}, \bibinfo {author} {\bibfnamefont {J.}~\bibnamefont {Wang}}, \bibinfo {author} {\bibfnamefont {J.}~\bibnamefont {Cano}}, \ and\ \bibinfo {author} {\bibfnamefont {A.~J.}\ \bibnamefont {Millis}},\ }\href {\doibase 10.1103/PhysRevB.104.075150} {\bibfield  {journal} {\bibinfo  {journal} {Phys. Rev. B}\ }\textbf {\bibinfo {volume} {104}},\ \bibinfo {pages} {075150} (\bibinfo {year} {2021})}\BibitemShut {NoStop}%
\bibitem [{\citenamefont {Zare}\ and\ \citenamefont {Mosadeq}(2021)}]{PhysRevB.104.115154}%
  \BibitemOpen
  \bibfield  {author} {\bibinfo {author} {\bibfnamefont {M.-H.}\ \bibnamefont {Zare}}\ and\ \bibinfo {author} {\bibfnamefont {H.}~\bibnamefont {Mosadeq}},\ }\href {\doibase 10.1103/PhysRevB.104.115154} {\bibfield  {journal} {\bibinfo  {journal} {Phys. Rev. B}\ }\textbf {\bibinfo {volume} {104}},\ \bibinfo {pages} {115154} (\bibinfo {year} {2021})}\BibitemShut {NoStop}%
\bibitem [{\citenamefont {Zang}\ \emph {et~al.}(2022)\citenamefont {Zang}, \citenamefont {Wang}, \citenamefont {Cano}, \citenamefont {Georges},\ and\ \citenamefont {Millis}}]{PhysRevX.12.021064}%
  \BibitemOpen
  \bibfield  {author} {\bibinfo {author} {\bibfnamefont {J.}~\bibnamefont {Zang}}, \bibinfo {author} {\bibfnamefont {J.}~\bibnamefont {Wang}}, \bibinfo {author} {\bibfnamefont {J.}~\bibnamefont {Cano}}, \bibinfo {author} {\bibfnamefont {A.}~\bibnamefont {Georges}}, \ and\ \bibinfo {author} {\bibfnamefont {A.~J.}\ \bibnamefont {Millis}},\ }\href {\doibase 10.1103/PhysRevX.12.021064} {\bibfield  {journal} {\bibinfo  {journal} {Phys. Rev. X}\ }\textbf {\bibinfo {volume} {12}},\ \bibinfo {pages} {021064} (\bibinfo {year} {2022})}\BibitemShut {NoStop}%
\bibitem [{\citenamefont {Wang}\ \emph {et~al.}(2023)\citenamefont {Wang}, \citenamefont {Zang}, \citenamefont {Cano},\ and\ \citenamefont {Millis}}]{PhysRevResearch.5.L012005}%
  \BibitemOpen
  \bibfield  {author} {\bibinfo {author} {\bibfnamefont {J.}~\bibnamefont {Wang}}, \bibinfo {author} {\bibfnamefont {J.}~\bibnamefont {Zang}}, \bibinfo {author} {\bibfnamefont {J.}~\bibnamefont {Cano}}, \ and\ \bibinfo {author} {\bibfnamefont {A.~J.}\ \bibnamefont {Millis}},\ }\href {\doibase 10.1103/PhysRevResearch.5.L012005} {\bibfield  {journal} {\bibinfo  {journal} {Phys. Rev. Res.}\ }\textbf {\bibinfo {volume} {5}},\ \bibinfo {pages} {L012005} (\bibinfo {year} {2023})}\BibitemShut {NoStop}%
\bibitem [{\citenamefont {Liu}\ \emph {et~al.}(2024)\citenamefont {Liu}, \citenamefont {He}, \citenamefont {Wang}, \citenamefont {Zhang}, \citenamefont {Cao},\ and\ \citenamefont {Xiao}}]{PhysRevLett.132.146401}%
  \BibitemOpen
  \bibfield  {author} {\bibinfo {author} {\bibfnamefont {X.}~\bibnamefont {Liu}}, \bibinfo {author} {\bibfnamefont {Y.}~\bibnamefont {He}}, \bibinfo {author} {\bibfnamefont {C.}~\bibnamefont {Wang}}, \bibinfo {author} {\bibfnamefont {X.-W.}\ \bibnamefont {Zhang}}, \bibinfo {author} {\bibfnamefont {T.}~\bibnamefont {Cao}}, \ and\ \bibinfo {author} {\bibfnamefont {D.}~\bibnamefont {Xiao}},\ }\href {\doibase 10.1103/PhysRevLett.132.146401} {\bibfield  {journal} {\bibinfo  {journal} {Phys. Rev. Lett.}\ }\textbf {\bibinfo {volume} {132}},\ \bibinfo {pages} {146401} (\bibinfo {year} {2024})}\BibitemShut {NoStop}%
\bibitem [{\citenamefont {Guerci}\ \emph {et~al.}(2024)\citenamefont {Guerci}, \citenamefont {Lucht}, \citenamefont {Cr\'epel}, \citenamefont {Cano}, \citenamefont {Pixley},\ and\ \citenamefont {Millis}}]{PhysRevB.110.165128}%
  \BibitemOpen
  \bibfield  {author} {\bibinfo {author} {\bibfnamefont {D.}~\bibnamefont {Guerci}}, \bibinfo {author} {\bibfnamefont {K.~P.}\ \bibnamefont {Lucht}}, \bibinfo {author} {\bibfnamefont {V.}~\bibnamefont {Cr\'epel}}, \bibinfo {author} {\bibfnamefont {J.}~\bibnamefont {Cano}}, \bibinfo {author} {\bibfnamefont {J.~H.}\ \bibnamefont {Pixley}}, \ and\ \bibinfo {author} {\bibfnamefont {A.}~\bibnamefont {Millis}},\ }\href {\doibase 10.1103/PhysRevB.110.165128} {\bibfield  {journal} {\bibinfo  {journal} {Phys. Rev. B}\ }\textbf {\bibinfo {volume} {110}},\ \bibinfo {pages} {165128} (\bibinfo {year} {2024})}\BibitemShut {NoStop}%
\bibitem [{\citenamefont {Venderley}\ and\ \citenamefont {Kim}(2019)}]{PhysRevB.100.060506}%
  \BibitemOpen
  \bibfield  {author} {\bibinfo {author} {\bibfnamefont {J.}~\bibnamefont {Venderley}}\ and\ \bibinfo {author} {\bibfnamefont {E.-A.}\ \bibnamefont {Kim}},\ }\href {\doibase 10.1103/PhysRevB.100.060506} {\bibfield  {journal} {\bibinfo  {journal} {Phys. Rev. B}\ }\textbf {\bibinfo {volume} {100}},\ \bibinfo {pages} {060506} (\bibinfo {year} {2019})}\BibitemShut {NoStop}%
\bibitem [{\citenamefont {B\'elanger}\ \emph {et~al.}(2022)\citenamefont {B\'elanger}, \citenamefont {Fournier},\ and\ \citenamefont {S\'en\'echal}}]{PhysRevB.106.235135}%
  \BibitemOpen
  \bibfield  {author} {\bibinfo {author} {\bibfnamefont {M.}~\bibnamefont {B\'elanger}}, \bibinfo {author} {\bibfnamefont {J.}~\bibnamefont {Fournier}}, \ and\ \bibinfo {author} {\bibfnamefont {D.}~\bibnamefont {S\'en\'echal}},\ }\href {\doibase 10.1103/PhysRevB.106.235135} {\bibfield  {journal} {\bibinfo  {journal} {Phys. Rev. B}\ }\textbf {\bibinfo {volume} {106}},\ \bibinfo {pages} {235135} (\bibinfo {year} {2022})}\BibitemShut {NoStop}%
\bibitem [{\citenamefont {Klebl}\ \emph {et~al.}(2023)\citenamefont {Klebl}, \citenamefont {Fischer}, \citenamefont {Classen}, \citenamefont {Scherer},\ and\ \citenamefont {Kennes}}]{PhysRevResearch.5.L012034}%
  \BibitemOpen
  \bibfield  {author} {\bibinfo {author} {\bibfnamefont {L.}~\bibnamefont {Klebl}}, \bibinfo {author} {\bibfnamefont {A.}~\bibnamefont {Fischer}}, \bibinfo {author} {\bibfnamefont {L.}~\bibnamefont {Classen}}, \bibinfo {author} {\bibfnamefont {M.~M.}\ \bibnamefont {Scherer}}, \ and\ \bibinfo {author} {\bibfnamefont {D.~M.}\ \bibnamefont {Kennes}},\ }\href {\doibase 10.1103/PhysRevResearch.5.L012034} {\bibfield  {journal} {\bibinfo  {journal} {Phys. Rev. Res.}\ }\textbf {\bibinfo {volume} {5}},\ \bibinfo {pages} {L012034} (\bibinfo {year} {2023})}\BibitemShut {NoStop}%
\bibitem [{\citenamefont {Wu}\ \emph {et~al.}(2023)\citenamefont {Wu}, \citenamefont {Wu},\ and\ \citenamefont {Yao}}]{PhysRevLett.130.126001}%
  \BibitemOpen
  \bibfield  {author} {\bibinfo {author} {\bibfnamefont {Y.-M.}\ \bibnamefont {Wu}}, \bibinfo {author} {\bibfnamefont {Z.}~\bibnamefont {Wu}}, \ and\ \bibinfo {author} {\bibfnamefont {H.}~\bibnamefont {Yao}},\ }\href {\doibase 10.1103/PhysRevLett.130.126001} {\bibfield  {journal} {\bibinfo  {journal} {Phys. Rev. Lett.}\ }\textbf {\bibinfo {volume} {130}},\ \bibinfo {pages} {126001} (\bibinfo {year} {2023})}\BibitemShut {NoStop}%
\bibitem [{\citenamefont {Cr\'epel}\ \emph {et~al.}(2023)\citenamefont {Cr\'epel}, \citenamefont {Guerci}, \citenamefont {Cano}, \citenamefont {Pixley},\ and\ \citenamefont {Millis}}]{PhysRevLett.131.056001}%
  \BibitemOpen
  \bibfield  {author} {\bibinfo {author} {\bibfnamefont {V.}~\bibnamefont {Cr\'epel}}, \bibinfo {author} {\bibfnamefont {D.}~\bibnamefont {Guerci}}, \bibinfo {author} {\bibfnamefont {J.}~\bibnamefont {Cano}}, \bibinfo {author} {\bibfnamefont {J.~H.}\ \bibnamefont {Pixley}}, \ and\ \bibinfo {author} {\bibfnamefont {A.}~\bibnamefont {Millis}},\ }\href {\doibase 10.1103/PhysRevLett.131.056001} {\bibfield  {journal} {\bibinfo  {journal} {Phys. Rev. Lett.}\ }\textbf {\bibinfo {volume} {131}},\ \bibinfo {pages} {056001} (\bibinfo {year} {2023})}\BibitemShut {NoStop}%
\bibitem [{\citenamefont {Zegrodnik}\ and\ \citenamefont {Biborski}(2023)}]{PhysRevB.108.064506}%
  \BibitemOpen
  \bibfield  {author} {\bibinfo {author} {\bibfnamefont {M.}~\bibnamefont {Zegrodnik}}\ and\ \bibinfo {author} {\bibfnamefont {A.}~\bibnamefont {Biborski}},\ }\href {\doibase 10.1103/PhysRevB.108.064506} {\bibfield  {journal} {\bibinfo  {journal} {Phys. Rev. B}\ }\textbf {\bibinfo {volume} {108}},\ \bibinfo {pages} {064506} (\bibinfo {year} {2023})}\BibitemShut {NoStop}%
\bibitem [{\citenamefont {Zhou}\ and\ \citenamefont {Zhang}(2023)}]{PhysRevB.108.155111}%
  \BibitemOpen
  \bibfield  {author} {\bibinfo {author} {\bibfnamefont {B.}~\bibnamefont {Zhou}}\ and\ \bibinfo {author} {\bibfnamefont {Y.-H.}\ \bibnamefont {Zhang}},\ }\href {\doibase 10.1103/PhysRevB.108.155111} {\bibfield  {journal} {\bibinfo  {journal} {Phys. Rev. B}\ }\textbf {\bibinfo {volume} {108}},\ \bibinfo {pages} {155111} (\bibinfo {year} {2023})}\BibitemShut {NoStop}%
\bibitem [{\citenamefont {Schrade}\ and\ \citenamefont {Fu}(2024)}]{PhysRevB.110.035143}%
  \BibitemOpen
  \bibfield  {author} {\bibinfo {author} {\bibfnamefont {C.}~\bibnamefont {Schrade}}\ and\ \bibinfo {author} {\bibfnamefont {L.}~\bibnamefont {Fu}},\ }\href {\doibase 10.1103/PhysRevB.110.035143} {\bibfield  {journal} {\bibinfo  {journal} {Phys. Rev. B}\ }\textbf {\bibinfo {volume} {110}},\ \bibinfo {pages} {035143} (\bibinfo {year} {2024})}\BibitemShut {NoStop}%
\bibitem [{\citenamefont {Kim}\ \emph {et~al.}(2025)\citenamefont {Kim}, \citenamefont {Mendez-Valderrama}, \citenamefont {Wang},\ and\ \citenamefont {Chowdhury}}]{kim2025theory}%
  \BibitemOpen
  \bibfield  {author} {\bibinfo {author} {\bibfnamefont {S.}~\bibnamefont {Kim}}, \bibinfo {author} {\bibfnamefont {J.~F.}\ \bibnamefont {Mendez-Valderrama}}, \bibinfo {author} {\bibfnamefont {X.}~\bibnamefont {Wang}}, \ and\ \bibinfo {author} {\bibfnamefont {D.}~\bibnamefont {Chowdhury}},\ }\href@noop {} {\bibfield  {journal} {\bibinfo  {journal} {Nature Communications}\ }\textbf {\bibinfo {volume} {16}},\ \bibinfo {pages} {1701} (\bibinfo {year} {2025})}\BibitemShut {NoStop}%
\bibitem [{\citenamefont {Hsu}\ \emph {et~al.}(2021)\citenamefont {Hsu}, \citenamefont {Wu},\ and\ \citenamefont {Das~Sarma}}]{PhysRevB.104.195134}%
  \BibitemOpen
  \bibfield  {author} {\bibinfo {author} {\bibfnamefont {Y.-T.}\ \bibnamefont {Hsu}}, \bibinfo {author} {\bibfnamefont {F.}~\bibnamefont {Wu}}, \ and\ \bibinfo {author} {\bibfnamefont {S.}~\bibnamefont {Das~Sarma}},\ }\href {\doibase 10.1103/PhysRevB.104.195134} {\bibfield  {journal} {\bibinfo  {journal} {Phys. Rev. B}\ }\textbf {\bibinfo {volume} {104}},\ \bibinfo {pages} {195134} (\bibinfo {year} {2021})}\BibitemShut {NoStop}%
\bibitem [{\citenamefont {Wang}\ \emph {et~al.}(2020)\citenamefont {Wang}, \citenamefont {Shih}, \citenamefont {Ghiotto}, \citenamefont {Xian}, \citenamefont {Rhodes}, \citenamefont {Tan}, \citenamefont {Claassen}, \citenamefont {Kennes}, \citenamefont {Bai}, \citenamefont {Kim} \emph {et~al.}}]{wang2020correlated}%
  \BibitemOpen
  \bibfield  {author} {\bibinfo {author} {\bibfnamefont {L.}~\bibnamefont {Wang}}, \bibinfo {author} {\bibfnamefont {E.-M.}\ \bibnamefont {Shih}}, \bibinfo {author} {\bibfnamefont {A.}~\bibnamefont {Ghiotto}}, \bibinfo {author} {\bibfnamefont {L.}~\bibnamefont {Xian}}, \bibinfo {author} {\bibfnamefont {D.~A.}\ \bibnamefont {Rhodes}}, \bibinfo {author} {\bibfnamefont {C.}~\bibnamefont {Tan}}, \bibinfo {author} {\bibfnamefont {M.}~\bibnamefont {Claassen}}, \bibinfo {author} {\bibfnamefont {D.~M.}\ \bibnamefont {Kennes}}, \bibinfo {author} {\bibfnamefont {Y.}~\bibnamefont {Bai}}, \bibinfo {author} {\bibfnamefont {B.}~\bibnamefont {Kim}},  \emph {et~al.},\ }\href@noop {} {\bibfield  {journal} {\bibinfo  {journal} {Nature materials}\ }\textbf {\bibinfo {volume} {19}},\ \bibinfo {pages} {861} (\bibinfo {year} {2020})}\BibitemShut {NoStop}%
\bibitem [{\citenamefont {An}\ \emph {et~al.}(2020)\citenamefont {An}, \citenamefont {Cai}, \citenamefont {Pei}, \citenamefont {Huang}, \citenamefont {Wu}, \citenamefont {Zhou}, \citenamefont {Lin}, \citenamefont {Ying}, \citenamefont {Ye}, \citenamefont {Feng}, \citenamefont {Gao}, \citenamefont {Cacho}, \citenamefont {Watson}, \citenamefont {Chen},\ and\ \citenamefont {Wang}}]{D0NH00248H}%
  \BibitemOpen
  \bibfield  {author} {\bibinfo {author} {\bibfnamefont {L.}~\bibnamefont {An}}, \bibinfo {author} {\bibfnamefont {X.}~\bibnamefont {Cai}}, \bibinfo {author} {\bibfnamefont {D.}~\bibnamefont {Pei}}, \bibinfo {author} {\bibfnamefont {M.}~\bibnamefont {Huang}}, \bibinfo {author} {\bibfnamefont {Z.}~\bibnamefont {Wu}}, \bibinfo {author} {\bibfnamefont {Z.}~\bibnamefont {Zhou}}, \bibinfo {author} {\bibfnamefont {J.}~\bibnamefont {Lin}}, \bibinfo {author} {\bibfnamefont {Z.}~\bibnamefont {Ying}}, \bibinfo {author} {\bibfnamefont {Z.}~\bibnamefont {Ye}}, \bibinfo {author} {\bibfnamefont {X.}~\bibnamefont {Feng}}, \bibinfo {author} {\bibfnamefont {R.}~\bibnamefont {Gao}}, \bibinfo {author} {\bibfnamefont {C.}~\bibnamefont {Cacho}}, \bibinfo {author} {\bibfnamefont {M.}~\bibnamefont {Watson}}, \bibinfo {author} {\bibfnamefont {Y.}~\bibnamefont {Chen}}, \ and\ \bibinfo {author} {\bibfnamefont {N.}~\bibnamefont {Wang}},\ }\href {\doibase 10.1039/D0NH00248H} {\bibfield  {journal} {\bibinfo  {journal} {Nanoscale Horiz.}\
  }\textbf {\bibinfo {volume} {5}},\ \bibinfo {pages} {1309} (\bibinfo {year} {2020})}\BibitemShut {NoStop}%
\bibitem [{\citenamefont {Xia}\ \emph {et~al.}(2025)\citenamefont {Xia}, \citenamefont {Han}, \citenamefont {Watanabe}, \citenamefont {Taniguchi}, \citenamefont {Shan},\ and\ \citenamefont {Mak}}]{xia2025superconductivity}%
  \BibitemOpen
  \bibfield  {author} {\bibinfo {author} {\bibfnamefont {Y.}~\bibnamefont {Xia}}, \bibinfo {author} {\bibfnamefont {Z.}~\bibnamefont {Han}}, \bibinfo {author} {\bibfnamefont {K.}~\bibnamefont {Watanabe}}, \bibinfo {author} {\bibfnamefont {T.}~\bibnamefont {Taniguchi}}, \bibinfo {author} {\bibfnamefont {J.}~\bibnamefont {Shan}}, \ and\ \bibinfo {author} {\bibfnamefont {K.~F.}\ \bibnamefont {Mak}},\ }\href@noop {} {\bibfield  {journal} {\bibinfo  {journal} {Nature}\ }\textbf {\bibinfo {volume} {637}},\ \bibinfo {pages} {833} (\bibinfo {year} {2025})}\BibitemShut {NoStop}%
\bibitem [{\citenamefont {Guo}\ \emph {et~al.}(2025)\citenamefont {Guo}, \citenamefont {Pack}, \citenamefont {Swann}, \citenamefont {Holtzman}, \citenamefont {Cothrine}, \citenamefont {Watanabe}, \citenamefont {Taniguchi}, \citenamefont {Mandrus}, \citenamefont {Barmak}, \citenamefont {Hone} \emph {et~al.}}]{guo2025superconductivity}%
  \BibitemOpen
  \bibfield  {author} {\bibinfo {author} {\bibfnamefont {Y.}~\bibnamefont {Guo}}, \bibinfo {author} {\bibfnamefont {J.}~\bibnamefont {Pack}}, \bibinfo {author} {\bibfnamefont {J.}~\bibnamefont {Swann}}, \bibinfo {author} {\bibfnamefont {L.}~\bibnamefont {Holtzman}}, \bibinfo {author} {\bibfnamefont {M.}~\bibnamefont {Cothrine}}, \bibinfo {author} {\bibfnamefont {K.}~\bibnamefont {Watanabe}}, \bibinfo {author} {\bibfnamefont {T.}~\bibnamefont {Taniguchi}}, \bibinfo {author} {\bibfnamefont {D.~G.}\ \bibnamefont {Mandrus}}, \bibinfo {author} {\bibfnamefont {K.}~\bibnamefont {Barmak}}, \bibinfo {author} {\bibfnamefont {J.}~\bibnamefont {Hone}},  \emph {et~al.},\ }\href@noop {} {\bibfield  {journal} {\bibinfo  {journal} {Nature}\ }\textbf {\bibinfo {volume} {637}},\ \bibinfo {pages} {839} (\bibinfo {year} {2025})}\BibitemShut {NoStop}%
\bibitem [{\citenamefont {Wu}\ \emph {et~al.}(2017)\citenamefont {Wu}, \citenamefont {Lovorn},\ and\ \citenamefont {MacDonald}}]{PhysRevLett.118.147401}%
  \BibitemOpen
  \bibfield  {author} {\bibinfo {author} {\bibfnamefont {F.}~\bibnamefont {Wu}}, \bibinfo {author} {\bibfnamefont {T.}~\bibnamefont {Lovorn}}, \ and\ \bibinfo {author} {\bibfnamefont {A.~H.}\ \bibnamefont {MacDonald}},\ }\href {\doibase 10.1103/PhysRevLett.118.147401} {\bibfield  {journal} {\bibinfo  {journal} {Phys. Rev. Lett.}\ }\textbf {\bibinfo {volume} {118}},\ \bibinfo {pages} {147401} (\bibinfo {year} {2017})}\BibitemShut {NoStop}%
\bibitem [{\citenamefont {Yu}\ \emph {et~al.}(2017)\citenamefont {Yu}, \citenamefont {Liu}, \citenamefont {Tang}, \citenamefont {Xu},\ and\ \citenamefont {Yao}}]{yu2017moire}%
  \BibitemOpen
  \bibfield  {author} {\bibinfo {author} {\bibfnamefont {H.}~\bibnamefont {Yu}}, \bibinfo {author} {\bibfnamefont {G.-B.}\ \bibnamefont {Liu}}, \bibinfo {author} {\bibfnamefont {J.}~\bibnamefont {Tang}}, \bibinfo {author} {\bibfnamefont {X.}~\bibnamefont {Xu}}, \ and\ \bibinfo {author} {\bibfnamefont {W.}~\bibnamefont {Yao}},\ }\href@noop {} {\bibfield  {journal} {\bibinfo  {journal} {Science advances}\ }\textbf {\bibinfo {volume} {3}},\ \bibinfo {pages} {e1701696} (\bibinfo {year} {2017})}\BibitemShut {NoStop}%
\bibitem [{\citenamefont {Wu}\ \emph {et~al.}(2018{\natexlab{b}})\citenamefont {Wu}, \citenamefont {Lovorn},\ and\ \citenamefont {MacDonald}}]{PhysRevB.97.035306}%
  \BibitemOpen
  \bibfield  {author} {\bibinfo {author} {\bibfnamefont {F.}~\bibnamefont {Wu}}, \bibinfo {author} {\bibfnamefont {T.}~\bibnamefont {Lovorn}}, \ and\ \bibinfo {author} {\bibfnamefont {A.~H.}\ \bibnamefont {MacDonald}},\ }\href {\doibase 10.1103/PhysRevB.97.035306} {\bibfield  {journal} {\bibinfo  {journal} {Phys. Rev. B}\ }\textbf {\bibinfo {volume} {97}},\ \bibinfo {pages} {035306} (\bibinfo {year} {2018}{\natexlab{b}})}\BibitemShut {NoStop}%
\bibitem [{\citenamefont {Seyler}\ \emph {et~al.}(2019)\citenamefont {Seyler}, \citenamefont {Rivera}, \citenamefont {Yu}, \citenamefont {Wilson}, \citenamefont {Ray}, \citenamefont {Mandrus}, \citenamefont {Yan}, \citenamefont {Yao},\ and\ \citenamefont {Xu}}]{seyler2019signatures}%
  \BibitemOpen
  \bibfield  {author} {\bibinfo {author} {\bibfnamefont {K.~L.}\ \bibnamefont {Seyler}}, \bibinfo {author} {\bibfnamefont {P.}~\bibnamefont {Rivera}}, \bibinfo {author} {\bibfnamefont {H.}~\bibnamefont {Yu}}, \bibinfo {author} {\bibfnamefont {N.~P.}\ \bibnamefont {Wilson}}, \bibinfo {author} {\bibfnamefont {E.~L.}\ \bibnamefont {Ray}}, \bibinfo {author} {\bibfnamefont {D.~G.}\ \bibnamefont {Mandrus}}, \bibinfo {author} {\bibfnamefont {J.}~\bibnamefont {Yan}}, \bibinfo {author} {\bibfnamefont {W.}~\bibnamefont {Yao}}, \ and\ \bibinfo {author} {\bibfnamefont {X.}~\bibnamefont {Xu}},\ }\href@noop {} {\bibfield  {journal} {\bibinfo  {journal} {Nature}\ }\textbf {\bibinfo {volume} {567}},\ \bibinfo {pages} {66} (\bibinfo {year} {2019})}\BibitemShut {NoStop}%
\bibitem [{\citenamefont {Jin}\ \emph {et~al.}(2019)\citenamefont {Jin}, \citenamefont {Regan}, \citenamefont {Yan}, \citenamefont {Iqbal Bakti~Utama}, \citenamefont {Wang}, \citenamefont {Zhao}, \citenamefont {Qin}, \citenamefont {Yang}, \citenamefont {Zheng}, \citenamefont {Shi} \emph {et~al.}}]{jin2019observation}%
  \BibitemOpen
  \bibfield  {author} {\bibinfo {author} {\bibfnamefont {C.}~\bibnamefont {Jin}}, \bibinfo {author} {\bibfnamefont {E.~C.}\ \bibnamefont {Regan}}, \bibinfo {author} {\bibfnamefont {A.}~\bibnamefont {Yan}}, \bibinfo {author} {\bibfnamefont {M.}~\bibnamefont {Iqbal Bakti~Utama}}, \bibinfo {author} {\bibfnamefont {D.}~\bibnamefont {Wang}}, \bibinfo {author} {\bibfnamefont {S.}~\bibnamefont {Zhao}}, \bibinfo {author} {\bibfnamefont {Y.}~\bibnamefont {Qin}}, \bibinfo {author} {\bibfnamefont {S.}~\bibnamefont {Yang}}, \bibinfo {author} {\bibfnamefont {Z.}~\bibnamefont {Zheng}}, \bibinfo {author} {\bibfnamefont {S.}~\bibnamefont {Shi}},  \emph {et~al.},\ }\href@noop {} {\bibfield  {journal} {\bibinfo  {journal} {Nature}\ }\textbf {\bibinfo {volume} {567}},\ \bibinfo {pages} {76} (\bibinfo {year} {2019})}\BibitemShut {NoStop}%
\bibitem [{\citenamefont {Tran}\ \emph {et~al.}(2019)\citenamefont {Tran}, \citenamefont {Moody}, \citenamefont {Wu}, \citenamefont {Lu}, \citenamefont {Choi}, \citenamefont {Kim}, \citenamefont {Rai}, \citenamefont {Sanchez}, \citenamefont {Quan}, \citenamefont {Singh} \emph {et~al.}}]{tran2019evidence}%
  \BibitemOpen
  \bibfield  {author} {\bibinfo {author} {\bibfnamefont {K.}~\bibnamefont {Tran}}, \bibinfo {author} {\bibfnamefont {G.}~\bibnamefont {Moody}}, \bibinfo {author} {\bibfnamefont {F.}~\bibnamefont {Wu}}, \bibinfo {author} {\bibfnamefont {X.}~\bibnamefont {Lu}}, \bibinfo {author} {\bibfnamefont {J.}~\bibnamefont {Choi}}, \bibinfo {author} {\bibfnamefont {K.}~\bibnamefont {Kim}}, \bibinfo {author} {\bibfnamefont {A.}~\bibnamefont {Rai}}, \bibinfo {author} {\bibfnamefont {D.~A.}\ \bibnamefont {Sanchez}}, \bibinfo {author} {\bibfnamefont {J.}~\bibnamefont {Quan}}, \bibinfo {author} {\bibfnamefont {A.}~\bibnamefont {Singh}},  \emph {et~al.},\ }\href@noop {} {\bibfield  {journal} {\bibinfo  {journal} {Nature}\ }\textbf {\bibinfo {volume} {567}},\ \bibinfo {pages} {71} (\bibinfo {year} {2019})}\BibitemShut {NoStop}%
\bibitem [{\citenamefont {Ruiz-Tijerina}\ and\ \citenamefont {Fal'ko}(2019)}]{PhysRevB.99.125424}%
  \BibitemOpen
  \bibfield  {author} {\bibinfo {author} {\bibfnamefont {D.~A.}\ \bibnamefont {Ruiz-Tijerina}}\ and\ \bibinfo {author} {\bibfnamefont {V.~I.}\ \bibnamefont {Fal'ko}},\ }\href {\doibase 10.1103/PhysRevB.99.125424} {\bibfield  {journal} {\bibinfo  {journal} {Phys. Rev. B}\ }\textbf {\bibinfo {volume} {99}},\ \bibinfo {pages} {125424} (\bibinfo {year} {2019})}\BibitemShut {NoStop}%
\bibitem [{\citenamefont {Huang}\ \emph {et~al.}(2022)\citenamefont {Huang}, \citenamefont {Choi}, \citenamefont {Shih},\ and\ \citenamefont {Li}}]{huang2022excitons}%
  \BibitemOpen
  \bibfield  {author} {\bibinfo {author} {\bibfnamefont {D.}~\bibnamefont {Huang}}, \bibinfo {author} {\bibfnamefont {J.}~\bibnamefont {Choi}}, \bibinfo {author} {\bibfnamefont {C.-K.}\ \bibnamefont {Shih}}, \ and\ \bibinfo {author} {\bibfnamefont {X.}~\bibnamefont {Li}},\ }\href@noop {} {\bibfield  {journal} {\bibinfo  {journal} {Nature nanotechnology}\ }\textbf {\bibinfo {volume} {17}},\ \bibinfo {pages} {227} (\bibinfo {year} {2022})}\BibitemShut {NoStop}%
\bibitem [{\citenamefont {Soltero}\ \emph {et~al.}(2024{\natexlab{a}})\citenamefont {Soltero}, \citenamefont {Kaliteevski}, \citenamefont {McHugh}, \citenamefont {Enaldiev},\ and\ \citenamefont {Fal’ko}}]{doi:10.1021/acs.nanolett.3c04427}%
  \BibitemOpen
  \bibfield  {author} {\bibinfo {author} {\bibfnamefont {I.}~\bibnamefont {Soltero}}, \bibinfo {author} {\bibfnamefont {M.~A.}\ \bibnamefont {Kaliteevski}}, \bibinfo {author} {\bibfnamefont {J.~G.}\ \bibnamefont {McHugh}}, \bibinfo {author} {\bibfnamefont {V.}~\bibnamefont {Enaldiev}}, \ and\ \bibinfo {author} {\bibfnamefont {V.~I.}\ \bibnamefont {Fal’ko}},\ }\href {\doibase 10.1021/acs.nanolett.3c04427} {\bibfield  {journal} {\bibinfo  {journal} {Nano Letters}\ }\textbf {\bibinfo {volume} {24}},\ \bibinfo {pages} {1996} (\bibinfo {year} {2024}{\natexlab{a}})},\ \bibinfo {note} {pMID: 38295286},\ \Eprint {http://arxiv.org/abs/https://doi.org/10.1021/acs.nanolett.3c04427} {https://doi.org/10.1021/acs.nanolett.3c04427} \BibitemShut {NoStop}%
\bibitem [{\citenamefont {Wu}\ \emph {et~al.}(2019)\citenamefont {Wu}, \citenamefont {Lovorn}, \citenamefont {Tutuc}, \citenamefont {Martin},\ and\ \citenamefont {MacDonald}}]{PhysRevLett.122.086402}%
  \BibitemOpen
  \bibfield  {author} {\bibinfo {author} {\bibfnamefont {F.}~\bibnamefont {Wu}}, \bibinfo {author} {\bibfnamefont {T.}~\bibnamefont {Lovorn}}, \bibinfo {author} {\bibfnamefont {E.}~\bibnamefont {Tutuc}}, \bibinfo {author} {\bibfnamefont {I.}~\bibnamefont {Martin}}, \ and\ \bibinfo {author} {\bibfnamefont {A.~H.}\ \bibnamefont {MacDonald}},\ }\href {\doibase 10.1103/PhysRevLett.122.086402} {\bibfield  {journal} {\bibinfo  {journal} {Phys. Rev. Lett.}\ }\textbf {\bibinfo {volume} {122}},\ \bibinfo {pages} {086402} (\bibinfo {year} {2019})}\BibitemShut {NoStop}%
\bibitem [{\citenamefont {Yu}\ \emph {et~al.}(2019)\citenamefont {Yu}, \citenamefont {Chen},\ and\ \citenamefont {Yao}}]{10.1093/nsr/nwz117}%
  \BibitemOpen
  \bibfield  {author} {\bibinfo {author} {\bibfnamefont {H.}~\bibnamefont {Yu}}, \bibinfo {author} {\bibfnamefont {M.}~\bibnamefont {Chen}}, \ and\ \bibinfo {author} {\bibfnamefont {W.}~\bibnamefont {Yao}},\ }\href {\doibase 10.1093/nsr/nwz117} {\bibfield  {journal} {\bibinfo  {journal} {National Science Review}\ }\textbf {\bibinfo {volume} {7}},\ \bibinfo {pages} {12} (\bibinfo {year} {2019})},\ \Eprint {http://arxiv.org/abs/https://academic.oup.com/nsr/article-pdf/7/1/12/40810220/nsr\_7\_1\_12.pdf} {https://academic.oup.com/nsr/article-pdf/7/1/12/40810220/nsr\_7\_1\_12.pdf} \BibitemShut {NoStop}%
\bibitem [{\citenamefont {Li}\ \emph {et~al.}(2021{\natexlab{a}})\citenamefont {Li}, \citenamefont {Jiang}, \citenamefont {Shen}, \citenamefont {Zhang}, \citenamefont {Li}, \citenamefont {Tao}, \citenamefont {Devakul}, \citenamefont {Watanabe}, \citenamefont {Taniguchi}, \citenamefont {Fu} \emph {et~al.}}]{li2021quantum}%
  \BibitemOpen
  \bibfield  {author} {\bibinfo {author} {\bibfnamefont {T.}~\bibnamefont {Li}}, \bibinfo {author} {\bibfnamefont {S.}~\bibnamefont {Jiang}}, \bibinfo {author} {\bibfnamefont {B.}~\bibnamefont {Shen}}, \bibinfo {author} {\bibfnamefont {Y.}~\bibnamefont {Zhang}}, \bibinfo {author} {\bibfnamefont {L.}~\bibnamefont {Li}}, \bibinfo {author} {\bibfnamefont {Z.}~\bibnamefont {Tao}}, \bibinfo {author} {\bibfnamefont {T.}~\bibnamefont {Devakul}}, \bibinfo {author} {\bibfnamefont {K.}~\bibnamefont {Watanabe}}, \bibinfo {author} {\bibfnamefont {T.}~\bibnamefont {Taniguchi}}, \bibinfo {author} {\bibfnamefont {L.}~\bibnamefont {Fu}},  \emph {et~al.},\ }\href@noop {} {\bibfield  {journal} {\bibinfo  {journal} {Nature}\ }\textbf {\bibinfo {volume} {600}},\ \bibinfo {pages} {641} (\bibinfo {year} {2021}{\natexlab{a}})}\BibitemShut {NoStop}%
\bibitem [{\citenamefont {Devakul}\ \emph {et~al.}(2021)\citenamefont {Devakul}, \citenamefont {Cr{\'e}pel}, \citenamefont {Zhang},\ and\ \citenamefont {Fu}}]{devakul2021magic}%
  \BibitemOpen
  \bibfield  {author} {\bibinfo {author} {\bibfnamefont {T.}~\bibnamefont {Devakul}}, \bibinfo {author} {\bibfnamefont {V.}~\bibnamefont {Cr{\'e}pel}}, \bibinfo {author} {\bibfnamefont {Y.}~\bibnamefont {Zhang}}, \ and\ \bibinfo {author} {\bibfnamefont {L.}~\bibnamefont {Fu}},\ }\href@noop {} {\bibfield  {journal} {\bibinfo  {journal} {Nature communications}\ }\textbf {\bibinfo {volume} {12}},\ \bibinfo {pages} {6730} (\bibinfo {year} {2021})}\BibitemShut {NoStop}%
\bibitem [{\citenamefont {Xie}\ \emph {et~al.}(2022)\citenamefont {Xie}, \citenamefont {Zhang}, \citenamefont {Hu}, \citenamefont {Mak},\ and\ \citenamefont {Law}}]{PhysRevLett.128.026402}%
  \BibitemOpen
  \bibfield  {author} {\bibinfo {author} {\bibfnamefont {Y.-M.}\ \bibnamefont {Xie}}, \bibinfo {author} {\bibfnamefont {C.-P.}\ \bibnamefont {Zhang}}, \bibinfo {author} {\bibfnamefont {J.-X.}\ \bibnamefont {Hu}}, \bibinfo {author} {\bibfnamefont {K.~F.}\ \bibnamefont {Mak}}, \ and\ \bibinfo {author} {\bibfnamefont {K.~T.}\ \bibnamefont {Law}},\ }\href {\doibase 10.1103/PhysRevLett.128.026402} {\bibfield  {journal} {\bibinfo  {journal} {Phys. Rev. Lett.}\ }\textbf {\bibinfo {volume} {128}},\ \bibinfo {pages} {026402} (\bibinfo {year} {2022})}\BibitemShut {NoStop}%
\bibitem [{\citenamefont {Cai}\ \emph {et~al.}(2023)\citenamefont {Cai}, \citenamefont {Anderson}, \citenamefont {Wang}, \citenamefont {Zhang}, \citenamefont {Liu}, \citenamefont {Holtzmann}, \citenamefont {Zhang}, \citenamefont {Fan}, \citenamefont {Taniguchi}, \citenamefont {Watanabe} \emph {et~al.}}]{cai2023signatures}%
  \BibitemOpen
  \bibfield  {author} {\bibinfo {author} {\bibfnamefont {J.}~\bibnamefont {Cai}}, \bibinfo {author} {\bibfnamefont {E.}~\bibnamefont {Anderson}}, \bibinfo {author} {\bibfnamefont {C.}~\bibnamefont {Wang}}, \bibinfo {author} {\bibfnamefont {X.}~\bibnamefont {Zhang}}, \bibinfo {author} {\bibfnamefont {X.}~\bibnamefont {Liu}}, \bibinfo {author} {\bibfnamefont {W.}~\bibnamefont {Holtzmann}}, \bibinfo {author} {\bibfnamefont {Y.}~\bibnamefont {Zhang}}, \bibinfo {author} {\bibfnamefont {F.}~\bibnamefont {Fan}}, \bibinfo {author} {\bibfnamefont {T.}~\bibnamefont {Taniguchi}}, \bibinfo {author} {\bibfnamefont {K.}~\bibnamefont {Watanabe}},  \emph {et~al.},\ }\href@noop {} {\bibfield  {journal} {\bibinfo  {journal} {Nature}\ }\textbf {\bibinfo {volume} {622}},\ \bibinfo {pages} {63} (\bibinfo {year} {2023})}\BibitemShut {NoStop}%
\bibitem [{\citenamefont {Zeng}\ \emph {et~al.}(2023)\citenamefont {Zeng}, \citenamefont {Xia}, \citenamefont {Kang}, \citenamefont {Zhu}, \citenamefont {Kn{\"u}ppel}, \citenamefont {Vaswani}, \citenamefont {Watanabe}, \citenamefont {Taniguchi}, \citenamefont {Mak},\ and\ \citenamefont {Shan}}]{zeng2023thermodynamic}%
  \BibitemOpen
  \bibfield  {author} {\bibinfo {author} {\bibfnamefont {Y.}~\bibnamefont {Zeng}}, \bibinfo {author} {\bibfnamefont {Z.}~\bibnamefont {Xia}}, \bibinfo {author} {\bibfnamefont {K.}~\bibnamefont {Kang}}, \bibinfo {author} {\bibfnamefont {J.}~\bibnamefont {Zhu}}, \bibinfo {author} {\bibfnamefont {P.}~\bibnamefont {Kn{\"u}ppel}}, \bibinfo {author} {\bibfnamefont {C.}~\bibnamefont {Vaswani}}, \bibinfo {author} {\bibfnamefont {K.}~\bibnamefont {Watanabe}}, \bibinfo {author} {\bibfnamefont {T.}~\bibnamefont {Taniguchi}}, \bibinfo {author} {\bibfnamefont {K.~F.}\ \bibnamefont {Mak}}, \ and\ \bibinfo {author} {\bibfnamefont {J.}~\bibnamefont {Shan}},\ }\href@noop {} {\bibfield  {journal} {\bibinfo  {journal} {Nature}\ }\textbf {\bibinfo {volume} {622}},\ \bibinfo {pages} {69} (\bibinfo {year} {2023})}\BibitemShut {NoStop}%
\bibitem [{\citenamefont {Park}\ \emph {et~al.}(2023)\citenamefont {Park}, \citenamefont {Cai}, \citenamefont {Anderson}, \citenamefont {Zhang}, \citenamefont {Zhu}, \citenamefont {Liu}, \citenamefont {Wang}, \citenamefont {Holtzmann}, \citenamefont {Hu}, \citenamefont {Liu} \emph {et~al.}}]{park2023observation}%
  \BibitemOpen
  \bibfield  {author} {\bibinfo {author} {\bibfnamefont {H.}~\bibnamefont {Park}}, \bibinfo {author} {\bibfnamefont {J.}~\bibnamefont {Cai}}, \bibinfo {author} {\bibfnamefont {E.}~\bibnamefont {Anderson}}, \bibinfo {author} {\bibfnamefont {Y.}~\bibnamefont {Zhang}}, \bibinfo {author} {\bibfnamefont {J.}~\bibnamefont {Zhu}}, \bibinfo {author} {\bibfnamefont {X.}~\bibnamefont {Liu}}, \bibinfo {author} {\bibfnamefont {C.}~\bibnamefont {Wang}}, \bibinfo {author} {\bibfnamefont {W.}~\bibnamefont {Holtzmann}}, \bibinfo {author} {\bibfnamefont {C.}~\bibnamefont {Hu}}, \bibinfo {author} {\bibfnamefont {Z.}~\bibnamefont {Liu}},  \emph {et~al.},\ }\href@noop {} {\bibfield  {journal} {\bibinfo  {journal} {Nature}\ }\textbf {\bibinfo {volume} {622}},\ \bibinfo {pages} {74} (\bibinfo {year} {2023})}\BibitemShut {NoStop}%
\bibitem [{\citenamefont {Xu}\ \emph {et~al.}(2023)\citenamefont {Xu}, \citenamefont {Sun}, \citenamefont {Jia}, \citenamefont {Liu}, \citenamefont {Xu}, \citenamefont {Li}, \citenamefont {Gu}, \citenamefont {Watanabe}, \citenamefont {Taniguchi}, \citenamefont {Tong}, \citenamefont {Jia}, \citenamefont {Shi}, \citenamefont {Jiang}, \citenamefont {Zhang}, \citenamefont {Liu},\ and\ \citenamefont {Li}}]{PhysRevX.13.031037}%
  \BibitemOpen
  \bibfield  {author} {\bibinfo {author} {\bibfnamefont {F.}~\bibnamefont {Xu}}, \bibinfo {author} {\bibfnamefont {Z.}~\bibnamefont {Sun}}, \bibinfo {author} {\bibfnamefont {T.}~\bibnamefont {Jia}}, \bibinfo {author} {\bibfnamefont {C.}~\bibnamefont {Liu}}, \bibinfo {author} {\bibfnamefont {C.}~\bibnamefont {Xu}}, \bibinfo {author} {\bibfnamefont {C.}~\bibnamefont {Li}}, \bibinfo {author} {\bibfnamefont {Y.}~\bibnamefont {Gu}}, \bibinfo {author} {\bibfnamefont {K.}~\bibnamefont {Watanabe}}, \bibinfo {author} {\bibfnamefont {T.}~\bibnamefont {Taniguchi}}, \bibinfo {author} {\bibfnamefont {B.}~\bibnamefont {Tong}}, \bibinfo {author} {\bibfnamefont {J.}~\bibnamefont {Jia}}, \bibinfo {author} {\bibfnamefont {Z.}~\bibnamefont {Shi}}, \bibinfo {author} {\bibfnamefont {S.}~\bibnamefont {Jiang}}, \bibinfo {author} {\bibfnamefont {Y.}~\bibnamefont {Zhang}}, \bibinfo {author} {\bibfnamefont {X.}~\bibnamefont {Liu}}, \ and\ \bibinfo {author} {\bibfnamefont {T.}~\bibnamefont {Li}},\ }\href {\doibase
  10.1103/PhysRevX.13.031037} {\bibfield  {journal} {\bibinfo  {journal} {Phys. Rev. X}\ }\textbf {\bibinfo {volume} {13}},\ \bibinfo {pages} {031037} (\bibinfo {year} {2023})}\BibitemShut {NoStop}%
\bibitem [{\citenamefont {Morales-Dur\'an}\ \emph {et~al.}(2024)\citenamefont {Morales-Dur\'an}, \citenamefont {Wei}, \citenamefont {Shi},\ and\ \citenamefont {MacDonald}}]{PhysRevLett.132.096602}%
  \BibitemOpen
  \bibfield  {author} {\bibinfo {author} {\bibfnamefont {N.}~\bibnamefont {Morales-Dur\'an}}, \bibinfo {author} {\bibfnamefont {N.}~\bibnamefont {Wei}}, \bibinfo {author} {\bibfnamefont {J.}~\bibnamefont {Shi}}, \ and\ \bibinfo {author} {\bibfnamefont {A.~H.}\ \bibnamefont {MacDonald}},\ }\href {\doibase 10.1103/PhysRevLett.132.096602} {\bibfield  {journal} {\bibinfo  {journal} {Phys. Rev. Lett.}\ }\textbf {\bibinfo {volume} {132}},\ \bibinfo {pages} {096602} (\bibinfo {year} {2024})}\BibitemShut {NoStop}%
\bibitem [{\citenamefont {Foutty}\ \emph {et~al.}(2024)\citenamefont {Foutty}, \citenamefont {Kometter}, \citenamefont {Devakul}, \citenamefont {Reddy}, \citenamefont {Watanabe}, \citenamefont {Taniguchi}, \citenamefont {Fu},\ and\ \citenamefont {Feldman}}]{doi:10.1126/science.adi4728}%
  \BibitemOpen
  \bibfield  {author} {\bibinfo {author} {\bibfnamefont {B.~A.}\ \bibnamefont {Foutty}}, \bibinfo {author} {\bibfnamefont {C.~R.}\ \bibnamefont {Kometter}}, \bibinfo {author} {\bibfnamefont {T.}~\bibnamefont {Devakul}}, \bibinfo {author} {\bibfnamefont {A.~P.}\ \bibnamefont {Reddy}}, \bibinfo {author} {\bibfnamefont {K.}~\bibnamefont {Watanabe}}, \bibinfo {author} {\bibfnamefont {T.}~\bibnamefont {Taniguchi}}, \bibinfo {author} {\bibfnamefont {L.}~\bibnamefont {Fu}}, \ and\ \bibinfo {author} {\bibfnamefont {B.~E.}\ \bibnamefont {Feldman}},\ }\href {\doibase 10.1126/science.adi4728} {\bibfield  {journal} {\bibinfo  {journal} {Science}\ }\textbf {\bibinfo {volume} {384}},\ \bibinfo {pages} {343} (\bibinfo {year} {2024})},\ \Eprint {http://arxiv.org/abs/https://www.science.org/doi/pdf/10.1126/science.adi4728} {https://www.science.org/doi/pdf/10.1126/science.adi4728} \BibitemShut {NoStop}%
\bibitem [{\citenamefont {Jia}\ \emph {et~al.}(2024)\citenamefont {Jia}, \citenamefont {Yu}, \citenamefont {Liu}, \citenamefont {Herzog-Arbeitman}, \citenamefont {Qi}, \citenamefont {Pi}, \citenamefont {Regnault}, \citenamefont {Weng}, \citenamefont {Bernevig},\ and\ \citenamefont {Wu}}]{PhysRevB.109.205121}%
  \BibitemOpen
  \bibfield  {author} {\bibinfo {author} {\bibfnamefont {Y.}~\bibnamefont {Jia}}, \bibinfo {author} {\bibfnamefont {J.}~\bibnamefont {Yu}}, \bibinfo {author} {\bibfnamefont {J.}~\bibnamefont {Liu}}, \bibinfo {author} {\bibfnamefont {J.}~\bibnamefont {Herzog-Arbeitman}}, \bibinfo {author} {\bibfnamefont {Z.}~\bibnamefont {Qi}}, \bibinfo {author} {\bibfnamefont {H.}~\bibnamefont {Pi}}, \bibinfo {author} {\bibfnamefont {N.}~\bibnamefont {Regnault}}, \bibinfo {author} {\bibfnamefont {H.}~\bibnamefont {Weng}}, \bibinfo {author} {\bibfnamefont {B.~A.}\ \bibnamefont {Bernevig}}, \ and\ \bibinfo {author} {\bibfnamefont {Q.}~\bibnamefont {Wu}},\ }\href {\doibase 10.1103/PhysRevB.109.205121} {\bibfield  {journal} {\bibinfo  {journal} {Phys. Rev. B}\ }\textbf {\bibinfo {volume} {109}},\ \bibinfo {pages} {205121} (\bibinfo {year} {2024})}\BibitemShut {NoStop}%
\bibitem [{\citenamefont {Zhang}\ \emph {et~al.}(2024)\citenamefont {Zhang}, \citenamefont {Wang}, \citenamefont {Liu}, \citenamefont {Fan}, \citenamefont {Cao},\ and\ \citenamefont {Xiao}}]{zhang2024polarization}%
  \BibitemOpen
  \bibfield  {author} {\bibinfo {author} {\bibfnamefont {X.-W.}\ \bibnamefont {Zhang}}, \bibinfo {author} {\bibfnamefont {C.}~\bibnamefont {Wang}}, \bibinfo {author} {\bibfnamefont {X.}~\bibnamefont {Liu}}, \bibinfo {author} {\bibfnamefont {Y.}~\bibnamefont {Fan}}, \bibinfo {author} {\bibfnamefont {T.}~\bibnamefont {Cao}}, \ and\ \bibinfo {author} {\bibfnamefont {D.}~\bibnamefont {Xiao}},\ }\href@noop {} {\bibfield  {journal} {\bibinfo  {journal} {Nature Communications}\ }\textbf {\bibinfo {volume} {15}},\ \bibinfo {pages} {4223} (\bibinfo {year} {2024})}\BibitemShut {NoStop}%
\bibitem [{\citenamefont {Shi}\ \emph {et~al.}(2024)\citenamefont {Shi}, \citenamefont {Morales-Dur\'an}, \citenamefont {Khalaf},\ and\ \citenamefont {MacDonald}}]{PhysRevB.110.035130}%
  \BibitemOpen
  \bibfield  {author} {\bibinfo {author} {\bibfnamefont {J.}~\bibnamefont {Shi}}, \bibinfo {author} {\bibfnamefont {N.}~\bibnamefont {Morales-Dur\'an}}, \bibinfo {author} {\bibfnamefont {E.}~\bibnamefont {Khalaf}}, \ and\ \bibinfo {author} {\bibfnamefont {A.~H.}\ \bibnamefont {MacDonald}},\ }\href {\doibase 10.1103/PhysRevB.110.035130} {\bibfield  {journal} {\bibinfo  {journal} {Phys. Rev. B}\ }\textbf {\bibinfo {volume} {110}},\ \bibinfo {pages} {035130} (\bibinfo {year} {2024})}\BibitemShut {NoStop}%
\bibitem [{\citenamefont {Li}\ \emph {et~al.}(2021{\natexlab{b}})\citenamefont {Li}, \citenamefont {Kumar}, \citenamefont {Sun},\ and\ \citenamefont {Lin}}]{PhysRevResearch.3.L032070}%
  \BibitemOpen
  \bibfield  {author} {\bibinfo {author} {\bibfnamefont {H.}~\bibnamefont {Li}}, \bibinfo {author} {\bibfnamefont {U.}~\bibnamefont {Kumar}}, \bibinfo {author} {\bibfnamefont {K.}~\bibnamefont {Sun}}, \ and\ \bibinfo {author} {\bibfnamefont {S.-Z.}\ \bibnamefont {Lin}},\ }\href {\doibase 10.1103/PhysRevResearch.3.L032070} {\bibfield  {journal} {\bibinfo  {journal} {Phys. Rev. Res.}\ }\textbf {\bibinfo {volume} {3}},\ \bibinfo {pages} {L032070} (\bibinfo {year} {2021}{\natexlab{b}})}\BibitemShut {NoStop}%
\bibitem [{\citenamefont {Cr\'epel}\ and\ \citenamefont {Fu}(2023)}]{PhysRevB.107.L201109}%
  \BibitemOpen
  \bibfield  {author} {\bibinfo {author} {\bibfnamefont {V.}~\bibnamefont {Cr\'epel}}\ and\ \bibinfo {author} {\bibfnamefont {L.}~\bibnamefont {Fu}},\ }\href {\doibase 10.1103/PhysRevB.107.L201109} {\bibfield  {journal} {\bibinfo  {journal} {Phys. Rev. B}\ }\textbf {\bibinfo {volume} {107}},\ \bibinfo {pages} {L201109} (\bibinfo {year} {2023})}\BibitemShut {NoStop}%
\bibitem [{\citenamefont {Reddy}\ \emph {et~al.}(2023)\citenamefont {Reddy}, \citenamefont {Alsallom}, \citenamefont {Zhang}, \citenamefont {Devakul},\ and\ \citenamefont {Fu}}]{PhysRevB.108.085117}%
  \BibitemOpen
  \bibfield  {author} {\bibinfo {author} {\bibfnamefont {A.~P.}\ \bibnamefont {Reddy}}, \bibinfo {author} {\bibfnamefont {F.}~\bibnamefont {Alsallom}}, \bibinfo {author} {\bibfnamefont {Y.}~\bibnamefont {Zhang}}, \bibinfo {author} {\bibfnamefont {T.}~\bibnamefont {Devakul}}, \ and\ \bibinfo {author} {\bibfnamefont {L.}~\bibnamefont {Fu}},\ }\href {\doibase 10.1103/PhysRevB.108.085117} {\bibfield  {journal} {\bibinfo  {journal} {Phys. Rev. B}\ }\textbf {\bibinfo {volume} {108}},\ \bibinfo {pages} {085117} (\bibinfo {year} {2023})}\BibitemShut {NoStop}%
\bibitem [{\citenamefont {Morales-Dur\'an}\ \emph {et~al.}(2023)\citenamefont {Morales-Dur\'an}, \citenamefont {Wang}, \citenamefont {Schleder}, \citenamefont {Angeli}, \citenamefont {Zhu}, \citenamefont {Kaxiras}, \citenamefont {Repellin},\ and\ \citenamefont {Cano}}]{PhysRevResearch.5.L032022}%
  \BibitemOpen
  \bibfield  {author} {\bibinfo {author} {\bibfnamefont {N.}~\bibnamefont {Morales-Dur\'an}}, \bibinfo {author} {\bibfnamefont {J.}~\bibnamefont {Wang}}, \bibinfo {author} {\bibfnamefont {G.~R.}\ \bibnamefont {Schleder}}, \bibinfo {author} {\bibfnamefont {M.}~\bibnamefont {Angeli}}, \bibinfo {author} {\bibfnamefont {Z.}~\bibnamefont {Zhu}}, \bibinfo {author} {\bibfnamefont {E.}~\bibnamefont {Kaxiras}}, \bibinfo {author} {\bibfnamefont {C.}~\bibnamefont {Repellin}}, \ and\ \bibinfo {author} {\bibfnamefont {J.}~\bibnamefont {Cano}},\ }\href {\doibase 10.1103/PhysRevResearch.5.L032022} {\bibfield  {journal} {\bibinfo  {journal} {Phys. Rev. Res.}\ }\textbf {\bibinfo {volume} {5}},\ \bibinfo {pages} {L032022} (\bibinfo {year} {2023})}\BibitemShut {NoStop}%
\bibitem [{\citenamefont {Dong}\ \emph {et~al.}(2023)\citenamefont {Dong}, \citenamefont {Wang}, \citenamefont {Ledwith}, \citenamefont {Vishwanath},\ and\ \citenamefont {Parker}}]{PhysRevLett.131.136502}%
  \BibitemOpen
  \bibfield  {author} {\bibinfo {author} {\bibfnamefont {J.}~\bibnamefont {Dong}}, \bibinfo {author} {\bibfnamefont {J.}~\bibnamefont {Wang}}, \bibinfo {author} {\bibfnamefont {P.~J.}\ \bibnamefont {Ledwith}}, \bibinfo {author} {\bibfnamefont {A.}~\bibnamefont {Vishwanath}}, \ and\ \bibinfo {author} {\bibfnamefont {D.~E.}\ \bibnamefont {Parker}},\ }\href {\doibase 10.1103/PhysRevLett.131.136502} {\bibfield  {journal} {\bibinfo  {journal} {Phys. Rev. Lett.}\ }\textbf {\bibinfo {volume} {131}},\ \bibinfo {pages} {136502} (\bibinfo {year} {2023})}\BibitemShut {NoStop}%
\bibitem [{\citenamefont {Wang}\ \emph {et~al.}(2024)\citenamefont {Wang}, \citenamefont {Zhang}, \citenamefont {Liu}, \citenamefont {He}, \citenamefont {Xu}, \citenamefont {Ran}, \citenamefont {Cao},\ and\ \citenamefont {Xiao}}]{PhysRevLett.132.036501}%
  \BibitemOpen
  \bibfield  {author} {\bibinfo {author} {\bibfnamefont {C.}~\bibnamefont {Wang}}, \bibinfo {author} {\bibfnamefont {X.-W.}\ \bibnamefont {Zhang}}, \bibinfo {author} {\bibfnamefont {X.}~\bibnamefont {Liu}}, \bibinfo {author} {\bibfnamefont {Y.}~\bibnamefont {He}}, \bibinfo {author} {\bibfnamefont {X.}~\bibnamefont {Xu}}, \bibinfo {author} {\bibfnamefont {Y.}~\bibnamefont {Ran}}, \bibinfo {author} {\bibfnamefont {T.}~\bibnamefont {Cao}}, \ and\ \bibinfo {author} {\bibfnamefont {D.}~\bibnamefont {Xiao}},\ }\href {\doibase 10.1103/PhysRevLett.132.036501} {\bibfield  {journal} {\bibinfo  {journal} {Phys. Rev. Lett.}\ }\textbf {\bibinfo {volume} {132}},\ \bibinfo {pages} {036501} (\bibinfo {year} {2024})}\BibitemShut {NoStop}%
\bibitem [{\citenamefont {Yu}\ \emph {et~al.}(2024)\citenamefont {Yu}, \citenamefont {Herzog-Arbeitman}, \citenamefont {Wang}, \citenamefont {Vafek}, \citenamefont {Bernevig},\ and\ \citenamefont {Regnault}}]{PhysRevB.109.045147}%
  \BibitemOpen
  \bibfield  {author} {\bibinfo {author} {\bibfnamefont {J.}~\bibnamefont {Yu}}, \bibinfo {author} {\bibfnamefont {J.}~\bibnamefont {Herzog-Arbeitman}}, \bibinfo {author} {\bibfnamefont {M.}~\bibnamefont {Wang}}, \bibinfo {author} {\bibfnamefont {O.}~\bibnamefont {Vafek}}, \bibinfo {author} {\bibfnamefont {B.~A.}\ \bibnamefont {Bernevig}}, \ and\ \bibinfo {author} {\bibfnamefont {N.}~\bibnamefont {Regnault}},\ }\href {\doibase 10.1103/PhysRevB.109.045147} {\bibfield  {journal} {\bibinfo  {journal} {Phys. Rev. B}\ }\textbf {\bibinfo {volume} {109}},\ \bibinfo {pages} {045147} (\bibinfo {year} {2024})}\BibitemShut {NoStop}%
\bibitem [{\citenamefont {Xu}\ \emph {et~al.}(2024)\citenamefont {Xu}, \citenamefont {Li}, \citenamefont {Xu}, \citenamefont {Bi},\ and\ \citenamefont {Zhang}}]{doi:10.1073/pnas.2316749121}%
  \BibitemOpen
  \bibfield  {author} {\bibinfo {author} {\bibfnamefont {C.}~\bibnamefont {Xu}}, \bibinfo {author} {\bibfnamefont {J.}~\bibnamefont {Li}}, \bibinfo {author} {\bibfnamefont {Y.}~\bibnamefont {Xu}}, \bibinfo {author} {\bibfnamefont {Z.}~\bibnamefont {Bi}}, \ and\ \bibinfo {author} {\bibfnamefont {Y.}~\bibnamefont {Zhang}},\ }\href {\doibase 10.1073/pnas.2316749121} {\bibfield  {journal} {\bibinfo  {journal} {Proceedings of the National Academy of Sciences}\ }\textbf {\bibinfo {volume} {121}},\ \bibinfo {pages} {e2316749121} (\bibinfo {year} {2024})},\ \Eprint {http://arxiv.org/abs/https://www.pnas.org/doi/pdf/10.1073/pnas.2316749121} {https://www.pnas.org/doi/pdf/10.1073/pnas.2316749121} \BibitemShut {NoStop}%
\bibitem [{\citenamefont {Song}\ \emph {et~al.}(2024)\citenamefont {Song}, \citenamefont {Zhang},\ and\ \citenamefont {Senthil}}]{PhysRevB.109.085143}%
  \BibitemOpen
  \bibfield  {author} {\bibinfo {author} {\bibfnamefont {X.-Y.}\ \bibnamefont {Song}}, \bibinfo {author} {\bibfnamefont {Y.-H.}\ \bibnamefont {Zhang}}, \ and\ \bibinfo {author} {\bibfnamefont {T.}~\bibnamefont {Senthil}},\ }\href {\doibase 10.1103/PhysRevB.109.085143} {\bibfield  {journal} {\bibinfo  {journal} {Phys. Rev. B}\ }\textbf {\bibinfo {volume} {109}},\ \bibinfo {pages} {085143} (\bibinfo {year} {2024})}\BibitemShut {NoStop}%
\bibitem [{\citenamefont {Abouelkomsan}\ \emph {et~al.}(2024)\citenamefont {Abouelkomsan}, \citenamefont {Reddy}, \citenamefont {Fu},\ and\ \citenamefont {Bergholtz}}]{PhysRevB.109.L121107}%
  \BibitemOpen
  \bibfield  {author} {\bibinfo {author} {\bibfnamefont {A.}~\bibnamefont {Abouelkomsan}}, \bibinfo {author} {\bibfnamefont {A.~P.}\ \bibnamefont {Reddy}}, \bibinfo {author} {\bibfnamefont {L.}~\bibnamefont {Fu}}, \ and\ \bibinfo {author} {\bibfnamefont {E.~J.}\ \bibnamefont {Bergholtz}},\ }\href {\doibase 10.1103/PhysRevB.109.L121107} {\bibfield  {journal} {\bibinfo  {journal} {Phys. Rev. B}\ }\textbf {\bibinfo {volume} {109}},\ \bibinfo {pages} {L121107} (\bibinfo {year} {2024})}\BibitemShut {NoStop}%
\bibitem [{\citenamefont {Chen}\ \emph {et~al.}(2025)\citenamefont {Chen}, \citenamefont {Luo}, \citenamefont {Zhu},\ and\ \citenamefont {Sheng}}]{chen2025robust}%
  \BibitemOpen
  \bibfield  {author} {\bibinfo {author} {\bibfnamefont {F.}~\bibnamefont {Chen}}, \bibinfo {author} {\bibfnamefont {W.-W.}\ \bibnamefont {Luo}}, \bibinfo {author} {\bibfnamefont {W.}~\bibnamefont {Zhu}}, \ and\ \bibinfo {author} {\bibfnamefont {D.}~\bibnamefont {Sheng}},\ }\href@noop {} {\bibfield  {journal} {\bibinfo  {journal} {Nature Communications}\ }\textbf {\bibinfo {volume} {16}},\ \bibinfo {pages} {2115} (\bibinfo {year} {2025})}\BibitemShut {NoStop}%
\bibitem [{\citenamefont {Li}\ and\ \citenamefont {Wu}(2025)}]{PhysRevB.111.125122}%
  \BibitemOpen
  \bibfield  {author} {\bibinfo {author} {\bibfnamefont {B.}~\bibnamefont {Li}}\ and\ \bibinfo {author} {\bibfnamefont {F.}~\bibnamefont {Wu}},\ }\href {\doibase 10.1103/PhysRevB.111.125122} {\bibfield  {journal} {\bibinfo  {journal} {Phys. Rev. B}\ }\textbf {\bibinfo {volume} {111}},\ \bibinfo {pages} {125122} (\bibinfo {year} {2025})}\BibitemShut {NoStop}%
\bibitem [{\citenamefont {Paul}\ \emph {et~al.}(2025)\citenamefont {Paul}, \citenamefont {Abouelkomsan}, \citenamefont {Reddy},\ and\ \citenamefont {Fu}}]{paul2025shining}%
  \BibitemOpen
  \bibfield  {author} {\bibinfo {author} {\bibfnamefont {N.}~\bibnamefont {Paul}}, \bibinfo {author} {\bibfnamefont {A.}~\bibnamefont {Abouelkomsan}}, \bibinfo {author} {\bibfnamefont {A.}~\bibnamefont {Reddy}}, \ and\ \bibinfo {author} {\bibfnamefont {L.}~\bibnamefont {Fu}},\ }\href@noop {} {\bibfield  {journal} {\bibinfo  {journal} {arXiv preprint arXiv:2502.17569}\ } (\bibinfo {year} {2025})}\BibitemShut {NoStop}%
\bibitem [{\citenamefont {Kang}\ \emph {et~al.}(2025)\citenamefont {Kang}, \citenamefont {Qiu}, \citenamefont {Shen}, \citenamefont {Lee}, \citenamefont {Xia}, \citenamefont {Zeng}, \citenamefont {Watanabe}, \citenamefont {Taniguchi}, \citenamefont {Shan},\ and\ \citenamefont {Mak}}]{kang2025time}%
  \BibitemOpen
  \bibfield  {author} {\bibinfo {author} {\bibfnamefont {K.}~\bibnamefont {Kang}}, \bibinfo {author} {\bibfnamefont {Y.}~\bibnamefont {Qiu}}, \bibinfo {author} {\bibfnamefont {B.}~\bibnamefont {Shen}}, \bibinfo {author} {\bibfnamefont {K.}~\bibnamefont {Lee}}, \bibinfo {author} {\bibfnamefont {Z.}~\bibnamefont {Xia}}, \bibinfo {author} {\bibfnamefont {Y.}~\bibnamefont {Zeng}}, \bibinfo {author} {\bibfnamefont {K.}~\bibnamefont {Watanabe}}, \bibinfo {author} {\bibfnamefont {T.}~\bibnamefont {Taniguchi}}, \bibinfo {author} {\bibfnamefont {J.}~\bibnamefont {Shan}}, \ and\ \bibinfo {author} {\bibfnamefont {K.~F.}\ \bibnamefont {Mak}},\ }\href@noop {} {\bibfield  {journal} {\bibinfo  {journal} {arXiv preprint arXiv:2501.02525}\ } (\bibinfo {year} {2025})}\BibitemShut {NoStop}%
\bibitem [{\citenamefont {Kousa}\ \emph {et~al.}(2025)\citenamefont {Kousa}, \citenamefont {Morales-Dur{\'a}n}, \citenamefont {Wolf}, \citenamefont {Khalaf},\ and\ \citenamefont {MacDonald}}]{kousa2025theory}%
  \BibitemOpen
  \bibfield  {author} {\bibinfo {author} {\bibfnamefont {B.~M.}\ \bibnamefont {Kousa}}, \bibinfo {author} {\bibfnamefont {N.}~\bibnamefont {Morales-Dur{\'a}n}}, \bibinfo {author} {\bibfnamefont {T.~M.}\ \bibnamefont {Wolf}}, \bibinfo {author} {\bibfnamefont {E.}~\bibnamefont {Khalaf}}, \ and\ \bibinfo {author} {\bibfnamefont {A.~H.}\ \bibnamefont {MacDonald}},\ }\href@noop {} {\bibfield  {journal} {\bibinfo  {journal} {arXiv preprint arXiv:2502.17574}\ } (\bibinfo {year} {2025})}\BibitemShut {NoStop}%
\bibitem [{\citenamefont {Shi}\ \emph {et~al.}(2025)\citenamefont {Shi}, \citenamefont {Cano},\ and\ \citenamefont {Morales-Dur{\'a}n}}]{shi2025effects}%
  \BibitemOpen
  \bibfield  {author} {\bibinfo {author} {\bibfnamefont {J.}~\bibnamefont {Shi}}, \bibinfo {author} {\bibfnamefont {J.}~\bibnamefont {Cano}}, \ and\ \bibinfo {author} {\bibfnamefont {N.}~\bibnamefont {Morales-Dur{\'a}n}},\ }\href@noop {} {\bibfield  {journal} {\bibinfo  {journal} {arXiv preprint arXiv:2503.15900}\ } (\bibinfo {year} {2025})}\BibitemShut {NoStop}%
\bibitem [{\citenamefont {Zaklama}\ \emph {et~al.}(2024)\citenamefont {Zaklama}, \citenamefont {Luo},\ and\ \citenamefont {Fu}}]{zaklama2024structure}%
  \BibitemOpen
  \bibfield  {author} {\bibinfo {author} {\bibfnamefont {T.}~\bibnamefont {Zaklama}}, \bibinfo {author} {\bibfnamefont {D.}~\bibnamefont {Luo}}, \ and\ \bibinfo {author} {\bibfnamefont {L.}~\bibnamefont {Fu}},\ }\href@noop {} {\bibfield  {journal} {\bibinfo  {journal} {arXiv preprint arXiv:2411.03496}\ } (\bibinfo {year} {2024})}\BibitemShut {NoStop}%
\bibitem [{\citenamefont {Reddy}\ \emph {et~al.}(2024)\citenamefont {Reddy}, \citenamefont {Sheng}, \citenamefont {Abouelkomsan}, \citenamefont {Bergholtz},\ and\ \citenamefont {Fu}}]{reddy2024anti}%
  \BibitemOpen
  \bibfield  {author} {\bibinfo {author} {\bibfnamefont {A.~P.}\ \bibnamefont {Reddy}}, \bibinfo {author} {\bibfnamefont {D.}~\bibnamefont {Sheng}}, \bibinfo {author} {\bibfnamefont {A.}~\bibnamefont {Abouelkomsan}}, \bibinfo {author} {\bibfnamefont {E.~J.}\ \bibnamefont {Bergholtz}}, \ and\ \bibinfo {author} {\bibfnamefont {L.}~\bibnamefont {Fu}},\ }\href@noop {} {\bibfield  {journal} {\bibinfo  {journal} {arXiv preprint arXiv:2411.19898}\ } (\bibinfo {year} {2024})}\BibitemShut {NoStop}%
\bibitem [{\citenamefont {Tang}\ \emph {et~al.}(2020)\citenamefont {Tang}, \citenamefont {Li}, \citenamefont {Li}, \citenamefont {Xu}, \citenamefont {Liu}, \citenamefont {Barmak}, \citenamefont {Watanabe}, \citenamefont {Taniguchi}, \citenamefont {MacDonald}, \citenamefont {Shan} \emph {et~al.}}]{tang2020simulation}%
  \BibitemOpen
  \bibfield  {author} {\bibinfo {author} {\bibfnamefont {Y.}~\bibnamefont {Tang}}, \bibinfo {author} {\bibfnamefont {L.}~\bibnamefont {Li}}, \bibinfo {author} {\bibfnamefont {T.}~\bibnamefont {Li}}, \bibinfo {author} {\bibfnamefont {Y.}~\bibnamefont {Xu}}, \bibinfo {author} {\bibfnamefont {S.}~\bibnamefont {Liu}}, \bibinfo {author} {\bibfnamefont {K.}~\bibnamefont {Barmak}}, \bibinfo {author} {\bibfnamefont {K.}~\bibnamefont {Watanabe}}, \bibinfo {author} {\bibfnamefont {T.}~\bibnamefont {Taniguchi}}, \bibinfo {author} {\bibfnamefont {A.~H.}\ \bibnamefont {MacDonald}}, \bibinfo {author} {\bibfnamefont {J.}~\bibnamefont {Shan}},  \emph {et~al.},\ }\href@noop {} {\bibfield  {journal} {\bibinfo  {journal} {Nature}\ }\textbf {\bibinfo {volume} {579}},\ \bibinfo {pages} {353} (\bibinfo {year} {2020})}\BibitemShut {NoStop}%
\bibitem [{\citenamefont {Regan}\ \emph {et~al.}(2020)\citenamefont {Regan}, \citenamefont {Wang}, \citenamefont {Jin}, \citenamefont {Bakti~Utama}, \citenamefont {Gao}, \citenamefont {Wei}, \citenamefont {Zhao}, \citenamefont {Zhao}, \citenamefont {Zhang}, \citenamefont {Yumigeta} \emph {et~al.}}]{regan2020mott}%
  \BibitemOpen
  \bibfield  {author} {\bibinfo {author} {\bibfnamefont {E.~C.}\ \bibnamefont {Regan}}, \bibinfo {author} {\bibfnamefont {D.}~\bibnamefont {Wang}}, \bibinfo {author} {\bibfnamefont {C.}~\bibnamefont {Jin}}, \bibinfo {author} {\bibfnamefont {M.~I.}\ \bibnamefont {Bakti~Utama}}, \bibinfo {author} {\bibfnamefont {B.}~\bibnamefont {Gao}}, \bibinfo {author} {\bibfnamefont {X.}~\bibnamefont {Wei}}, \bibinfo {author} {\bibfnamefont {S.}~\bibnamefont {Zhao}}, \bibinfo {author} {\bibfnamefont {W.}~\bibnamefont {Zhao}}, \bibinfo {author} {\bibfnamefont {Z.}~\bibnamefont {Zhang}}, \bibinfo {author} {\bibfnamefont {K.}~\bibnamefont {Yumigeta}},  \emph {et~al.},\ }\href@noop {} {\bibfield  {journal} {\bibinfo  {journal} {Nature}\ }\textbf {\bibinfo {volume} {579}},\ \bibinfo {pages} {359} (\bibinfo {year} {2020})}\BibitemShut {NoStop}%
\bibitem [{\citenamefont {Zhang}\ \emph {et~al.}(2020)\citenamefont {Zhang}, \citenamefont {Wang}, \citenamefont {Watanabe}, \citenamefont {Taniguchi}, \citenamefont {Ueno}, \citenamefont {Tutuc},\ and\ \citenamefont {LeRoy}}]{zhang2020flat}%
  \BibitemOpen
  \bibfield  {author} {\bibinfo {author} {\bibfnamefont {Z.}~\bibnamefont {Zhang}}, \bibinfo {author} {\bibfnamefont {Y.}~\bibnamefont {Wang}}, \bibinfo {author} {\bibfnamefont {K.}~\bibnamefont {Watanabe}}, \bibinfo {author} {\bibfnamefont {T.}~\bibnamefont {Taniguchi}}, \bibinfo {author} {\bibfnamefont {K.}~\bibnamefont {Ueno}}, \bibinfo {author} {\bibfnamefont {E.}~\bibnamefont {Tutuc}}, \ and\ \bibinfo {author} {\bibfnamefont {B.~J.}\ \bibnamefont {LeRoy}},\ }\href@noop {} {\bibfield  {journal} {\bibinfo  {journal} {Nature Physics}\ }\textbf {\bibinfo {volume} {16}},\ \bibinfo {pages} {1093} (\bibinfo {year} {2020})}\BibitemShut {NoStop}%
\bibitem [{\citenamefont {Enaldiev}\ \emph {et~al.}(2020)\citenamefont {Enaldiev}, \citenamefont {Z\'olyomi}, \citenamefont {Yelgel}, \citenamefont {Magorrian},\ and\ \citenamefont {Fal'ko}}]{PhysRevLett.124.206101}%
  \BibitemOpen
  \bibfield  {author} {\bibinfo {author} {\bibfnamefont {V.~V.}\ \bibnamefont {Enaldiev}}, \bibinfo {author} {\bibfnamefont {V.}~\bibnamefont {Z\'olyomi}}, \bibinfo {author} {\bibfnamefont {C.}~\bibnamefont {Yelgel}}, \bibinfo {author} {\bibfnamefont {S.~J.}\ \bibnamefont {Magorrian}}, \ and\ \bibinfo {author} {\bibfnamefont {V.~I.}\ \bibnamefont {Fal'ko}},\ }\href {\doibase 10.1103/PhysRevLett.124.206101} {\bibfield  {journal} {\bibinfo  {journal} {Phys. Rev. Lett.}\ }\textbf {\bibinfo {volume} {124}},\ \bibinfo {pages} {206101} (\bibinfo {year} {2020})}\BibitemShut {NoStop}%
\bibitem [{\citenamefont {Xu}\ \emph {et~al.}(2020)\citenamefont {Xu}, \citenamefont {Liu}, \citenamefont {Rhodes}, \citenamefont {Watanabe}, \citenamefont {Taniguchi}, \citenamefont {Hone}, \citenamefont {Elser}, \citenamefont {Mak},\ and\ \citenamefont {Shan}}]{xu2020correlated}%
  \BibitemOpen
  \bibfield  {author} {\bibinfo {author} {\bibfnamefont {Y.}~\bibnamefont {Xu}}, \bibinfo {author} {\bibfnamefont {S.}~\bibnamefont {Liu}}, \bibinfo {author} {\bibfnamefont {D.~A.}\ \bibnamefont {Rhodes}}, \bibinfo {author} {\bibfnamefont {K.}~\bibnamefont {Watanabe}}, \bibinfo {author} {\bibfnamefont {T.}~\bibnamefont {Taniguchi}}, \bibinfo {author} {\bibfnamefont {J.}~\bibnamefont {Hone}}, \bibinfo {author} {\bibfnamefont {V.}~\bibnamefont {Elser}}, \bibinfo {author} {\bibfnamefont {K.~F.}\ \bibnamefont {Mak}}, \ and\ \bibinfo {author} {\bibfnamefont {J.}~\bibnamefont {Shan}},\ }\href@noop {} {\bibfield  {journal} {\bibinfo  {journal} {Nature}\ }\textbf {\bibinfo {volume} {587}},\ \bibinfo {pages} {214} (\bibinfo {year} {2020})}\BibitemShut {NoStop}%
\bibitem [{\citenamefont {Angeli}\ and\ \citenamefont {MacDonald}(2021)}]{doi:10.1073/pnas.2021826118}%
  \BibitemOpen
  \bibfield  {author} {\bibinfo {author} {\bibfnamefont {M.}~\bibnamefont {Angeli}}\ and\ \bibinfo {author} {\bibfnamefont {A.~H.}\ \bibnamefont {MacDonald}},\ }\href {\doibase 10.1073/pnas.2021826118} {\bibfield  {journal} {\bibinfo  {journal} {Proceedings of the National Academy of Sciences}\ }\textbf {\bibinfo {volume} {118}},\ \bibinfo {pages} {e2021826118} (\bibinfo {year} {2021})},\ \Eprint {http://arxiv.org/abs/https://www.pnas.org/doi/pdf/10.1073/pnas.2021826118} {https://www.pnas.org/doi/pdf/10.1073/pnas.2021826118} \BibitemShut {NoStop}%
\bibitem [{\citenamefont {Li}\ \emph {et~al.}(2021{\natexlab{c}})\citenamefont {Li}, \citenamefont {Li}, \citenamefont {Regan}, \citenamefont {Wang}, \citenamefont {Zhao}, \citenamefont {Kahn}, \citenamefont {Yumigeta}, \citenamefont {Blei}, \citenamefont {Taniguchi}, \citenamefont {Watanabe} \emph {et~al.}}]{li2021imaging}%
  \BibitemOpen
  \bibfield  {author} {\bibinfo {author} {\bibfnamefont {H.}~\bibnamefont {Li}}, \bibinfo {author} {\bibfnamefont {S.}~\bibnamefont {Li}}, \bibinfo {author} {\bibfnamefont {E.~C.}\ \bibnamefont {Regan}}, \bibinfo {author} {\bibfnamefont {D.}~\bibnamefont {Wang}}, \bibinfo {author} {\bibfnamefont {W.}~\bibnamefont {Zhao}}, \bibinfo {author} {\bibfnamefont {S.}~\bibnamefont {Kahn}}, \bibinfo {author} {\bibfnamefont {K.}~\bibnamefont {Yumigeta}}, \bibinfo {author} {\bibfnamefont {M.}~\bibnamefont {Blei}}, \bibinfo {author} {\bibfnamefont {T.}~\bibnamefont {Taniguchi}}, \bibinfo {author} {\bibfnamefont {K.}~\bibnamefont {Watanabe}},  \emph {et~al.},\ }\href@noop {} {\bibfield  {journal} {\bibinfo  {journal} {Nature}\ }\textbf {\bibinfo {volume} {597}},\ \bibinfo {pages} {650} (\bibinfo {year} {2021}{\natexlab{c}})}\BibitemShut {NoStop}%
\bibitem [{\citenamefont {Wang}\ \emph {et~al.}(2022)\citenamefont {Wang}, \citenamefont {Yasuda}, \citenamefont {Zhang}, \citenamefont {Liu}, \citenamefont {Watanabe}, \citenamefont {Taniguchi}, \citenamefont {Hone}, \citenamefont {Fu},\ and\ \citenamefont {Jarillo-Herrero}}]{wang2022interfacial}%
  \BibitemOpen
  \bibfield  {author} {\bibinfo {author} {\bibfnamefont {X.}~\bibnamefont {Wang}}, \bibinfo {author} {\bibfnamefont {K.}~\bibnamefont {Yasuda}}, \bibinfo {author} {\bibfnamefont {Y.}~\bibnamefont {Zhang}}, \bibinfo {author} {\bibfnamefont {S.}~\bibnamefont {Liu}}, \bibinfo {author} {\bibfnamefont {K.}~\bibnamefont {Watanabe}}, \bibinfo {author} {\bibfnamefont {T.}~\bibnamefont {Taniguchi}}, \bibinfo {author} {\bibfnamefont {J.}~\bibnamefont {Hone}}, \bibinfo {author} {\bibfnamefont {L.}~\bibnamefont {Fu}}, \ and\ \bibinfo {author} {\bibfnamefont {P.}~\bibnamefont {Jarillo-Herrero}},\ }\href@noop {} {\bibfield  {journal} {\bibinfo  {journal} {Nature nanotechnology}\ }\textbf {\bibinfo {volume} {17}},\ \bibinfo {pages} {367} (\bibinfo {year} {2022})}\BibitemShut {NoStop}%
\bibitem [{\citenamefont {Enaldiev}\ \emph {et~al.}(2022{\natexlab{a}})\citenamefont {Enaldiev}, \citenamefont {Ferreira},\ and\ \citenamefont {Fal’ko}}]{enaldiev2022scalable}%
  \BibitemOpen
  \bibfield  {author} {\bibinfo {author} {\bibfnamefont {V.~V.}\ \bibnamefont {Enaldiev}}, \bibinfo {author} {\bibfnamefont {F.}~\bibnamefont {Ferreira}}, \ and\ \bibinfo {author} {\bibfnamefont {V.~I.}\ \bibnamefont {Fal’ko}},\ }\href@noop {} {\bibfield  {journal} {\bibinfo  {journal} {Nano Letters}\ }\textbf {\bibinfo {volume} {22}},\ \bibinfo {pages} {1534} (\bibinfo {year} {2022}{\natexlab{a}})}\BibitemShut {NoStop}%
\bibitem [{\citenamefont {Anderson}\ \emph {et~al.}(2023)\citenamefont {Anderson}, \citenamefont {Fan}, \citenamefont {Cai}, \citenamefont {Holtzmann}, \citenamefont {Taniguchi}, \citenamefont {Watanabe}, \citenamefont {Xiao}, \citenamefont {Yao},\ and\ \citenamefont {Xu}}]{doi:10.1126/science.adg4268}%
  \BibitemOpen
  \bibfield  {author} {\bibinfo {author} {\bibfnamefont {E.}~\bibnamefont {Anderson}}, \bibinfo {author} {\bibfnamefont {F.-R.}\ \bibnamefont {Fan}}, \bibinfo {author} {\bibfnamefont {J.}~\bibnamefont {Cai}}, \bibinfo {author} {\bibfnamefont {W.}~\bibnamefont {Holtzmann}}, \bibinfo {author} {\bibfnamefont {T.}~\bibnamefont {Taniguchi}}, \bibinfo {author} {\bibfnamefont {K.}~\bibnamefont {Watanabe}}, \bibinfo {author} {\bibfnamefont {D.}~\bibnamefont {Xiao}}, \bibinfo {author} {\bibfnamefont {W.}~\bibnamefont {Yao}}, \ and\ \bibinfo {author} {\bibfnamefont {X.}~\bibnamefont {Xu}},\ }\href {\doibase 10.1126/science.adg4268} {\bibfield  {journal} {\bibinfo  {journal} {Science}\ }\textbf {\bibinfo {volume} {381}},\ \bibinfo {pages} {325} (\bibinfo {year} {2023})},\ \Eprint {http://arxiv.org/abs/https://www.science.org/doi/pdf/10.1126/science.adg4268} {https://www.science.org/doi/pdf/10.1126/science.adg4268} \BibitemShut {NoStop}%
\bibitem [{\citenamefont {Kol\'a\ifmmode~\check{r}\else \v{r}\fi{}}\ \emph {et~al.}(2024)\citenamefont {Kol\'a\ifmmode~\check{r}\else \v{r}\fi{}}, \citenamefont {Yang}, \citenamefont {von Oppen},\ and\ \citenamefont {Mora}}]{PhysRevB.110.115114}%
  \BibitemOpen
  \bibfield  {author} {\bibinfo {author} {\bibfnamefont {K.~c.~v.}\ \bibnamefont {Kol\'a\ifmmode~\check{r}\else \v{r}\fi{}}}, \bibinfo {author} {\bibfnamefont {K.}~\bibnamefont {Yang}}, \bibinfo {author} {\bibfnamefont {F.}~\bibnamefont {von Oppen}}, \ and\ \bibinfo {author} {\bibfnamefont {C.}~\bibnamefont {Mora}},\ }\href {\doibase 10.1103/PhysRevB.110.115114} {\bibfield  {journal} {\bibinfo  {journal} {Phys. Rev. B}\ }\textbf {\bibinfo {volume} {110}},\ \bibinfo {pages} {115114} (\bibinfo {year} {2024})}\BibitemShut {NoStop}%
\bibitem [{\citenamefont {Song}\ \emph {et~al.}(2022)\citenamefont {Song}, \citenamefont {Wang}, \citenamefont {Zheng}, \citenamefont {Narang},\ and\ \citenamefont {Wang}}]{song2022deep}%
  \BibitemOpen
  \bibfield  {author} {\bibinfo {author} {\bibfnamefont {Z.}~\bibnamefont {Song}}, \bibinfo {author} {\bibfnamefont {Y.}~\bibnamefont {Wang}}, \bibinfo {author} {\bibfnamefont {H.}~\bibnamefont {Zheng}}, \bibinfo {author} {\bibfnamefont {P.}~\bibnamefont {Narang}}, \ and\ \bibinfo {author} {\bibfnamefont {L.-W.}\ \bibnamefont {Wang}},\ }\href@noop {} {\bibfield  {journal} {\bibinfo  {journal} {Journal of the American Chemical Society}\ }\textbf {\bibinfo {volume} {144}},\ \bibinfo {pages} {14657} (\bibinfo {year} {2022})}\BibitemShut {NoStop}%
\bibitem [{\citenamefont {Enaldiev}\ \emph {et~al.}(2022{\natexlab{b}})\citenamefont {Enaldiev}, \citenamefont {Ferreira}, \citenamefont {McHugh},\ and\ \citenamefont {Fal’ko}}]{enaldiev2022self}%
  \BibitemOpen
  \bibfield  {author} {\bibinfo {author} {\bibfnamefont {V.}~\bibnamefont {Enaldiev}}, \bibinfo {author} {\bibfnamefont {F.}~\bibnamefont {Ferreira}}, \bibinfo {author} {\bibfnamefont {J.}~\bibnamefont {McHugh}}, \ and\ \bibinfo {author} {\bibfnamefont {V.~I.}\ \bibnamefont {Fal’ko}},\ }\href@noop {} {\bibfield  {journal} {\bibinfo  {journal} {npj 2D Materials and Applications}\ }\textbf {\bibinfo {volume} {6}},\ \bibinfo {pages} {74} (\bibinfo {year} {2022}{\natexlab{b}})}\BibitemShut {NoStop}%
\bibitem [{\citenamefont {Pasek}\ \emph {et~al.}(2023)\citenamefont {Pasek}, \citenamefont {Kupczynski},\ and\ \citenamefont {Potasz}}]{PhysRevB.108.165152}%
  \BibitemOpen
  \bibfield  {author} {\bibinfo {author} {\bibfnamefont {W.}~\bibnamefont {Pasek}}, \bibinfo {author} {\bibfnamefont {M.}~\bibnamefont {Kupczynski}}, \ and\ \bibinfo {author} {\bibfnamefont {P.}~\bibnamefont {Potasz}},\ }\href {\doibase 10.1103/PhysRevB.108.165152} {\bibfield  {journal} {\bibinfo  {journal} {Phys. Rev. B}\ }\textbf {\bibinfo {volume} {108}},\ \bibinfo {pages} {165152} (\bibinfo {year} {2023})}\BibitemShut {NoStop}%
\bibitem [{\citenamefont {Soltero}\ \emph {et~al.}(2024{\natexlab{b}})\citenamefont {Soltero}, \citenamefont {Kaliteevski}, \citenamefont {McHugh}, \citenamefont {Enaldiev},\ and\ \citenamefont {Fal’ko}}]{soltero2024competition}%
  \BibitemOpen
  \bibfield  {author} {\bibinfo {author} {\bibfnamefont {I.}~\bibnamefont {Soltero}}, \bibinfo {author} {\bibfnamefont {M.~A.}\ \bibnamefont {Kaliteevski}}, \bibinfo {author} {\bibfnamefont {J.~G.}\ \bibnamefont {McHugh}}, \bibinfo {author} {\bibfnamefont {V.}~\bibnamefont {Enaldiev}}, \ and\ \bibinfo {author} {\bibfnamefont {V.~I.}\ \bibnamefont {Fal’ko}},\ }\href@noop {} {\bibfield  {journal} {\bibinfo  {journal} {Nano Letters}\ }\textbf {\bibinfo {volume} {24}},\ \bibinfo {pages} {1996} (\bibinfo {year} {2024}{\natexlab{b}})}\BibitemShut {NoStop}%
\bibitem [{\citenamefont {Cavicchi}\ \emph {et~al.}(2024)\citenamefont {Cavicchi}, \citenamefont {Reijnders}, \citenamefont {Katsnelson},\ and\ \citenamefont {Polini}}]{cavicchi2024optical}%
  \BibitemOpen
  \bibfield  {author} {\bibinfo {author} {\bibfnamefont {L.}~\bibnamefont {Cavicchi}}, \bibinfo {author} {\bibfnamefont {K.~J.}\ \bibnamefont {Reijnders}}, \bibinfo {author} {\bibfnamefont {M.~I.}\ \bibnamefont {Katsnelson}}, \ and\ \bibinfo {author} {\bibfnamefont {M.}~\bibnamefont {Polini}},\ }\href@noop {} {\bibfield  {journal} {\bibinfo  {journal} {arXiv preprint arXiv:2410.18025}\ } (\bibinfo {year} {2024})}\BibitemShut {NoStop}%
\bibitem [{\citenamefont {Mora}\ \emph {et~al.}(2019)\citenamefont {Mora}, \citenamefont {Regnault},\ and\ \citenamefont {Bernevig}}]{PhysRevLett.123.026402}%
  \BibitemOpen
  \bibfield  {author} {\bibinfo {author} {\bibfnamefont {C.}~\bibnamefont {Mora}}, \bibinfo {author} {\bibfnamefont {N.}~\bibnamefont {Regnault}}, \ and\ \bibinfo {author} {\bibfnamefont {B.~A.}\ \bibnamefont {Bernevig}},\ }\href {\doibase 10.1103/PhysRevLett.123.026402} {\bibfield  {journal} {\bibinfo  {journal} {Phys. Rev. Lett.}\ }\textbf {\bibinfo {volume} {123}},\ \bibinfo {pages} {026402} (\bibinfo {year} {2019})}\BibitemShut {NoStop}%
\bibitem [{\citenamefont {Li}\ \emph {et~al.}(2019)\citenamefont {Li}, \citenamefont {Wu},\ and\ \citenamefont {MacDonald}}]{li2019electronic}%
  \BibitemOpen
  \bibfield  {author} {\bibinfo {author} {\bibfnamefont {X.}~\bibnamefont {Li}}, \bibinfo {author} {\bibfnamefont {F.}~\bibnamefont {Wu}}, \ and\ \bibinfo {author} {\bibfnamefont {A.~H.}\ \bibnamefont {MacDonald}},\ }\href@noop {} {\enquote {\bibinfo {title} {Electronic structure of single-twist trilayer graphene},}\ } (\bibinfo {year} {2019}),\ \Eprint {http://arxiv.org/abs/1907.12338} {arXiv:1907.12338 [cond-mat.mtrl-sci]} \BibitemShut {NoStop}%
\bibitem [{\citenamefont {Carr}\ \emph {et~al.}(2020)\citenamefont {Carr}, \citenamefont {Li}, \citenamefont {Zhu}, \citenamefont {Kaxiras}, \citenamefont {Sachdev},\ and\ \citenamefont {Kruchkov}}]{doi:10.1021/acs.nanolett.9b04979}%
  \BibitemOpen
  \bibfield  {author} {\bibinfo {author} {\bibfnamefont {S.}~\bibnamefont {Carr}}, \bibinfo {author} {\bibfnamefont {C.}~\bibnamefont {Li}}, \bibinfo {author} {\bibfnamefont {Z.}~\bibnamefont {Zhu}}, \bibinfo {author} {\bibfnamefont {E.}~\bibnamefont {Kaxiras}}, \bibinfo {author} {\bibfnamefont {S.}~\bibnamefont {Sachdev}}, \ and\ \bibinfo {author} {\bibfnamefont {A.}~\bibnamefont {Kruchkov}},\ }\href {\doibase 10.1021/acs.nanolett.9b04979} {\bibfield  {journal} {\bibinfo  {journal} {Nano Letters}\ }\textbf {\bibinfo {volume} {20}},\ \bibinfo {pages} {3030} (\bibinfo {year} {2020})},\ \bibinfo {note} {pMID: 32208724},\ \Eprint {http://arxiv.org/abs/https://doi.org/10.1021/acs.nanolett.9b04979} {https://doi.org/10.1021/acs.nanolett.9b04979} \BibitemShut {NoStop}%
\bibitem [{\citenamefont {Tritsaris}\ \emph {et~al.}(2020)\citenamefont {Tritsaris}, \citenamefont {Carr}, \citenamefont {Zhu}, \citenamefont {Xie}, \citenamefont {Torrisi}, \citenamefont {Tang}, \citenamefont {Mattheakis}, \citenamefont {Larson},\ and\ \citenamefont {Kaxiras}}]{tritsaris2020electronic}%
  \BibitemOpen
  \bibfield  {author} {\bibinfo {author} {\bibfnamefont {G.~A.}\ \bibnamefont {Tritsaris}}, \bibinfo {author} {\bibfnamefont {S.}~\bibnamefont {Carr}}, \bibinfo {author} {\bibfnamefont {Z.}~\bibnamefont {Zhu}}, \bibinfo {author} {\bibfnamefont {Y.}~\bibnamefont {Xie}}, \bibinfo {author} {\bibfnamefont {S.~B.}\ \bibnamefont {Torrisi}}, \bibinfo {author} {\bibfnamefont {J.}~\bibnamefont {Tang}}, \bibinfo {author} {\bibfnamefont {M.}~\bibnamefont {Mattheakis}}, \bibinfo {author} {\bibfnamefont {D.~T.}\ \bibnamefont {Larson}}, \ and\ \bibinfo {author} {\bibfnamefont {E.}~\bibnamefont {Kaxiras}},\ }\href@noop {} {\bibfield  {journal} {\bibinfo  {journal} {2D Materials}\ }\textbf {\bibinfo {volume} {7}},\ \bibinfo {pages} {035028} (\bibinfo {year} {2020})}\BibitemShut {NoStop}%
\bibitem [{\citenamefont {Lopez-Bezanilla}\ and\ \citenamefont {Lado}(2020)}]{PhysRevResearch.2.033357}%
  \BibitemOpen
  \bibfield  {author} {\bibinfo {author} {\bibfnamefont {A.}~\bibnamefont {Lopez-Bezanilla}}\ and\ \bibinfo {author} {\bibfnamefont {J.~L.}\ \bibnamefont {Lado}},\ }\href {\doibase 10.1103/PhysRevResearch.2.033357} {\bibfield  {journal} {\bibinfo  {journal} {Phys. Rev. Res.}\ }\textbf {\bibinfo {volume} {2}},\ \bibinfo {pages} {033357} (\bibinfo {year} {2020})}\BibitemShut {NoStop}%
\bibitem [{\citenamefont {Ramires}\ and\ \citenamefont {Lado}(2021)}]{PhysRevLett.127.026401}%
  \BibitemOpen
  \bibfield  {author} {\bibinfo {author} {\bibfnamefont {A.}~\bibnamefont {Ramires}}\ and\ \bibinfo {author} {\bibfnamefont {J.~L.}\ \bibnamefont {Lado}},\ }\href {\doibase 10.1103/PhysRevLett.127.026401} {\bibfield  {journal} {\bibinfo  {journal} {Phys. Rev. Lett.}\ }\textbf {\bibinfo {volume} {127}},\ \bibinfo {pages} {026401} (\bibinfo {year} {2021})}\BibitemShut {NoStop}%
\bibitem [{\citenamefont {Lei}\ \emph {et~al.}(2021)\citenamefont {Lei}, \citenamefont {Linhart}, \citenamefont {Qin}, \citenamefont {Libisch},\ and\ \citenamefont {MacDonald}}]{PhysRevB.104.035139}%
  \BibitemOpen
  \bibfield  {author} {\bibinfo {author} {\bibfnamefont {C.}~\bibnamefont {Lei}}, \bibinfo {author} {\bibfnamefont {L.}~\bibnamefont {Linhart}}, \bibinfo {author} {\bibfnamefont {W.}~\bibnamefont {Qin}}, \bibinfo {author} {\bibfnamefont {F.}~\bibnamefont {Libisch}}, \ and\ \bibinfo {author} {\bibfnamefont {A.~H.}\ \bibnamefont {MacDonald}},\ }\href {\doibase 10.1103/PhysRevB.104.035139} {\bibfield  {journal} {\bibinfo  {journal} {Phys. Rev. B}\ }\textbf {\bibinfo {volume} {104}},\ \bibinfo {pages} {035139} (\bibinfo {year} {2021})}\BibitemShut {NoStop}%
\bibitem [{\citenamefont {Phong}\ \emph {et~al.}(2021)\citenamefont {Phong}, \citenamefont {Pantale\'on}, \citenamefont {Cea},\ and\ \citenamefont {Guinea}}]{PhysRevB.104.L121116}%
  \BibitemOpen
  \bibfield  {author} {\bibinfo {author} {\bibfnamefont {V.~o.~T.}\ \bibnamefont {Phong}}, \bibinfo {author} {\bibfnamefont {P.~A.}\ \bibnamefont {Pantale\'on}}, \bibinfo {author} {\bibfnamefont {T.}~\bibnamefont {Cea}}, \ and\ \bibinfo {author} {\bibfnamefont {F.}~\bibnamefont {Guinea}},\ }\href {\doibase 10.1103/PhysRevB.104.L121116} {\bibfield  {journal} {\bibinfo  {journal} {Phys. Rev. B}\ }\textbf {\bibinfo {volume} {104}},\ \bibinfo {pages} {L121116} (\bibinfo {year} {2021})}\BibitemShut {NoStop}%
\bibitem [{\citenamefont {C\ifmmode \u{a}\else \u{a}\fi{}lug\ifmmode~\u{a}\else \u{a}\fi{}ru}\ \emph {et~al.}(2021)\citenamefont {C\ifmmode \u{a}\else \u{a}\fi{}lug\ifmmode~\u{a}\else \u{a}\fi{}ru}, \citenamefont {Xie}, \citenamefont {Song}, \citenamefont {Lian}, \citenamefont {Regnault},\ and\ \citenamefont {Bernevig}}]{PhysRevB.103.195411}%
  \BibitemOpen
  \bibfield  {author} {\bibinfo {author} {\bibfnamefont {D.}~\bibnamefont {C\ifmmode \u{a}\else \u{a}\fi{}lug\ifmmode~\u{a}\else \u{a}\fi{}ru}}, \bibinfo {author} {\bibfnamefont {F.}~\bibnamefont {Xie}}, \bibinfo {author} {\bibfnamefont {Z.-D.}\ \bibnamefont {Song}}, \bibinfo {author} {\bibfnamefont {B.}~\bibnamefont {Lian}}, \bibinfo {author} {\bibfnamefont {N.}~\bibnamefont {Regnault}}, \ and\ \bibinfo {author} {\bibfnamefont {B.~A.}\ \bibnamefont {Bernevig}},\ }\href {\doibase 10.1103/PhysRevB.103.195411} {\bibfield  {journal} {\bibinfo  {journal} {Phys. Rev. B}\ }\textbf {\bibinfo {volume} {103}},\ \bibinfo {pages} {195411} (\bibinfo {year} {2021})}\BibitemShut {NoStop}%
\bibitem [{\citenamefont {Xie}\ \emph {et~al.}(2021)\citenamefont {Xie}, \citenamefont {Regnault}, \citenamefont {C\ifmmode \u{a}\else \u{a}\fi{}lug\ifmmode~\u{a}\else \u{a}\fi{}ru}, \citenamefont {Bernevig},\ and\ \citenamefont {Lian}}]{PhysRevB.104.115167}%
  \BibitemOpen
  \bibfield  {author} {\bibinfo {author} {\bibfnamefont {F.}~\bibnamefont {Xie}}, \bibinfo {author} {\bibfnamefont {N.}~\bibnamefont {Regnault}}, \bibinfo {author} {\bibfnamefont {D.}~\bibnamefont {C\ifmmode \u{a}\else \u{a}\fi{}lug\ifmmode~\u{a}\else \u{a}\fi{}ru}}, \bibinfo {author} {\bibfnamefont {B.~A.}\ \bibnamefont {Bernevig}}, \ and\ \bibinfo {author} {\bibfnamefont {B.}~\bibnamefont {Lian}},\ }\href {\doibase 10.1103/PhysRevB.104.115167} {\bibfield  {journal} {\bibinfo  {journal} {Phys. Rev. B}\ }\textbf {\bibinfo {volume} {104}},\ \bibinfo {pages} {115167} (\bibinfo {year} {2021})}\BibitemShut {NoStop}%
\bibitem [{\citenamefont {Guerci}\ \emph {et~al.}(2022)\citenamefont {Guerci}, \citenamefont {Simon},\ and\ \citenamefont {Mora}}]{PhysRevResearch.4.L012013}%
  \BibitemOpen
  \bibfield  {author} {\bibinfo {author} {\bibfnamefont {D.}~\bibnamefont {Guerci}}, \bibinfo {author} {\bibfnamefont {P.}~\bibnamefont {Simon}}, \ and\ \bibinfo {author} {\bibfnamefont {C.}~\bibnamefont {Mora}},\ }\href {\doibase 10.1103/PhysRevResearch.4.L012013} {\bibfield  {journal} {\bibinfo  {journal} {Phys. Rev. Res.}\ }\textbf {\bibinfo {volume} {4}},\ \bibinfo {pages} {L012013} (\bibinfo {year} {2022})}\BibitemShut {NoStop}%
\bibitem [{\citenamefont {Devakul}\ \emph {et~al.}(2023)\citenamefont {Devakul}, \citenamefont {Ledwith}, \citenamefont {Xia}, \citenamefont {Uri}, \citenamefont {de~la Barrera}, \citenamefont {Jarillo-Herrero},\ and\ \citenamefont {Fu}}]{doi:10.1126/sciadv.adi6063}%
  \BibitemOpen
  \bibfield  {author} {\bibinfo {author} {\bibfnamefont {T.}~\bibnamefont {Devakul}}, \bibinfo {author} {\bibfnamefont {P.~J.}\ \bibnamefont {Ledwith}}, \bibinfo {author} {\bibfnamefont {L.-Q.}\ \bibnamefont {Xia}}, \bibinfo {author} {\bibfnamefont {A.}~\bibnamefont {Uri}}, \bibinfo {author} {\bibfnamefont {S.~C.}\ \bibnamefont {de~la Barrera}}, \bibinfo {author} {\bibfnamefont {P.}~\bibnamefont {Jarillo-Herrero}}, \ and\ \bibinfo {author} {\bibfnamefont {L.}~\bibnamefont {Fu}},\ }\href {\doibase 10.1126/sciadv.adi6063} {\bibfield  {journal} {\bibinfo  {journal} {Science Advances}\ }\textbf {\bibinfo {volume} {9}},\ \bibinfo {pages} {eadi6063} (\bibinfo {year} {2023})},\ \Eprint {http://arxiv.org/abs/https://www.science.org/doi/pdf/10.1126/sciadv.adi6063} {https://www.science.org/doi/pdf/10.1126/sciadv.adi6063} \BibitemShut {NoStop}%
\bibitem [{\citenamefont {Christos}\ \emph {et~al.}(2022)\citenamefont {Christos}, \citenamefont {Sachdev},\ and\ \citenamefont {Scheurer}}]{PhysRevX.12.021018}%
  \BibitemOpen
  \bibfield  {author} {\bibinfo {author} {\bibfnamefont {M.}~\bibnamefont {Christos}}, \bibinfo {author} {\bibfnamefont {S.}~\bibnamefont {Sachdev}}, \ and\ \bibinfo {author} {\bibfnamefont {M.~S.}\ \bibnamefont {Scheurer}},\ }\href {\doibase 10.1103/PhysRevX.12.021018} {\bibfield  {journal} {\bibinfo  {journal} {Phys. Rev. X}\ }\textbf {\bibinfo {volume} {12}},\ \bibinfo {pages} {021018} (\bibinfo {year} {2022})}\BibitemShut {NoStop}%
\bibitem [{\citenamefont {Zhang}\ \emph {et~al.}(2022)\citenamefont {Zhang}, \citenamefont {Polski}, \citenamefont {Lewandowski}, \citenamefont {Thomson}, \citenamefont {Peng}, \citenamefont {Choi}, \citenamefont {Kim}, \citenamefont {Watanabe}, \citenamefont {Taniguchi}, \citenamefont {Alicea}, \citenamefont {von Oppen}, \citenamefont {Refael},\ and\ \citenamefont {Nadj-Perge}}]{doi:10.1126/science.abn8585}%
  \BibitemOpen
  \bibfield  {author} {\bibinfo {author} {\bibfnamefont {Y.}~\bibnamefont {Zhang}}, \bibinfo {author} {\bibfnamefont {R.}~\bibnamefont {Polski}}, \bibinfo {author} {\bibfnamefont {C.}~\bibnamefont {Lewandowski}}, \bibinfo {author} {\bibfnamefont {A.}~\bibnamefont {Thomson}}, \bibinfo {author} {\bibfnamefont {Y.}~\bibnamefont {Peng}}, \bibinfo {author} {\bibfnamefont {Y.}~\bibnamefont {Choi}}, \bibinfo {author} {\bibfnamefont {H.}~\bibnamefont {Kim}}, \bibinfo {author} {\bibfnamefont {K.}~\bibnamefont {Watanabe}}, \bibinfo {author} {\bibfnamefont {T.}~\bibnamefont {Taniguchi}}, \bibinfo {author} {\bibfnamefont {J.}~\bibnamefont {Alicea}}, \bibinfo {author} {\bibfnamefont {F.}~\bibnamefont {von Oppen}}, \bibinfo {author} {\bibfnamefont {G.}~\bibnamefont {Refael}}, \ and\ \bibinfo {author} {\bibfnamefont {S.}~\bibnamefont {Nadj-Perge}},\ }\href {\doibase 10.1126/science.abn8585} {\bibfield  {journal} {\bibinfo  {journal} {Science}\ }\textbf {\bibinfo {volume} {377}},\ \bibinfo {pages} {1538} (\bibinfo {year}
  {2022})},\ \Eprint {http://arxiv.org/abs/https://www.science.org/doi/pdf/10.1126/science.abn8585} {https://www.science.org/doi/pdf/10.1126/science.abn8585} \BibitemShut {NoStop}%
\bibitem [{\citenamefont {Hao}\ \emph {et~al.}(2021)\citenamefont {Hao}, \citenamefont {Zimmerman}, \citenamefont {Ledwith}, \citenamefont {Khalaf}, \citenamefont {Najafabadi}, \citenamefont {Watanabe}, \citenamefont {Taniguchi}, \citenamefont {Vishwanath},\ and\ \citenamefont {Kim}}]{doi:10.1126/science.abg0399}%
  \BibitemOpen
  \bibfield  {author} {\bibinfo {author} {\bibfnamefont {Z.}~\bibnamefont {Hao}}, \bibinfo {author} {\bibfnamefont {A.~M.}\ \bibnamefont {Zimmerman}}, \bibinfo {author} {\bibfnamefont {P.}~\bibnamefont {Ledwith}}, \bibinfo {author} {\bibfnamefont {E.}~\bibnamefont {Khalaf}}, \bibinfo {author} {\bibfnamefont {D.~H.}\ \bibnamefont {Najafabadi}}, \bibinfo {author} {\bibfnamefont {K.}~\bibnamefont {Watanabe}}, \bibinfo {author} {\bibfnamefont {T.}~\bibnamefont {Taniguchi}}, \bibinfo {author} {\bibfnamefont {A.}~\bibnamefont {Vishwanath}}, \ and\ \bibinfo {author} {\bibfnamefont {P.}~\bibnamefont {Kim}},\ }\href {\doibase 10.1126/science.abg0399} {\bibfield  {journal} {\bibinfo  {journal} {Science}\ }\textbf {\bibinfo {volume} {371}},\ \bibinfo {pages} {1133} (\bibinfo {year} {2021})},\ \Eprint {http://arxiv.org/abs/https://www.science.org/doi/pdf/10.1126/science.abg0399} {https://www.science.org/doi/pdf/10.1126/science.abg0399} \BibitemShut {NoStop}%
\bibitem [{\citenamefont {Park}\ \emph {et~al.}(2021)\citenamefont {Park}, \citenamefont {Cao}, \citenamefont {Watanabe}, \citenamefont {Taniguchi},\ and\ \citenamefont {Jarillo-Herrero}}]{park2021tunable}%
  \BibitemOpen
  \bibfield  {author} {\bibinfo {author} {\bibfnamefont {J.~M.}\ \bibnamefont {Park}}, \bibinfo {author} {\bibfnamefont {Y.}~\bibnamefont {Cao}}, \bibinfo {author} {\bibfnamefont {K.}~\bibnamefont {Watanabe}}, \bibinfo {author} {\bibfnamefont {T.}~\bibnamefont {Taniguchi}}, \ and\ \bibinfo {author} {\bibfnamefont {P.}~\bibnamefont {Jarillo-Herrero}},\ }\href@noop {} {\bibfield  {journal} {\bibinfo  {journal} {Nature}\ }\textbf {\bibinfo {volume} {590}},\ \bibinfo {pages} {249} (\bibinfo {year} {2021})}\BibitemShut {NoStop}%
\bibitem [{\citenamefont {Cao}\ \emph {et~al.}(2021)\citenamefont {Cao}, \citenamefont {Park}, \citenamefont {Watanabe}, \citenamefont {Taniguchi},\ and\ \citenamefont {Jarillo-Herrero}}]{Cao2021}%
  \BibitemOpen
  \bibfield  {author} {\bibinfo {author} {\bibfnamefont {Y.}~\bibnamefont {Cao}}, \bibinfo {author} {\bibfnamefont {J.~M.}\ \bibnamefont {Park}}, \bibinfo {author} {\bibfnamefont {K.}~\bibnamefont {Watanabe}}, \bibinfo {author} {\bibfnamefont {T.}~\bibnamefont {Taniguchi}}, \ and\ \bibinfo {author} {\bibfnamefont {P.}~\bibnamefont {Jarillo-Herrero}},\ }\href {\doibase 10.1038/s41586-021-03685-y} {\bibfield  {journal} {\bibinfo  {journal} {Nature}\ }\textbf {\bibinfo {volume} {595}},\ \bibinfo {pages} {526} (\bibinfo {year} {2021})}\BibitemShut {NoStop}%
\bibitem [{\citenamefont {Kim}\ \emph {et~al.}(2022)\citenamefont {Kim}, \citenamefont {Choi}, \citenamefont {Lewandowski}, \citenamefont {Thomson}, \citenamefont {Zhang}, \citenamefont {Polski}, \citenamefont {Watanabe}, \citenamefont {Taniguchi}, \citenamefont {Alicea},\ and\ \citenamefont {Nadj-Perge}}]{Kim2022}%
  \BibitemOpen
  \bibfield  {author} {\bibinfo {author} {\bibfnamefont {H.}~\bibnamefont {Kim}}, \bibinfo {author} {\bibfnamefont {Y.}~\bibnamefont {Choi}}, \bibinfo {author} {\bibfnamefont {C.}~\bibnamefont {Lewandowski}}, \bibinfo {author} {\bibfnamefont {A.}~\bibnamefont {Thomson}}, \bibinfo {author} {\bibfnamefont {Y.}~\bibnamefont {Zhang}}, \bibinfo {author} {\bibfnamefont {R.}~\bibnamefont {Polski}}, \bibinfo {author} {\bibfnamefont {K.}~\bibnamefont {Watanabe}}, \bibinfo {author} {\bibfnamefont {T.}~\bibnamefont {Taniguchi}}, \bibinfo {author} {\bibfnamefont {J.}~\bibnamefont {Alicea}}, \ and\ \bibinfo {author} {\bibfnamefont {S.}~\bibnamefont {Nadj-Perge}},\ }\href {\doibase 10.1038/s41586-022-04715-z} {\bibfield  {journal} {\bibinfo  {journal} {Nature}\ }\textbf {\bibinfo {volume} {606}},\ \bibinfo {pages} {494} (\bibinfo {year} {2022})}\BibitemShut {NoStop}%
\bibitem [{\citenamefont {Zhu}\ \emph {et~al.}(2020{\natexlab{a}})\citenamefont {Zhu}, \citenamefont {Cazeaux}, \citenamefont {Luskin},\ and\ \citenamefont {Kaxiras}}]{PhysRevB.101.224107}%
  \BibitemOpen
  \bibfield  {author} {\bibinfo {author} {\bibfnamefont {Z.}~\bibnamefont {Zhu}}, \bibinfo {author} {\bibfnamefont {P.}~\bibnamefont {Cazeaux}}, \bibinfo {author} {\bibfnamefont {M.}~\bibnamefont {Luskin}}, \ and\ \bibinfo {author} {\bibfnamefont {E.}~\bibnamefont {Kaxiras}},\ }\href {\doibase 10.1103/PhysRevB.101.224107} {\bibfield  {journal} {\bibinfo  {journal} {Phys. Rev. B}\ }\textbf {\bibinfo {volume} {101}},\ \bibinfo {pages} {224107} (\bibinfo {year} {2020}{\natexlab{a}})}\BibitemShut {NoStop}%
\bibitem [{\citenamefont {Zhu}\ \emph {et~al.}(2020{\natexlab{b}})\citenamefont {Zhu}, \citenamefont {Carr}, \citenamefont {Massatt}, \citenamefont {Luskin},\ and\ \citenamefont {Kaxiras}}]{PhysRevLett.125.116404}%
  \BibitemOpen
  \bibfield  {author} {\bibinfo {author} {\bibfnamefont {Z.}~\bibnamefont {Zhu}}, \bibinfo {author} {\bibfnamefont {S.}~\bibnamefont {Carr}}, \bibinfo {author} {\bibfnamefont {D.}~\bibnamefont {Massatt}}, \bibinfo {author} {\bibfnamefont {M.}~\bibnamefont {Luskin}}, \ and\ \bibinfo {author} {\bibfnamefont {E.}~\bibnamefont {Kaxiras}},\ }\href {\doibase 10.1103/PhysRevLett.125.116404} {\bibfield  {journal} {\bibinfo  {journal} {Phys. Rev. Lett.}\ }\textbf {\bibinfo {volume} {125}},\ \bibinfo {pages} {116404} (\bibinfo {year} {2020}{\natexlab{b}})}\BibitemShut {NoStop}%
\bibitem [{\citenamefont {Lin}\ \emph {et~al.}(2020)\citenamefont {Lin}, \citenamefont {Qiao}, \citenamefont {Huang}, \citenamefont {Liu}, \citenamefont {Fu}, \citenamefont {Mayorov}, \citenamefont {Chen}, \citenamefont {Mukherjee}, \citenamefont {Qu}, \citenamefont {Sow} \emph {et~al.}}]{lin2020heteromoire}%
  \BibitemOpen
  \bibfield  {author} {\bibinfo {author} {\bibfnamefont {F.}~\bibnamefont {Lin}}, \bibinfo {author} {\bibfnamefont {J.}~\bibnamefont {Qiao}}, \bibinfo {author} {\bibfnamefont {J.}~\bibnamefont {Huang}}, \bibinfo {author} {\bibfnamefont {J.}~\bibnamefont {Liu}}, \bibinfo {author} {\bibfnamefont {D.}~\bibnamefont {Fu}}, \bibinfo {author} {\bibfnamefont {A.~S.}\ \bibnamefont {Mayorov}}, \bibinfo {author} {\bibfnamefont {H.}~\bibnamefont {Chen}}, \bibinfo {author} {\bibfnamefont {P.}~\bibnamefont {Mukherjee}}, \bibinfo {author} {\bibfnamefont {T.}~\bibnamefont {Qu}}, \bibinfo {author} {\bibfnamefont {C.-H.}\ \bibnamefont {Sow}},  \emph {et~al.},\ }\href@noop {} {\bibfield  {journal} {\bibinfo  {journal} {Nano Letters}\ }\textbf {\bibinfo {volume} {20}},\ \bibinfo {pages} {7572} (\bibinfo {year} {2020})}\BibitemShut {NoStop}%
\bibitem [{\citenamefont {Zhang}\ \emph {et~al.}(2021)\citenamefont {Zhang}, \citenamefont {Tsai}, \citenamefont {Zhu}, \citenamefont {Ren}, \citenamefont {Luo}, \citenamefont {Carr}, \citenamefont {Luskin}, \citenamefont {Kaxiras},\ and\ \citenamefont {Wang}}]{PhysRevLett.127.166802}%
  \BibitemOpen
  \bibfield  {author} {\bibinfo {author} {\bibfnamefont {X.}~\bibnamefont {Zhang}}, \bibinfo {author} {\bibfnamefont {K.-T.}\ \bibnamefont {Tsai}}, \bibinfo {author} {\bibfnamefont {Z.}~\bibnamefont {Zhu}}, \bibinfo {author} {\bibfnamefont {W.}~\bibnamefont {Ren}}, \bibinfo {author} {\bibfnamefont {Y.}~\bibnamefont {Luo}}, \bibinfo {author} {\bibfnamefont {S.}~\bibnamefont {Carr}}, \bibinfo {author} {\bibfnamefont {M.}~\bibnamefont {Luskin}}, \bibinfo {author} {\bibfnamefont {E.}~\bibnamefont {Kaxiras}}, \ and\ \bibinfo {author} {\bibfnamefont {K.}~\bibnamefont {Wang}},\ }\href {\doibase 10.1103/PhysRevLett.127.166802} {\bibfield  {journal} {\bibinfo  {journal} {Phys. Rev. Lett.}\ }\textbf {\bibinfo {volume} {127}},\ \bibinfo {pages} {166802} (\bibinfo {year} {2021})}\BibitemShut {NoStop}%
\bibitem [{\citenamefont {Gao}\ and\ \citenamefont {Khalaf}(2022)}]{gao2022symmetry}%
  \BibitemOpen
  \bibfield  {author} {\bibinfo {author} {\bibfnamefont {Q.}~\bibnamefont {Gao}}\ and\ \bibinfo {author} {\bibfnamefont {E.}~\bibnamefont {Khalaf}},\ }\href@noop {} {\bibfield  {journal} {\bibinfo  {journal} {Physical Review B}\ }\textbf {\bibinfo {volume} {106}},\ \bibinfo {pages} {075420} (\bibinfo {year} {2022})}\BibitemShut {NoStop}%
\bibitem [{\citenamefont {Ma}\ \emph {et~al.}(2022)\citenamefont {Ma}, \citenamefont {Li}, \citenamefont {Lu}, \citenamefont {Xu}, \citenamefont {Gao},\ and\ \citenamefont {Xie}}]{Ma2022}%
  \BibitemOpen
  \bibfield  {author} {\bibinfo {author} {\bibfnamefont {Z.}~\bibnamefont {Ma}}, \bibinfo {author} {\bibfnamefont {S.}~\bibnamefont {Li}}, \bibinfo {author} {\bibfnamefont {M.}~\bibnamefont {Lu}}, \bibinfo {author} {\bibfnamefont {D.-H.}\ \bibnamefont {Xu}}, \bibinfo {author} {\bibfnamefont {J.-H.}\ \bibnamefont {Gao}}, \ and\ \bibinfo {author} {\bibfnamefont {X.}~\bibnamefont {Xie}},\ }\href {\doibase 10.1007/s11433-022-1993-7} {\bibfield  {journal} {\bibinfo  {journal} {Science China Physics, Mechanics {\&} Astronomy}\ }\textbf {\bibinfo {volume} {66}},\ \bibinfo {pages} {227211} (\bibinfo {year} {2022})}\BibitemShut {NoStop}%
\bibitem [{\citenamefont {Liang}\ \emph {et~al.}(2022)\citenamefont {Liang}, \citenamefont {Xiao}, \citenamefont {Ma},\ and\ \citenamefont {Gao}}]{PhysRevB.105.195422}%
  \BibitemOpen
  \bibfield  {author} {\bibinfo {author} {\bibfnamefont {M.}~\bibnamefont {Liang}}, \bibinfo {author} {\bibfnamefont {M.-M.}\ \bibnamefont {Xiao}}, \bibinfo {author} {\bibfnamefont {Z.}~\bibnamefont {Ma}}, \ and\ \bibinfo {author} {\bibfnamefont {J.-H.}\ \bibnamefont {Gao}},\ }\href {\doibase 10.1103/PhysRevB.105.195422} {\bibfield  {journal} {\bibinfo  {journal} {Phys. Rev. B}\ }\textbf {\bibinfo {volume} {105}},\ \bibinfo {pages} {195422} (\bibinfo {year} {2022})}\BibitemShut {NoStop}%
\bibitem [{\citenamefont {Uri}\ \emph {et~al.}(2023)\citenamefont {Uri}, \citenamefont {de~la Barrera}, \citenamefont {Randeria}, \citenamefont {Rodan-Legrain}, \citenamefont {Devakul}, \citenamefont {Crowley}, \citenamefont {Paul}, \citenamefont {Watanabe}, \citenamefont {Taniguchi}, \citenamefont {Lifshitz} \emph {et~al.}}]{uri2023superconductivity}%
  \BibitemOpen
  \bibfield  {author} {\bibinfo {author} {\bibfnamefont {A.}~\bibnamefont {Uri}}, \bibinfo {author} {\bibfnamefont {S.~C.}\ \bibnamefont {de~la Barrera}}, \bibinfo {author} {\bibfnamefont {M.~T.}\ \bibnamefont {Randeria}}, \bibinfo {author} {\bibfnamefont {D.}~\bibnamefont {Rodan-Legrain}}, \bibinfo {author} {\bibfnamefont {T.}~\bibnamefont {Devakul}}, \bibinfo {author} {\bibfnamefont {P.~J.}\ \bibnamefont {Crowley}}, \bibinfo {author} {\bibfnamefont {N.}~\bibnamefont {Paul}}, \bibinfo {author} {\bibfnamefont {K.}~\bibnamefont {Watanabe}}, \bibinfo {author} {\bibfnamefont {T.}~\bibnamefont {Taniguchi}}, \bibinfo {author} {\bibfnamefont {R.}~\bibnamefont {Lifshitz}},  \emph {et~al.},\ }\href@noop {} {\bibfield  {journal} {\bibinfo  {journal} {Nature}\ }\textbf {\bibinfo {volume} {620}},\ \bibinfo {pages} {762} (\bibinfo {year} {2023})}\BibitemShut {NoStop}%
\bibitem [{\citenamefont {Mao}\ \emph {et~al.}(2023)\citenamefont {Mao}, \citenamefont {Guerci},\ and\ \citenamefont {Mora}}]{PhysRevB.107.125423}%
  \BibitemOpen
  \bibfield  {author} {\bibinfo {author} {\bibfnamefont {Y.}~\bibnamefont {Mao}}, \bibinfo {author} {\bibfnamefont {D.}~\bibnamefont {Guerci}}, \ and\ \bibinfo {author} {\bibfnamefont {C.}~\bibnamefont {Mora}},\ }\href {\doibase 10.1103/PhysRevB.107.125423} {\bibfield  {journal} {\bibinfo  {journal} {Phys. Rev. B}\ }\textbf {\bibinfo {volume} {107}},\ \bibinfo {pages} {125423} (\bibinfo {year} {2023})}\BibitemShut {NoStop}%
\bibitem [{\citenamefont {Popov}\ and\ \citenamefont {Tarnopolsky}(2023)}]{PhysRevB.108.L081124}%
  \BibitemOpen
  \bibfield  {author} {\bibinfo {author} {\bibfnamefont {F.~K.}\ \bibnamefont {Popov}}\ and\ \bibinfo {author} {\bibfnamefont {G.}~\bibnamefont {Tarnopolsky}},\ }\href {\doibase 10.1103/PhysRevB.108.L081124} {\bibfield  {journal} {\bibinfo  {journal} {Phys. Rev. B}\ }\textbf {\bibinfo {volume} {108}},\ \bibinfo {pages} {L081124} (\bibinfo {year} {2023})}\BibitemShut {NoStop}%
\bibitem [{\citenamefont {Lin}\ \emph {et~al.}(2022)\citenamefont {Lin}, \citenamefont {Li}, \citenamefont {Su},\ and\ \citenamefont {Ni}}]{PhysRevB.106.075423}%
  \BibitemOpen
  \bibfield  {author} {\bibinfo {author} {\bibfnamefont {X.}~\bibnamefont {Lin}}, \bibinfo {author} {\bibfnamefont {C.}~\bibnamefont {Li}}, \bibinfo {author} {\bibfnamefont {K.}~\bibnamefont {Su}}, \ and\ \bibinfo {author} {\bibfnamefont {J.}~\bibnamefont {Ni}},\ }\href {\doibase 10.1103/PhysRevB.106.075423} {\bibfield  {journal} {\bibinfo  {journal} {Phys. Rev. B}\ }\textbf {\bibinfo {volume} {106}},\ \bibinfo {pages} {075423} (\bibinfo {year} {2022})}\BibitemShut {NoStop}%
\bibitem [{\citenamefont {Meng}\ \emph {et~al.}(2023)\citenamefont {Meng}, \citenamefont {Zhan},\ and\ \citenamefont {Yuan}}]{meng2023commensurate}%
  \BibitemOpen
  \bibfield  {author} {\bibinfo {author} {\bibfnamefont {H.}~\bibnamefont {Meng}}, \bibinfo {author} {\bibfnamefont {Z.}~\bibnamefont {Zhan}}, \ and\ \bibinfo {author} {\bibfnamefont {S.}~\bibnamefont {Yuan}},\ }\href@noop {} {\bibfield  {journal} {\bibinfo  {journal} {Phys. Rev. B}\ }\textbf {\bibinfo {volume} {107}},\ \bibinfo {pages} {035109} (\bibinfo {year} {2023})}\BibitemShut {NoStop}%
\bibitem [{\citenamefont {Turkel}\ \emph {et~al.}(2022)\citenamefont {Turkel}, \citenamefont {Swann}, \citenamefont {Zhu}, \citenamefont {Christos}, \citenamefont {Watanabe}, \citenamefont {Taniguchi}, \citenamefont {Sachdev}, \citenamefont {Scheurer}, \citenamefont {Kaxiras}, \citenamefont {Dean},\ and\ \citenamefont {Pasupathy}}]{doi:10.1126/science.abk1895}%
  \BibitemOpen
  \bibfield  {author} {\bibinfo {author} {\bibfnamefont {S.}~\bibnamefont {Turkel}}, \bibinfo {author} {\bibfnamefont {J.}~\bibnamefont {Swann}}, \bibinfo {author} {\bibfnamefont {Z.}~\bibnamefont {Zhu}}, \bibinfo {author} {\bibfnamefont {M.}~\bibnamefont {Christos}}, \bibinfo {author} {\bibfnamefont {K.}~\bibnamefont {Watanabe}}, \bibinfo {author} {\bibfnamefont {T.}~\bibnamefont {Taniguchi}}, \bibinfo {author} {\bibfnamefont {S.}~\bibnamefont {Sachdev}}, \bibinfo {author} {\bibfnamefont {M.~S.}\ \bibnamefont {Scheurer}}, \bibinfo {author} {\bibfnamefont {E.}~\bibnamefont {Kaxiras}}, \bibinfo {author} {\bibfnamefont {C.~R.}\ \bibnamefont {Dean}}, \ and\ \bibinfo {author} {\bibfnamefont {A.~N.}\ \bibnamefont {Pasupathy}},\ }\href {\doibase 10.1126/science.abk1895} {\bibfield  {journal} {\bibinfo  {journal} {Science}\ }\textbf {\bibinfo {volume} {376}},\ \bibinfo {pages} {193} (\bibinfo {year} {2022})},\ \Eprint {http://arxiv.org/abs/https://www.science.org/doi/pdf/10.1126/science.abk1895}
  {https://www.science.org/doi/pdf/10.1126/science.abk1895} \BibitemShut {NoStop}%
\bibitem [{\citenamefont {Craig}\ \emph {et~al.}(2024)\citenamefont {Craig}, \citenamefont {Van~Winkle}, \citenamefont {Groschner}, \citenamefont {Zhang}, \citenamefont {Dowlatshahi}, \citenamefont {Zhu}, \citenamefont {Taniguchi}, \citenamefont {Watanabe}, \citenamefont {Griffin},\ and\ \citenamefont {Bediako}}]{craig2024local}%
  \BibitemOpen
  \bibfield  {author} {\bibinfo {author} {\bibfnamefont {I.~M.}\ \bibnamefont {Craig}}, \bibinfo {author} {\bibfnamefont {M.}~\bibnamefont {Van~Winkle}}, \bibinfo {author} {\bibfnamefont {C.}~\bibnamefont {Groschner}}, \bibinfo {author} {\bibfnamefont {K.}~\bibnamefont {Zhang}}, \bibinfo {author} {\bibfnamefont {N.}~\bibnamefont {Dowlatshahi}}, \bibinfo {author} {\bibfnamefont {Z.}~\bibnamefont {Zhu}}, \bibinfo {author} {\bibfnamefont {T.}~\bibnamefont {Taniguchi}}, \bibinfo {author} {\bibfnamefont {K.}~\bibnamefont {Watanabe}}, \bibinfo {author} {\bibfnamefont {S.~M.}\ \bibnamefont {Griffin}}, \ and\ \bibinfo {author} {\bibfnamefont {D.~K.}\ \bibnamefont {Bediako}},\ }\href@noop {} {\bibfield  {journal} {\bibinfo  {journal} {Nature Materials}\ }\textbf {\bibinfo {volume} {23}},\ \bibinfo {pages} {323} (\bibinfo {year} {2024})}\BibitemShut {NoStop}%
\bibitem [{\citenamefont {Li}\ \emph {et~al.}(2022)\citenamefont {Li}, \citenamefont {Xue}, \citenamefont {Fan}, \citenamefont {Gao}, \citenamefont {Shi}, \citenamefont {Liu}, \citenamefont {Watanabe}, \citenamefont {Tanguchi}, \citenamefont {Zhao}, \citenamefont {Wu} \emph {et~al.}}]{li2022symmetry}%
  \BibitemOpen
  \bibfield  {author} {\bibinfo {author} {\bibfnamefont {Y.}~\bibnamefont {Li}}, \bibinfo {author} {\bibfnamefont {M.}~\bibnamefont {Xue}}, \bibinfo {author} {\bibfnamefont {H.}~\bibnamefont {Fan}}, \bibinfo {author} {\bibfnamefont {C.-F.}\ \bibnamefont {Gao}}, \bibinfo {author} {\bibfnamefont {Y.}~\bibnamefont {Shi}}, \bibinfo {author} {\bibfnamefont {Y.}~\bibnamefont {Liu}}, \bibinfo {author} {\bibfnamefont {K.}~\bibnamefont {Watanabe}}, \bibinfo {author} {\bibfnamefont {T.}~\bibnamefont {Tanguchi}}, \bibinfo {author} {\bibfnamefont {Y.}~\bibnamefont {Zhao}}, \bibinfo {author} {\bibfnamefont {F.}~\bibnamefont {Wu}},  \emph {et~al.},\ }\href@noop {} {\bibfield  {journal} {\bibinfo  {journal} {Nano Letters}\ }\textbf {\bibinfo {volume} {22}},\ \bibinfo {pages} {6215} (\bibinfo {year} {2022})}\BibitemShut {NoStop}%
\bibitem [{\citenamefont {Shin}\ \emph {et~al.}(2021)\citenamefont {Shin}, \citenamefont {Park}, \citenamefont {Chittari}, \citenamefont {Sun},\ and\ \citenamefont {Jung}}]{shin2021electron}%
  \BibitemOpen
  \bibfield  {author} {\bibinfo {author} {\bibfnamefont {J.}~\bibnamefont {Shin}}, \bibinfo {author} {\bibfnamefont {Y.}~\bibnamefont {Park}}, \bibinfo {author} {\bibfnamefont {B.~L.}\ \bibnamefont {Chittari}}, \bibinfo {author} {\bibfnamefont {J.-H.}\ \bibnamefont {Sun}}, \ and\ \bibinfo {author} {\bibfnamefont {J.}~\bibnamefont {Jung}},\ }\href@noop {} {\bibfield  {journal} {\bibinfo  {journal} {Physical Review B}\ }\textbf {\bibinfo {volume} {103}},\ \bibinfo {pages} {075423} (\bibinfo {year} {2021})}\BibitemShut {NoStop}%
\bibitem [{\citenamefont {Nakatsuji}\ \emph {et~al.}(2023)\citenamefont {Nakatsuji}, \citenamefont {Kawakami},\ and\ \citenamefont {Koshino}}]{PhysRevX.13.041007}%
  \BibitemOpen
  \bibfield  {author} {\bibinfo {author} {\bibfnamefont {N.}~\bibnamefont {Nakatsuji}}, \bibinfo {author} {\bibfnamefont {T.}~\bibnamefont {Kawakami}}, \ and\ \bibinfo {author} {\bibfnamefont {M.}~\bibnamefont {Koshino}},\ }\href {\doibase 10.1103/PhysRevX.13.041007} {\bibfield  {journal} {\bibinfo  {journal} {Phys. Rev. X}\ }\textbf {\bibinfo {volume} {13}},\ \bibinfo {pages} {041007} (\bibinfo {year} {2023})}\BibitemShut {NoStop}%
\bibitem [{\citenamefont {Park}\ \emph {et~al.}(2024)\citenamefont {Park}, \citenamefont {Park}, \citenamefont {Ko}, \citenamefont {Yananose}, \citenamefont {Engelke}, \citenamefont {Zhang}, \citenamefont {Davydov}, \citenamefont {Green}, \citenamefont {Park}, \citenamefont {Lee} \emph {et~al.}}]{park2024tunable}%
  \BibitemOpen
  \bibfield  {author} {\bibinfo {author} {\bibfnamefont {D.}~\bibnamefont {Park}}, \bibinfo {author} {\bibfnamefont {C.}~\bibnamefont {Park}}, \bibinfo {author} {\bibfnamefont {E.}~\bibnamefont {Ko}}, \bibinfo {author} {\bibfnamefont {K.}~\bibnamefont {Yananose}}, \bibinfo {author} {\bibfnamefont {R.}~\bibnamefont {Engelke}}, \bibinfo {author} {\bibfnamefont {X.}~\bibnamefont {Zhang}}, \bibinfo {author} {\bibfnamefont {K.}~\bibnamefont {Davydov}}, \bibinfo {author} {\bibfnamefont {M.}~\bibnamefont {Green}}, \bibinfo {author} {\bibfnamefont {S.~H.}\ \bibnamefont {Park}}, \bibinfo {author} {\bibfnamefont {J.~H.}\ \bibnamefont {Lee}},  \emph {et~al.},\ }\href@noop {} {\bibfield  {journal} {\bibinfo  {journal} {arXiv preprint arXiv:2402.15760}\ } (\bibinfo {year} {2024})}\BibitemShut {NoStop}%
\bibitem [{\citenamefont {AlBuhairan}\ and\ \citenamefont {Vogl}(2023)}]{PhysRevB.108.155106}%
  \BibitemOpen
  \bibfield  {author} {\bibinfo {author} {\bibfnamefont {H.}~\bibnamefont {AlBuhairan}}\ and\ \bibinfo {author} {\bibfnamefont {M.}~\bibnamefont {Vogl}},\ }\href {\doibase 10.1103/PhysRevB.108.155106} {\bibfield  {journal} {\bibinfo  {journal} {Phys. Rev. B}\ }\textbf {\bibinfo {volume} {108}},\ \bibinfo {pages} {155106} (\bibinfo {year} {2023})}\BibitemShut {NoStop}%
\bibitem [{\citenamefont {Liang}\ \emph {et~al.}(2025)\citenamefont {Liang}, \citenamefont {Ding}, \citenamefont {Wu}, \citenamefont {Zhao},\ and\ \citenamefont {Gao}}]{PhysRevB.111.085412}%
  \BibitemOpen
  \bibfield  {author} {\bibinfo {author} {\bibfnamefont {M.}~\bibnamefont {Liang}}, \bibinfo {author} {\bibfnamefont {S.-P.}\ \bibnamefont {Ding}}, \bibinfo {author} {\bibfnamefont {M.}~\bibnamefont {Wu}}, \bibinfo {author} {\bibfnamefont {C.}~\bibnamefont {Zhao}}, \ and\ \bibinfo {author} {\bibfnamefont {J.-H.}\ \bibnamefont {Gao}},\ }\href {\doibase 10.1103/PhysRevB.111.085412} {\bibfield  {journal} {\bibinfo  {journal} {Phys. Rev. B}\ }\textbf {\bibinfo {volume} {111}},\ \bibinfo {pages} {085412} (\bibinfo {year} {2025})}\BibitemShut {NoStop}%
\bibitem [{\citenamefont {He}\ \emph {et~al.}(2025)\citenamefont {He}, \citenamefont {Gong}, \citenamefont {Tong}, \citenamefont {Zhai}, \citenamefont {Yao},\ and\ \citenamefont {An}}]{PhysRevB.111.125410}%
  \BibitemOpen
  \bibfield  {author} {\bibinfo {author} {\bibfnamefont {H.}~\bibnamefont {He}}, \bibinfo {author} {\bibfnamefont {Z.}~\bibnamefont {Gong}}, \bibinfo {author} {\bibfnamefont {Q.-J.}\ \bibnamefont {Tong}}, \bibinfo {author} {\bibfnamefont {D.}~\bibnamefont {Zhai}}, \bibinfo {author} {\bibfnamefont {W.}~\bibnamefont {Yao}}, \ and\ \bibinfo {author} {\bibfnamefont {X.-T.}\ \bibnamefont {An}},\ }\href {\doibase 10.1103/PhysRevB.111.125410} {\bibfield  {journal} {\bibinfo  {journal} {Phys. Rev. B}\ }\textbf {\bibinfo {volume} {111}},\ \bibinfo {pages} {125410} (\bibinfo {year} {2025})}\BibitemShut {NoStop}%
\bibitem [{\citenamefont {Fedorko}\ \emph {et~al.}(2025)\citenamefont {Fedorko}, \citenamefont {Liu},\ and\ \citenamefont {Bi}}]{fedorko2025engineeringmoirekagomesuperlattices}%
  \BibitemOpen
  \bibfield  {author} {\bibinfo {author} {\bibfnamefont {A.}~\bibnamefont {Fedorko}}, \bibinfo {author} {\bibfnamefont {C.-X.}\ \bibnamefont {Liu}}, \ and\ \bibinfo {author} {\bibfnamefont {Z.}~\bibnamefont {Bi}},\ }\href {https://arxiv.org/abs/2503.13422} {\enquote {\bibinfo {title} {Engineering moir\'e kagome superlattices in twisted transition metal dichalcogenides},}\ } (\bibinfo {year} {2025}),\ \Eprint {http://arxiv.org/abs/2503.13422} {arXiv:2503.13422 [cond-mat.mes-hall]} \BibitemShut {NoStop}%
\bibitem [{\citenamefont {Mounet}\ \emph {et~al.}(2018)\citenamefont {Mounet}, \citenamefont {Gibertini}, \citenamefont {Schwaller}, \citenamefont {Campi}, \citenamefont {Merkys}, \citenamefont {Marrazzo}, \citenamefont {Sohier}, \citenamefont {Castelli}, \citenamefont {Cepellotti}, \citenamefont {Pizzi} \emph {et~al.}}]{mounet2018two}%
  \BibitemOpen
  \bibfield  {author} {\bibinfo {author} {\bibfnamefont {N.}~\bibnamefont {Mounet}}, \bibinfo {author} {\bibfnamefont {M.}~\bibnamefont {Gibertini}}, \bibinfo {author} {\bibfnamefont {P.}~\bibnamefont {Schwaller}}, \bibinfo {author} {\bibfnamefont {D.}~\bibnamefont {Campi}}, \bibinfo {author} {\bibfnamefont {A.}~\bibnamefont {Merkys}}, \bibinfo {author} {\bibfnamefont {A.}~\bibnamefont {Marrazzo}}, \bibinfo {author} {\bibfnamefont {T.}~\bibnamefont {Sohier}}, \bibinfo {author} {\bibfnamefont {I.~E.}\ \bibnamefont {Castelli}}, \bibinfo {author} {\bibfnamefont {A.}~\bibnamefont {Cepellotti}}, \bibinfo {author} {\bibfnamefont {G.}~\bibnamefont {Pizzi}},  \emph {et~al.},\ }\href@noop {} {\bibfield  {journal} {\bibinfo  {journal} {Nature nanotechnology}\ }\textbf {\bibinfo {volume} {13}},\ \bibinfo {pages} {246} (\bibinfo {year} {2018})}\BibitemShut {NoStop}%
\bibitem [{\citenamefont {Moon}\ and\ \citenamefont {Koshino}(2013)}]{moon2013opticalabsorption}%
  \BibitemOpen
  \bibfield  {author} {\bibinfo {author} {\bibfnamefont {P.}~\bibnamefont {Moon}}\ and\ \bibinfo {author} {\bibfnamefont {M.}~\bibnamefont {Koshino}},\ }\href {\doibase 10.1103/PhysRevB.87.205404} {\bibfield  {journal} {\bibinfo  {journal} {Phys. Rev. B}\ }\textbf {\bibinfo {volume} {87}},\ \bibinfo {pages} {205404} (\bibinfo {year} {2013})}\BibitemShut {NoStop}%
\bibitem [{\citenamefont {Nam}\ and\ \citenamefont {Koshino}(2017)}]{nam2017lattice}%
  \BibitemOpen
  \bibfield  {author} {\bibinfo {author} {\bibfnamefont {N.~N.~T.}\ \bibnamefont {Nam}}\ and\ \bibinfo {author} {\bibfnamefont {M.}~\bibnamefont {Koshino}},\ }\href@noop {} {\bibfield  {journal} {\bibinfo  {journal} {Phys. Rev. B}\ }\textbf {\bibinfo {volume} {96}},\ \bibinfo {pages} {075311} (\bibinfo {year} {2017})},\ \bibinfo {note} {errata ibid {\bf 101}, 099901 (2020)}\BibitemShut {NoStop}%
\bibitem [{\citenamefont {Koshino}\ and\ \citenamefont {Nam}(2020)}]{koshino2020effective}%
  \BibitemOpen
  \bibfield  {author} {\bibinfo {author} {\bibfnamefont {M.}~\bibnamefont {Koshino}}\ and\ \bibinfo {author} {\bibfnamefont {N.~N.}\ \bibnamefont {Nam}},\ }\href@noop {} {\bibfield  {journal} {\bibinfo  {journal} {Phys. Rev. B}\ }\textbf {\bibinfo {volume} {101}},\ \bibinfo {pages} {195425} (\bibinfo {year} {2020})}\BibitemShut {NoStop}%
\bibitem [{\citenamefont {Suzuura}\ and\ \citenamefont {Ando}(2002)}]{PhysRevB.65.235412}%
  \BibitemOpen
  \bibfield  {author} {\bibinfo {author} {\bibfnamefont {H.}~\bibnamefont {Suzuura}}\ and\ \bibinfo {author} {\bibfnamefont {T.}~\bibnamefont {Ando}},\ }\href {\doibase 10.1103/PhysRevB.65.235412} {\bibfield  {journal} {\bibinfo  {journal} {Phys. Rev. B}\ }\textbf {\bibinfo {volume} {65}},\ \bibinfo {pages} {235412} (\bibinfo {year} {2002})}\BibitemShut {NoStop}%
\bibitem [{\citenamefont {San-Jose}\ \emph {et~al.}(2014)\citenamefont {San-Jose}, \citenamefont {Guti\'errez-Rubio}, \citenamefont {Sturla},\ and\ \citenamefont {Guinea}}]{PhysRevB.90.115152}%
  \BibitemOpen
  \bibfield  {author} {\bibinfo {author} {\bibfnamefont {P.}~\bibnamefont {San-Jose}}, \bibinfo {author} {\bibfnamefont {A.}~\bibnamefont {Guti\'errez-Rubio}}, \bibinfo {author} {\bibfnamefont {M.}~\bibnamefont {Sturla}}, \ and\ \bibinfo {author} {\bibfnamefont {F.}~\bibnamefont {Guinea}},\ }\href {\doibase 10.1103/PhysRevB.90.115152} {\bibfield  {journal} {\bibinfo  {journal} {Phys. Rev. B}\ }\textbf {\bibinfo {volume} {90}},\ \bibinfo {pages} {115152} (\bibinfo {year} {2014})}\BibitemShut {NoStop}%
\bibitem [{\citenamefont {Yananose}\ \emph {et~al.}(2025)\citenamefont {Yananose}, \citenamefont {Park},\ and\ \citenamefont {Son}}]{yananose2025metamorphicquantumdotarrays}%
  \BibitemOpen
  \bibfield  {author} {\bibinfo {author} {\bibfnamefont {K.}~\bibnamefont {Yananose}}, \bibinfo {author} {\bibfnamefont {C.}~\bibnamefont {Park}}, \ and\ \bibinfo {author} {\bibfnamefont {Y.-W.}\ \bibnamefont {Son}},\ }\href {https://arxiv.org/abs/2504.14925} {\enquote {\bibinfo {title} {Metamorphic quantum dot arrays in twisted trilayer hexagonal boron nitride},}\ } (\bibinfo {year} {2025}),\ \Eprint {http://arxiv.org/abs/2504.14925} {arXiv:2504.14925 [cond-mat.mes-hall]} \BibitemShut {NoStop}%
\bibitem [{\citenamefont {Fang}\ \emph {et~al.}(2018)\citenamefont {Fang}, \citenamefont {Carr}, \citenamefont {Cazalilla},\ and\ \citenamefont {Kaxiras}}]{PhysRevB.98.075106}%
  \BibitemOpen
  \bibfield  {author} {\bibinfo {author} {\bibfnamefont {S.}~\bibnamefont {Fang}}, \bibinfo {author} {\bibfnamefont {S.}~\bibnamefont {Carr}}, \bibinfo {author} {\bibfnamefont {M.~A.}\ \bibnamefont {Cazalilla}}, \ and\ \bibinfo {author} {\bibfnamefont {E.}~\bibnamefont {Kaxiras}},\ }\href {\doibase 10.1103/PhysRevB.98.075106} {\bibfield  {journal} {\bibinfo  {journal} {Phys. Rev. B}\ }\textbf {\bibinfo {volume} {98}},\ \bibinfo {pages} {075106} (\bibinfo {year} {2018})}\BibitemShut {NoStop}%
\bibitem [{\citenamefont {Fallahazad}\ \emph {et~al.}(2016)\citenamefont {Fallahazad}, \citenamefont {Movva}, \citenamefont {Kim}, \citenamefont {Larentis}, \citenamefont {Taniguchi}, \citenamefont {Watanabe}, \citenamefont {Banerjee},\ and\ \citenamefont {Tutuc}}]{PhysRevLett.116.086601}%
  \BibitemOpen
  \bibfield  {author} {\bibinfo {author} {\bibfnamefont {B.}~\bibnamefont {Fallahazad}}, \bibinfo {author} {\bibfnamefont {H.~C.~P.}\ \bibnamefont {Movva}}, \bibinfo {author} {\bibfnamefont {K.}~\bibnamefont {Kim}}, \bibinfo {author} {\bibfnamefont {S.}~\bibnamefont {Larentis}}, \bibinfo {author} {\bibfnamefont {T.}~\bibnamefont {Taniguchi}}, \bibinfo {author} {\bibfnamefont {K.}~\bibnamefont {Watanabe}}, \bibinfo {author} {\bibfnamefont {S.~K.}\ \bibnamefont {Banerjee}}, \ and\ \bibinfo {author} {\bibfnamefont {E.}~\bibnamefont {Tutuc}},\ }\href {\doibase 10.1103/PhysRevLett.116.086601} {\bibfield  {journal} {\bibinfo  {journal} {Phys. Rev. Lett.}\ }\textbf {\bibinfo {volume} {116}},\ \bibinfo {pages} {086601} (\bibinfo {year} {2016})}\BibitemShut {NoStop}%
\bibitem [{\citenamefont {Rasmussen}\ and\ \citenamefont {Thygesen}(2015)}]{rasmussen2015computational}%
  \BibitemOpen
  \bibfield  {author} {\bibinfo {author} {\bibfnamefont {F.~A.}\ \bibnamefont {Rasmussen}}\ and\ \bibinfo {author} {\bibfnamefont {K.~S.}\ \bibnamefont {Thygesen}},\ }\href@noop {} {\bibfield  {journal} {\bibinfo  {journal} {The Journal of Physical Chemistry C}\ }\textbf {\bibinfo {volume} {119}},\ \bibinfo {pages} {13169} (\bibinfo {year} {2015})}\BibitemShut {NoStop}%
\bibitem [{\citenamefont {Magorrian}\ \emph {et~al.}(2021)\citenamefont {Magorrian}, \citenamefont {Enaldiev}, \citenamefont {Z\'olyomi}, \citenamefont {Ferreira}, \citenamefont {Fal'ko},\ and\ \citenamefont {Ruiz-Tijerina}}]{PhysRevB.104.125440}%
  \BibitemOpen
  \bibfield  {author} {\bibinfo {author} {\bibfnamefont {S.~J.}\ \bibnamefont {Magorrian}}, \bibinfo {author} {\bibfnamefont {V.~V.}\ \bibnamefont {Enaldiev}}, \bibinfo {author} {\bibfnamefont {V.}~\bibnamefont {Z\'olyomi}}, \bibinfo {author} {\bibfnamefont {F.}~\bibnamefont {Ferreira}}, \bibinfo {author} {\bibfnamefont {V.~I.}\ \bibnamefont {Fal'ko}}, \ and\ \bibinfo {author} {\bibfnamefont {D.~A.}\ \bibnamefont {Ruiz-Tijerina}},\ }\href {\doibase 10.1103/PhysRevB.104.125440} {\bibfield  {journal} {\bibinfo  {journal} {Phys. Rev. B}\ }\textbf {\bibinfo {volume} {104}},\ \bibinfo {pages} {125440} (\bibinfo {year} {2021})}\BibitemShut {NoStop}%
\bibitem [{\citenamefont {Towle}\ \emph {et~al.}(1966)\citenamefont {Towle}, \citenamefont {Oberbeck}, \citenamefont {Brown†},\ and\ \citenamefont {Stajdohar}}]{doi:10.1126/science.154.3751.895}%
  \BibitemOpen
  \bibfield  {author} {\bibinfo {author} {\bibfnamefont {L.~C.}\ \bibnamefont {Towle}}, \bibinfo {author} {\bibfnamefont {V.}~\bibnamefont {Oberbeck}}, \bibinfo {author} {\bibfnamefont {B.~E.}\ \bibnamefont {Brown†}}, \ and\ \bibinfo {author} {\bibfnamefont {R.~E.}\ \bibnamefont {Stajdohar}},\ }\href {\doibase 10.1126/science.154.3751.895} {\bibfield  {journal} {\bibinfo  {journal} {Science}\ }\textbf {\bibinfo {volume} {154}},\ \bibinfo {pages} {895} (\bibinfo {year} {1966})},\ \Eprint {http://arxiv.org/abs/https://www.science.org/doi/pdf/10.1126/science.154.3751.895} {https://www.science.org/doi/pdf/10.1126/science.154.3751.895} \BibitemShut {NoStop}%
\bibitem [{\citenamefont {Suzuki}\ \emph {et~al.}(2014)\citenamefont {Suzuki}, \citenamefont {Sakano}, \citenamefont {Zhang}, \citenamefont {Akashi}, \citenamefont {Morikawa}, \citenamefont {Harasawa}, \citenamefont {Yaji}, \citenamefont {Kuroda}, \citenamefont {Miyamoto}, \citenamefont {Okuda} \emph {et~al.}}]{suzuki2014valley}%
  \BibitemOpen
  \bibfield  {author} {\bibinfo {author} {\bibfnamefont {R.}~\bibnamefont {Suzuki}}, \bibinfo {author} {\bibfnamefont {M.}~\bibnamefont {Sakano}}, \bibinfo {author} {\bibfnamefont {Y.}~\bibnamefont {Zhang}}, \bibinfo {author} {\bibfnamefont {R.}~\bibnamefont {Akashi}}, \bibinfo {author} {\bibfnamefont {D.}~\bibnamefont {Morikawa}}, \bibinfo {author} {\bibfnamefont {A.}~\bibnamefont {Harasawa}}, \bibinfo {author} {\bibfnamefont {K.}~\bibnamefont {Yaji}}, \bibinfo {author} {\bibfnamefont {K.}~\bibnamefont {Kuroda}}, \bibinfo {author} {\bibfnamefont {K.}~\bibnamefont {Miyamoto}}, \bibinfo {author} {\bibfnamefont {T.}~\bibnamefont {Okuda}},  \emph {et~al.},\ }\href@noop {} {\bibfield  {journal} {\bibinfo  {journal} {Nature nanotechnology}\ }\textbf {\bibinfo {volume} {9}},\ \bibinfo {pages} {611} (\bibinfo {year} {2014})}\BibitemShut {NoStop}%
\bibitem [{\citenamefont {Jiang}\ \emph {et~al.}(2014)\citenamefont {Jiang}, \citenamefont {Liu}, \citenamefont {Huang}, \citenamefont {Zhang}, \citenamefont {Li}, \citenamefont {Gong}, \citenamefont {Shen}, \citenamefont {Liu},\ and\ \citenamefont {Wu}}]{jiang2014valley}%
  \BibitemOpen
  \bibfield  {author} {\bibinfo {author} {\bibfnamefont {T.}~\bibnamefont {Jiang}}, \bibinfo {author} {\bibfnamefont {H.}~\bibnamefont {Liu}}, \bibinfo {author} {\bibfnamefont {D.}~\bibnamefont {Huang}}, \bibinfo {author} {\bibfnamefont {S.}~\bibnamefont {Zhang}}, \bibinfo {author} {\bibfnamefont {Y.}~\bibnamefont {Li}}, \bibinfo {author} {\bibfnamefont {X.}~\bibnamefont {Gong}}, \bibinfo {author} {\bibfnamefont {Y.-R.}\ \bibnamefont {Shen}}, \bibinfo {author} {\bibfnamefont {W.-T.}\ \bibnamefont {Liu}}, \ and\ \bibinfo {author} {\bibfnamefont {S.}~\bibnamefont {Wu}},\ }\href@noop {} {\bibfield  {journal} {\bibinfo  {journal} {Nature nanotechnology}\ }\textbf {\bibinfo {volume} {9}},\ \bibinfo {pages} {825} (\bibinfo {year} {2014})}\BibitemShut {NoStop}%
\bibitem [{\citenamefont {Haldane}(1988)}]{PhysRevLett.61.2015}%
  \BibitemOpen
  \bibfield  {author} {\bibinfo {author} {\bibfnamefont {F.~D.~M.}\ \bibnamefont {Haldane}},\ }\href {\doibase 10.1103/PhysRevLett.61.2015} {\bibfield  {journal} {\bibinfo  {journal} {Phys. Rev. Lett.}\ }\textbf {\bibinfo {volume} {61}},\ \bibinfo {pages} {2015} (\bibinfo {year} {1988})}\BibitemShut {NoStop}%
\bibitem [{\citenamefont {Sticlet}\ and\ \citenamefont {Pi\'echon}(2013)}]{PhysRevB.87.115402}%
  \BibitemOpen
  \bibfield  {author} {\bibinfo {author} {\bibfnamefont {D.}~\bibnamefont {Sticlet}}\ and\ \bibinfo {author} {\bibfnamefont {F.}~\bibnamefont {Pi\'echon}},\ }\href {\doibase 10.1103/PhysRevB.87.115402} {\bibfield  {journal} {\bibinfo  {journal} {Phys. Rev. B}\ }\textbf {\bibinfo {volume} {87}},\ \bibinfo {pages} {115402} (\bibinfo {year} {2013})}\BibitemShut {NoStop}%
\bibitem [{\citenamefont {Chang}\ \emph {et~al.}(2022)\citenamefont {Chang}, \citenamefont {Hao},\ and\ \citenamefont {Liu}}]{Chang_2022}%
  \BibitemOpen
  \bibfield  {author} {\bibinfo {author} {\bibfnamefont {Z.-W.}\ \bibnamefont {Chang}}, \bibinfo {author} {\bibfnamefont {W.-C.}\ \bibnamefont {Hao}}, \ and\ \bibinfo {author} {\bibfnamefont {X.}~\bibnamefont {Liu}},\ }\href {\doibase 10.1088/1361-648X/ac98fc} {\bibfield  {journal} {\bibinfo  {journal} {Journal of Physics: Condensed Matter}\ }\textbf {\bibinfo {volume} {34}},\ \bibinfo {pages} {485502} (\bibinfo {year} {2022})}\BibitemShut {NoStop}%
\bibitem [{\citenamefont {Chang}\ \emph {et~al.}(2017)\citenamefont {Chang}, \citenamefont {Xu}, \citenamefont {Wieder}, \citenamefont {Sanchez}, \citenamefont {Huang}, \citenamefont {Belopolski}, \citenamefont {Chang}, \citenamefont {Zhang}, \citenamefont {Bansil}, \citenamefont {Lin},\ and\ \citenamefont {Hasan}}]{PhysRevLett.119.206401}%
  \BibitemOpen
  \bibfield  {author} {\bibinfo {author} {\bibfnamefont {G.}~\bibnamefont {Chang}}, \bibinfo {author} {\bibfnamefont {S.-Y.}\ \bibnamefont {Xu}}, \bibinfo {author} {\bibfnamefont {B.~J.}\ \bibnamefont {Wieder}}, \bibinfo {author} {\bibfnamefont {D.~S.}\ \bibnamefont {Sanchez}}, \bibinfo {author} {\bibfnamefont {S.-M.}\ \bibnamefont {Huang}}, \bibinfo {author} {\bibfnamefont {I.}~\bibnamefont {Belopolski}}, \bibinfo {author} {\bibfnamefont {T.-R.}\ \bibnamefont {Chang}}, \bibinfo {author} {\bibfnamefont {S.}~\bibnamefont {Zhang}}, \bibinfo {author} {\bibfnamefont {A.}~\bibnamefont {Bansil}}, \bibinfo {author} {\bibfnamefont {H.}~\bibnamefont {Lin}}, \ and\ \bibinfo {author} {\bibfnamefont {M.~Z.}\ \bibnamefont {Hasan}},\ }\href {\doibase 10.1103/PhysRevLett.119.206401} {\bibfield  {journal} {\bibinfo  {journal} {Phys. Rev. Lett.}\ }\textbf {\bibinfo {volume} {119}},\ \bibinfo {pages} {206401} (\bibinfo {year} {2017})}\BibitemShut {NoStop}%
\bibitem [{\citenamefont {Tang}\ \emph {et~al.}(2017)\citenamefont {Tang}, \citenamefont {Zhou},\ and\ \citenamefont {Zhang}}]{PhysRevLett.119.206402}%
  \BibitemOpen
  \bibfield  {author} {\bibinfo {author} {\bibfnamefont {P.}~\bibnamefont {Tang}}, \bibinfo {author} {\bibfnamefont {Q.}~\bibnamefont {Zhou}}, \ and\ \bibinfo {author} {\bibfnamefont {S.-C.}\ \bibnamefont {Zhang}},\ }\href {\doibase 10.1103/PhysRevLett.119.206402} {\bibfield  {journal} {\bibinfo  {journal} {Phys. Rev. Lett.}\ }\textbf {\bibinfo {volume} {119}},\ \bibinfo {pages} {206402} (\bibinfo {year} {2017})}\BibitemShut {NoStop}%
\bibitem [{\citenamefont {Takane}\ \emph {et~al.}(2019)\citenamefont {Takane}, \citenamefont {Wang}, \citenamefont {Souma}, \citenamefont {Nakayama}, \citenamefont {Nakamura}, \citenamefont {Oinuma}, \citenamefont {Nakata}, \citenamefont {Iwasawa}, \citenamefont {Cacho}, \citenamefont {Kim}, \citenamefont {Horiba}, \citenamefont {Kumigashira}, \citenamefont {Takahashi}, \citenamefont {Ando},\ and\ \citenamefont {Sato}}]{PhysRevLett.122.076402}%
  \BibitemOpen
  \bibfield  {author} {\bibinfo {author} {\bibfnamefont {D.}~\bibnamefont {Takane}}, \bibinfo {author} {\bibfnamefont {Z.}~\bibnamefont {Wang}}, \bibinfo {author} {\bibfnamefont {S.}~\bibnamefont {Souma}}, \bibinfo {author} {\bibfnamefont {K.}~\bibnamefont {Nakayama}}, \bibinfo {author} {\bibfnamefont {T.}~\bibnamefont {Nakamura}}, \bibinfo {author} {\bibfnamefont {H.}~\bibnamefont {Oinuma}}, \bibinfo {author} {\bibfnamefont {Y.}~\bibnamefont {Nakata}}, \bibinfo {author} {\bibfnamefont {H.}~\bibnamefont {Iwasawa}}, \bibinfo {author} {\bibfnamefont {C.}~\bibnamefont {Cacho}}, \bibinfo {author} {\bibfnamefont {T.}~\bibnamefont {Kim}}, \bibinfo {author} {\bibfnamefont {K.}~\bibnamefont {Horiba}}, \bibinfo {author} {\bibfnamefont {H.}~\bibnamefont {Kumigashira}}, \bibinfo {author} {\bibfnamefont {T.}~\bibnamefont {Takahashi}}, \bibinfo {author} {\bibfnamefont {Y.}~\bibnamefont {Ando}}, \ and\ \bibinfo {author} {\bibfnamefont {T.}~\bibnamefont {Sato}},\ }\href {\doibase 10.1103/PhysRevLett.122.076402} {\bibfield
  {journal} {\bibinfo  {journal} {Phys. Rev. Lett.}\ }\textbf {\bibinfo {volume} {122}},\ \bibinfo {pages} {076402} (\bibinfo {year} {2019})}\BibitemShut {NoStop}%
\bibitem [{\citenamefont {Yang}\ \emph {et~al.}(2019)\citenamefont {Yang}, \citenamefont {Sun}, \citenamefont {Xia}, \citenamefont {Xue}, \citenamefont {Gao}, \citenamefont {Ge}, \citenamefont {Jia}, \citenamefont {Yuan}, \citenamefont {Chong},\ and\ \citenamefont {Zhang}}]{yang2019topological}%
  \BibitemOpen
  \bibfield  {author} {\bibinfo {author} {\bibfnamefont {Y.}~\bibnamefont {Yang}}, \bibinfo {author} {\bibfnamefont {H.-x.}\ \bibnamefont {Sun}}, \bibinfo {author} {\bibfnamefont {J.-p.}\ \bibnamefont {Xia}}, \bibinfo {author} {\bibfnamefont {H.}~\bibnamefont {Xue}}, \bibinfo {author} {\bibfnamefont {Z.}~\bibnamefont {Gao}}, \bibinfo {author} {\bibfnamefont {Y.}~\bibnamefont {Ge}}, \bibinfo {author} {\bibfnamefont {D.}~\bibnamefont {Jia}}, \bibinfo {author} {\bibfnamefont {S.-q.}\ \bibnamefont {Yuan}}, \bibinfo {author} {\bibfnamefont {Y.}~\bibnamefont {Chong}}, \ and\ \bibinfo {author} {\bibfnamefont {B.}~\bibnamefont {Zhang}},\ }\href@noop {} {\bibfield  {journal} {\bibinfo  {journal} {Nature Physics}\ }\textbf {\bibinfo {volume} {15}},\ \bibinfo {pages} {645} (\bibinfo {year} {2019})}\BibitemShut {NoStop}%
\bibitem [{\citenamefont {Sanchez}\ \emph {et~al.}(2019)\citenamefont {Sanchez}, \citenamefont {Belopolski}, \citenamefont {Cochran}, \citenamefont {Xu}, \citenamefont {Yin}, \citenamefont {Chang}, \citenamefont {Xie}, \citenamefont {Manna}, \citenamefont {S{\"u}{\ss}}, \citenamefont {Huang} \emph {et~al.}}]{sanchez2019topological}%
  \BibitemOpen
  \bibfield  {author} {\bibinfo {author} {\bibfnamefont {D.~S.}\ \bibnamefont {Sanchez}}, \bibinfo {author} {\bibfnamefont {I.}~\bibnamefont {Belopolski}}, \bibinfo {author} {\bibfnamefont {T.~A.}\ \bibnamefont {Cochran}}, \bibinfo {author} {\bibfnamefont {X.}~\bibnamefont {Xu}}, \bibinfo {author} {\bibfnamefont {J.-X.}\ \bibnamefont {Yin}}, \bibinfo {author} {\bibfnamefont {G.}~\bibnamefont {Chang}}, \bibinfo {author} {\bibfnamefont {W.}~\bibnamefont {Xie}}, \bibinfo {author} {\bibfnamefont {K.}~\bibnamefont {Manna}}, \bibinfo {author} {\bibfnamefont {V.}~\bibnamefont {S{\"u}{\ss}}}, \bibinfo {author} {\bibfnamefont {C.-Y.}\ \bibnamefont {Huang}},  \emph {et~al.},\ }\href@noop {} {\bibfield  {journal} {\bibinfo  {journal} {Nature}\ }\textbf {\bibinfo {volume} {567}},\ \bibinfo {pages} {500} (\bibinfo {year} {2019})}\BibitemShut {NoStop}%
\bibitem [{\citenamefont {Rao}\ \emph {et~al.}(2019)\citenamefont {Rao}, \citenamefont {Li}, \citenamefont {Zhang}, \citenamefont {Tian}, \citenamefont {Li}, \citenamefont {Fu}, \citenamefont {Tang}, \citenamefont {Wang}, \citenamefont {Li}, \citenamefont {Fan} \emph {et~al.}}]{rao2019observation}%
  \BibitemOpen
  \bibfield  {author} {\bibinfo {author} {\bibfnamefont {Z.}~\bibnamefont {Rao}}, \bibinfo {author} {\bibfnamefont {H.}~\bibnamefont {Li}}, \bibinfo {author} {\bibfnamefont {T.}~\bibnamefont {Zhang}}, \bibinfo {author} {\bibfnamefont {S.}~\bibnamefont {Tian}}, \bibinfo {author} {\bibfnamefont {C.}~\bibnamefont {Li}}, \bibinfo {author} {\bibfnamefont {B.}~\bibnamefont {Fu}}, \bibinfo {author} {\bibfnamefont {C.}~\bibnamefont {Tang}}, \bibinfo {author} {\bibfnamefont {L.}~\bibnamefont {Wang}}, \bibinfo {author} {\bibfnamefont {Z.}~\bibnamefont {Li}}, \bibinfo {author} {\bibfnamefont {W.}~\bibnamefont {Fan}},  \emph {et~al.},\ }\href@noop {} {\bibfield  {journal} {\bibinfo  {journal} {Nature}\ }\textbf {\bibinfo {volume} {567}},\ \bibinfo {pages} {496} (\bibinfo {year} {2019})}\BibitemShut {NoStop}%
\bibitem [{\citenamefont {Li}\ and\ \citenamefont {Wu}(2017)}]{li2017binary}%
  \BibitemOpen
  \bibfield  {author} {\bibinfo {author} {\bibfnamefont {L.}~\bibnamefont {Li}}\ and\ \bibinfo {author} {\bibfnamefont {M.}~\bibnamefont {Wu}},\ }\href@noop {} {\bibfield  {journal} {\bibinfo  {journal} {ACS nano}\ }\textbf {\bibinfo {volume} {11}},\ \bibinfo {pages} {6382} (\bibinfo {year} {2017})}\BibitemShut {NoStop}%
\bibitem [{\citenamefont {Tong}\ \emph {et~al.}(2020)\citenamefont {Tong}, \citenamefont {Chen}, \citenamefont {Xiao}, \citenamefont {Yu},\ and\ \citenamefont {Yao}}]{tong2020interferences}%
  \BibitemOpen
  \bibfield  {author} {\bibinfo {author} {\bibfnamefont {Q.}~\bibnamefont {Tong}}, \bibinfo {author} {\bibfnamefont {M.}~\bibnamefont {Chen}}, \bibinfo {author} {\bibfnamefont {F.}~\bibnamefont {Xiao}}, \bibinfo {author} {\bibfnamefont {H.}~\bibnamefont {Yu}}, \ and\ \bibinfo {author} {\bibfnamefont {W.}~\bibnamefont {Yao}},\ }\href@noop {} {\bibfield  {journal} {\bibinfo  {journal} {2D Materials}\ }\textbf {\bibinfo {volume} {8}},\ \bibinfo {pages} {025007} (\bibinfo {year} {2020})}\BibitemShut {NoStop}%
\bibitem [{\citenamefont {Ferreira}\ \emph {et~al.}(2021)\citenamefont {Ferreira}, \citenamefont {Enaldiev}, \citenamefont {Fal’ko},\ and\ \citenamefont {Magorrian}}]{ferreira2021weak}%
  \BibitemOpen
  \bibfield  {author} {\bibinfo {author} {\bibfnamefont {F.}~\bibnamefont {Ferreira}}, \bibinfo {author} {\bibfnamefont {V.~V.}\ \bibnamefont {Enaldiev}}, \bibinfo {author} {\bibfnamefont {V.~I.}\ \bibnamefont {Fal’ko}}, \ and\ \bibinfo {author} {\bibfnamefont {S.~J.}\ \bibnamefont {Magorrian}},\ }\href@noop {} {\bibfield  {journal} {\bibinfo  {journal} {Scientific Reports}\ }\textbf {\bibinfo {volume} {11}},\ \bibinfo {pages} {13422} (\bibinfo {year} {2021})}\BibitemShut {NoStop}%
\bibitem [{\citenamefont {Kresse}\ and\ \citenamefont {Furthmüller}(1996)}]{KRESSE199615}%
  \BibitemOpen
  \bibfield  {author} {\bibinfo {author} {\bibfnamefont {G.}~\bibnamefont {Kresse}}\ and\ \bibinfo {author} {\bibfnamefont {J.}~\bibnamefont {Furthmüller}},\ }\href {\doibase https://doi.org/10.1016/0927-0256(96)00008-0} {\bibfield  {journal} {\bibinfo  {journal} {Computational Materials Science}\ }\textbf {\bibinfo {volume} {6}},\ \bibinfo {pages} {15} (\bibinfo {year} {1996})}\BibitemShut {NoStop}%
\bibitem [{\citenamefont {Kresse}\ and\ \citenamefont {Furthm\"uller}(1996)}]{PhysRevB.54.11169}%
  \BibitemOpen
  \bibfield  {author} {\bibinfo {author} {\bibfnamefont {G.}~\bibnamefont {Kresse}}\ and\ \bibinfo {author} {\bibfnamefont {J.}~\bibnamefont {Furthm\"uller}},\ }\href {\doibase 10.1103/PhysRevB.54.11169} {\bibfield  {journal} {\bibinfo  {journal} {Phys. Rev. B}\ }\textbf {\bibinfo {volume} {54}},\ \bibinfo {pages} {11169} (\bibinfo {year} {1996})}\BibitemShut {NoStop}%
\bibitem [{\citenamefont {Perdew}\ \emph {et~al.}(1996)\citenamefont {Perdew}, \citenamefont {Burke},\ and\ \citenamefont {Ernzerhof}}]{PhysRevLett.77.3865}%
  \BibitemOpen
  \bibfield  {author} {\bibinfo {author} {\bibfnamefont {J.~P.}\ \bibnamefont {Perdew}}, \bibinfo {author} {\bibfnamefont {K.}~\bibnamefont {Burke}}, \ and\ \bibinfo {author} {\bibfnamefont {M.}~\bibnamefont {Ernzerhof}},\ }\href {\doibase 10.1103/PhysRevLett.77.3865} {\bibfield  {journal} {\bibinfo  {journal} {Phys. Rev. Lett.}\ }\textbf {\bibinfo {volume} {77}},\ \bibinfo {pages} {3865} (\bibinfo {year} {1996})}\BibitemShut {NoStop}%
\bibitem [{\citenamefont {Grimme}\ \emph {et~al.}(2010)\citenamefont {Grimme}, \citenamefont {Antony}, \citenamefont {Ehrlich},\ and\ \citenamefont {Krieg}}]{10.1063/1.3382344}%
  \BibitemOpen
  \bibfield  {author} {\bibinfo {author} {\bibfnamefont {S.}~\bibnamefont {Grimme}}, \bibinfo {author} {\bibfnamefont {J.}~\bibnamefont {Antony}}, \bibinfo {author} {\bibfnamefont {S.}~\bibnamefont {Ehrlich}}, \ and\ \bibinfo {author} {\bibfnamefont {H.}~\bibnamefont {Krieg}},\ }\href {\doibase 10.1063/1.3382344} {\bibfield  {journal} {\bibinfo  {journal} {The Journal of Chemical Physics}\ }\textbf {\bibinfo {volume} {132}},\ \bibinfo {pages} {154104} (\bibinfo {year} {2010})},\ \Eprint {http://arxiv.org/abs/https://pubs.aip.org/aip/jcp/article-pdf/doi/10.1063/1.3382344/15684000/154104\_1\_online.pdf} {https://pubs.aip.org/aip/jcp/article-pdf/doi/10.1063/1.3382344/15684000/154104\_1\_online.pdf} \BibitemShut {NoStop}%
\bibitem [{\citenamefont {Steiner}\ \emph {et~al.}(2016)\citenamefont {Steiner}, \citenamefont {Khmelevskyi}, \citenamefont {Marsmann},\ and\ \citenamefont {Kresse}}]{PhysRevB.93.224425}%
  \BibitemOpen
  \bibfield  {author} {\bibinfo {author} {\bibfnamefont {S.}~\bibnamefont {Steiner}}, \bibinfo {author} {\bibfnamefont {S.}~\bibnamefont {Khmelevskyi}}, \bibinfo {author} {\bibfnamefont {M.}~\bibnamefont {Marsmann}}, \ and\ \bibinfo {author} {\bibfnamefont {G.}~\bibnamefont {Kresse}},\ }\href {\doibase 10.1103/PhysRevB.93.224425} {\bibfield  {journal} {\bibinfo  {journal} {Phys. Rev. B}\ }\textbf {\bibinfo {volume} {93}},\ \bibinfo {pages} {224425} (\bibinfo {year} {2016})}\BibitemShut {NoStop}%
\bibitem [{\citenamefont {Wu}\ \emph {et~al.}(2007)\citenamefont {Wu}, \citenamefont {Bergman}, \citenamefont {Balents},\ and\ \citenamefont {Das~Sarma}}]{PhysRevLett.99.070401}%
  \BibitemOpen
  \bibfield  {author} {\bibinfo {author} {\bibfnamefont {C.}~\bibnamefont {Wu}}, \bibinfo {author} {\bibfnamefont {D.}~\bibnamefont {Bergman}}, \bibinfo {author} {\bibfnamefont {L.}~\bibnamefont {Balents}}, \ and\ \bibinfo {author} {\bibfnamefont {S.}~\bibnamefont {Das~Sarma}},\ }\href {\doibase 10.1103/PhysRevLett.99.070401} {\bibfield  {journal} {\bibinfo  {journal} {Phys. Rev. Lett.}\ }\textbf {\bibinfo {volume} {99}},\ \bibinfo {pages} {070401} (\bibinfo {year} {2007})}\BibitemShut {NoStop}%
\bibitem [{\citenamefont {Wu}\ and\ \citenamefont {Das~Sarma}(2008)}]{PhysRevB.77.235107}%
  \BibitemOpen
  \bibfield  {author} {\bibinfo {author} {\bibfnamefont {C.}~\bibnamefont {Wu}}\ and\ \bibinfo {author} {\bibfnamefont {S.}~\bibnamefont {Das~Sarma}},\ }\href {\doibase 10.1103/PhysRevB.77.235107} {\bibfield  {journal} {\bibinfo  {journal} {Phys. Rev. B}\ }\textbf {\bibinfo {volume} {77}},\ \bibinfo {pages} {235107} (\bibinfo {year} {2008})}\BibitemShut {NoStop}%
\bibitem [{\citenamefont {Tamaki}\ \emph {et~al.}(2020)\citenamefont {Tamaki}, \citenamefont {Kawakami},\ and\ \citenamefont {Koshino}}]{PhysRevB.101.205311}%
  \BibitemOpen
  \bibfield  {author} {\bibinfo {author} {\bibfnamefont {G.}~\bibnamefont {Tamaki}}, \bibinfo {author} {\bibfnamefont {T.}~\bibnamefont {Kawakami}}, \ and\ \bibinfo {author} {\bibfnamefont {M.}~\bibnamefont {Koshino}},\ }\href {\doibase 10.1103/PhysRevB.101.205311} {\bibfield  {journal} {\bibinfo  {journal} {Phys. Rev. B}\ }\textbf {\bibinfo {volume} {101}},\ \bibinfo {pages} {205311} (\bibinfo {year} {2020})}\BibitemShut {NoStop}%
\end{thebibliography}%
%\printbibliography %Prints bibliography
\end{document}